\documentclass[a4paper, 12pt]{article}

\usepackage{tikz}
\usepackage{latexsym,amsmath,amsfonts,amssymb}
\usepackage[latin1]{inputenc}
\usepackage[american]{babel}
\usepackage{bbm}
\usepackage{hyperref}
\usepackage[nosort]{cite}

\renewcommand{\baselinestretch}{1.2} 

\newcommand{\ud}[2]{^{#1}_{\phantom{#1}#2}}
\newcommand{\du}[2]{_{#1}^{\phantom{#1}#2}}

\newcommand{\rep}[1]{\ensuremath{\mathbf{#1}}}

\newcommand{\wt}{\widetilde}
\newcommand{\wh}{\widehat}
\newcommand{\wb}{\overline}
\newcommand{\matht}[1]{\ensuremath{\boldsymbol{#1}}}

\newcommand{\eg}{\textit{e.g.}}

\newcommand{\ie}{\textit{i.e.}}

\numberwithin{equation}{section}

\newcommand{\Dslash}{D\!\!\!\!\slash\,}
\newcommand{\nn}{\nonumber}
\newcommand{\mat}[1]{\begin{pmatrix} #1 \end{pmatrix}}
\newcommand{\smat}[1]{\big( \begin{smallmatrix} #1 \end{smallmatrix} \big)}
\newcommand{\be}{\begin{equation}} \newcommand{\ee}{\end{equation}}
\newcommand{\bea}{\begin{equation} \begin{aligned}} \newcommand{\eea}{\end{aligned} \end{equation}}

\newcommand{\cB}{\mathcal{B}}
\newcommand{\cC}{\mathcal{C}}
\newcommand{\cD}{\mathcal{D}}

\newcommand{\cL}{\mathcal{L}}
\newcommand{\cM}{\mathcal{M}}
\newcommand{\cN}{\mathcal{N}}
\newcommand{\cO}{\mathcal{O}}
\newcommand{\cP}{\mathcal{P}}

\newcommand{\cT}{\mathcal{T}}

\newcommand{\cX}{\mathcal{X}}

\newcommand{\bC}{\mathbb{C}}

\newcommand{\bP}{\mathbb{P}}

\newcommand{\bR}{\mathbb{R}}

\newcommand{\bZ}{\mathbb{Z}}

\newcommand{\sT}{{\sf{T}}}
\newcommand{\unit}{\mathbbm{1}}

\DeclareMathOperator{\Tr}{Tr}

\begin{document}

\thispagestyle{empty}
\begin{flushright}
SISSA  57/2017/FISI
\end{flushright}
\vspace{10mm}
\begin{center}
{\huge  Three-dimensional dualities \\[.5em] with bosons and fermions} 
\\[15mm]
{Francesco Benini}\footnote{\it e-mail: fbenini@sissa.it}
\vskip 6mm
 
\bigskip
{\it
SISSA, via Bonomea 265, 34136 Trieste, Italy \\[.5em]
INFN, Sezione di Trieste, via Valerio 2, 34127 Trieste, Italy \\[.5em]
Institute for Advanced Study, Princeton, NJ 08540, USA
}
\vskip 6 mm

\bigskip
\bigskip

{\bf Abstract}\\[5mm]
{\parbox{14cm}{\hspace{5mm}

We propose new infinite families of non-supersymmetric IR dualities in three space-time dimensions, between Chern-Simons gauge theories (with classical gauge groups) with both scalars and fermions in the fundamental representation. In all cases we study the phase diagram as we vary two relevant couplings, finding interesting lines of phase transitions. In various cases the dualities lead to predictions about multi-critical fixed points and the emergence of IR quantum symmetries. For unitary groups we also discuss the coupling to background gauge fields and the map of simple monopole operators.
}
}
\end{center}
\newpage
\pagenumbering{arabic}
\setcounter{page}{1}
\setcounter{footnote}{0}
\renewcommand{\thefootnote}{\arabic{footnote}}

{\renewcommand{\baselinestretch}{1} \parskip=0pt
\setcounter{tocdepth}{2}
\tableofcontents}

\section{Introduction}

Infra-red (IR) dualities are the phenomenon by which two different quantum field theories (QFTs) describe the same physics at long distances. When non-trivial, dualities are an extremely powerful tool to understand the non-perturbative dynamics of QFTs. For instance, one theory could flow to strong coupling in the IR, while the other could be weakly coupled or even IR-free: in this case the latter solves the IR physics of the former. When both theories flow to the same interacting conformal field theory (CFT), dualities realize the idea of universality. In this case, one QFT could develop quantum symmetries at long distances because of strong coupling, and such emergent symmetries could be revealed by the second QFT in which they are manifest at all energies.

Dualities are familiar in two space-time dimensions, and are abundant among supersymmetric (SUSY) theories in two, three and four dimensions. On the other hand, dualities become rare without supersymmetry in more than two space-time dimensions (not because they do not exist, but rather because they are difficult to find and to corroborate).

However, the state of the art in three space-time dimensions has drastically changed in the last few years. A convergence of ideas from the condensed matter literature (\eg{}  \cite{Peskin:1977kp, Dasgupta:1981zz, Barkeshli:2014ida, Son:2015xqa, Wang:2015qmt, Potter:2015cdn, Wang:2016gqj}), the study of Chern-Simons-matter theories in the large $N$ limit (\eg{} \cite{Aharony:2011jz, Giombi:2011kc, Aharony:2012nh, Jain:2013gza}), the bulk of knowledge about SUSY dualities (\eg{} \cite{Intriligator:1996ex, deBoer:1996mp, Aharony:1997bx, Aharony:1997gp, Giveon:2008zn, Benini:2011mf}) and the careful analysis of Abelian Chern-Simons-matter theories \cite{Karch:2016sxi, Murugan:2016zal, Seiberg:2016gmd}, has led to the proposal of infinite families of non-supersymmetric dualities \cite{Aharony:2015mjs, Hsin:2016blu, Metlitski:2016dht, Aharony:2016jvv, Komargodski:2017keh} between Chern-Simons (CS) gauge theories coupled to matter fields in the fundamental representation---bosonic on one side and fermionic on the other side. For that reason they are sometimes called ``bosonization dualities''. Various other dualities have been found as well, including multiple gauge groups \cite{Karch:2016aux, Jensen:2017dso} or matter in other representations \cite{Gomis:2017ixy}. Other works elaborating on the dualities are \cite{Gur-Ari:2015pca, Kachru:2016aon, Benini:2017dus, Wang:2017txt, Chen:2017lkr, Gaiotto:2017tne, Jensen:2017xbs, Armoni:2017jkl, Cordova:2017vab, Aharony:toappear}.

In this paper we propose new non-supersymmetric IR dualities in three space-time dimensions, between Chern-Simons-matter theories with both scalar and fermionic matter fields in the fundamental representation.%
\footnote{Dualities between CS-matter theories with a single fundamental scalar and fermion in the large $N$ limit were proposed and analyzed in \cite{Jain:2013gza}.}
Succinctly, we propose the following dualities:
\bea
\label{summary dualities I}
SU(N)_{k-\frac{N_f}2} \text{ with $N_s$ $\phi$, $N_f$ $\psi$} &\qquad\longleftrightarrow\qquad U(k)_{-N + \frac{N_s}2} \text{ with $N_f$ $\phi$, $N_s$ $\psi$} \\
USp(2N)_{k-\frac{N_f}2} \text{ with $N_s$ $\phi$, $N_f$ $\psi$} &\qquad\longleftrightarrow\qquad USp(2k)_{-N + \frac{N_s}2} \text{ with $N_f$ $\phi$, $N_s$ $\psi$} \\
SO(N)_{k-\frac{N_f}2} \text{ with $N_s$ $\phi$, $N_f$ $\psi$} &\qquad\longleftrightarrow\qquad SO(k)_{-N + \frac{N_s}2} \text{ with $N_f$ $\phi$, $N_s$ $\psi$}
\eea
as well as
\be
\label{summary dualities II}
U(N)_{k-\frac{N_f}2, k-\frac{N_f}2 \pm N} \text{ with $N_s$ $\phi$, $N_f$ $\psi$} \qquad\longleftrightarrow\qquad U(k)_{-N + \frac{N_s}2, -N + \frac{N_s}2 \mp k} \text{ with $N_f$ $\phi$, $N_s$ $\psi$} \;.
\ee
We propose the $SU/U$ and $U/U$ dualities in the range of parameters $N_s \leq N$, $N_f \leq k$ and $(N_s, N_f) \neq (N,k)$; the $USp$ dualities in the range $N_s \leq N$, $N_f \leq k$; the $SO$ dualities in the range $N_s \leq N$, $N_f \leq k$ and $N_s + N_f + 3 \leq N+k$. Although we have not analyzed this point in full details, it appears that these ranges can be extended, along the lines of \cite{Komargodski:2017keh}, by invoking quantum phases with a condensate of the fermion bilinear and spontaneous symmetry breaking.

Let us explain our notation in (\ref{summary dualities I})-(\ref{summary dualities II}). We indicate the gauge group and in subscript its CS level. The latter gets contribution from the bare CS level in the Lagrangian, which is always integer, and the regularization of the fermion determinant. As in \cite{Seiberg:2016rsg}, we write the Lagrangian of a complex Dirac fermion $\psi$ coupled to a $U(1)$ gauge field $A$ as
\be
i \bar\psi \Dslash_A \psi
\ee
and use a regularization of the fermion determinant%
\footnote{Specifically, the regularized fermion determinant is the exponential of the eta invariant, see the very clear exposition in \cite{Witten:2015aba}. On the other hand, we implicitly use a Yang-Mills regulator for the gauge sector.}
such that, when integrating out the fermion with a positive mass we are left with a vanishing Lagrangian, while negative mass leads to a CS term at level $-1$ (as well as a gravitational CS term, defined in Appendix~\ref{app: level-rank dualities}):
\be
- \frac1{4\pi} AdA -2 \text{CS}_g \;.
\ee
Thus, the bare CS level is $k$ on the left-hand-side (LHS) and $-N+N_s$ on the right-hand-side (RHS) of (\ref{summary dualities I}). In (\ref{summary dualities II}) the first and second subscripts refer to the CS levels for the $SU$ and $U(1)$ part of the gauge group---see (\ref{U_k1,k2 Lagrangian})---and $U(N)_k \equiv U(N)_{k,k}$.

We indicate scalar fields as $\phi$ and fermionic fields as $\psi$; in all cases they transform in the fundamental representation of the gauge group, which is complex $N$-dimensional for $SU(N)$ and $U(N)$, pseudo-real $2N$-dimensional for $USp(2N)$ and real $N$-dimensional for $SO(N)$. On the LHS there are $N_s$ scalars and $N_f$ fermions,%
\footnote{According to the property of the gauge representation, we count Dirac fermions for $SU$, $U$ and $USp$, and Majorana fermions for $SO$.}
while the opposite is true on the RHS. The theories include all relevant couplings that are compatible with the global symmetries preserved by the gauging (specified in the corresponding Sections), in particular they include quartic scalar couplings as well as mixed couplings quadratic both in the scalars and in the fermions.

There are two obvious quadratic relevant deformations that are compatible with all symmetries: a ``diagonal'' mass term for all scalars, schematically $m_\phi^2 |\phi|^2$, and a diagonal mass term for all fermions, $m_\psi \bar\psi \psi$. For generic values of the masses the theories are either completely gapped, possibly with topological order described by a topological quantum field theory (TQFT), or can develop Goldstone bosons. As we tune the couplings we find lines with interesting phase transitions, that we can conjecture be described by conformal field theories (CFTs). Those lines will meet at one or more multi-critical fixed points. On the other hand, the transitions could be first order instead of second order---then the dualities are less interesting. This, however, would not change much our discussion. A schematic structure of the phase diagrams is in Figure \ref{fig: masks}. There, a grey area covers the deep quantum region of the phase diagram where it is hard to understand the detailed structure;%
\footnote{For instance, when four lines come together, one could expect two tri-critical points connected by an intermediate transition line. Given the structure of the phase diagram presented in this paper, one could try to identify those tri-critical points. On the other hand, one could envision the possibility that the required two tri-critical points and intermediate gapless line do not exist, and thus the four lines are forced to meet at a single point. Or first-order transitions could be involved. We leave this question for future work.}
only in a few cases we will be able to make sharp predictions.

The conjectured dualities have very interesting implications. In various cases they predict the emergence of time-reversal and parity invariance quantum-mechanically in the IR, or the emergence of other internal global symmetries (assuming the transitions are second order). In some cases the dualities predict that the IR physics decouples into two or more separate CFTs (typically a Wilson-Fisher fixed point and some free fermions).

We subject the dualities to various checks. We study their consistency under massive deformations, and verify that they reduce to the dualities with a single matter species \cite{Aharony:2015mjs, Hsin:2016blu, Aharony:2016jvv} and to the level-rank dualities of spin-Chern-Simons theories. We couple the theories to background gauge fields and keep into account their counterterms---as well as the counterterm for the gravitational field. This allows us to gauge part of the global symmetry and generate new dualities, as well as to test the proposed ones. We spell out the map of the simplest monopole operators in the unitary case.

In the last Section we derive new Abelian dualities combining the dualities in \cite{Seiberg:2016gmd}. We find the following:
\bea
U(1)_0 \stackrel{\rule{0pt}{.7em}^\text{\Large$\curvearrowright$}}{\text{ with 2 $\phi$ and $V_\text{EP}$}} &\qquad\longleftrightarrow\qquad \stackrel{\rule{0pt}{.7em}^\text{\Large$\curvearrowright$}}{U(1)_0 \text{ with 2 $\psi$}} \\
U(1)_1 \text{ with 2 $\phi$ and $V_\text{EP}$} &\qquad\longleftrightarrow\qquad \stackrel{\rule{0pt}{.7em}^\text{\Large$\curvearrowright$}}{U(1)_{-\frac12} \text{ with $\phi$, $\psi$}} \\[.3em]
U(1)_2 \text{ with 2 $\phi$ and $V_\text{EP}$} &\qquad\longleftrightarrow\qquad U(1)_{-1} \text{ with 2 $\psi$} \\[.6em]
U(1)_\frac32 \text{ with $\phi$, $\psi$} &\qquad\longleftrightarrow\qquad U(1)_{-\frac32} \text{ with $\phi$, $\psi$}
\eea
where $V_\text{EP}$ is an ``easy plane'' quartic scalar potential that further breaks the global symmetry. The circular arrows indicate a self-duality.
The dualities in the first line were already reported in \cite{Motrunich:2003fz, Xu:2015lxa, Karch:2016sxi, Hsin:2016blu, Benini:2017dus, Wang:2017txt}. More details are in Section \ref{sec: Abelian dualities}.

The paper is organized as follows. In Sections \ref{sec: SU/U duality}, \ref{sec: U/U duality}, \ref{sec: USp duality} and \ref{sec: SO duality} we present the $SU/U$, $U/U$, $USp$ and $SO$ dualities, respectively. We describe the faithful global symmetry, the couplings, the phase diagram, the coupling to background fields and the map of monopole operators. We also give some simple examples in each case. In Section \ref{sec: Abelian dualities} we present new Abelian dualities. We conclude in Section \ref{sec: conclusions}. In Appendix \ref{app: summary} we summarize the dualities with a single matter species, while in Appendices \ref{app: charge conjugation} and \ref{app: spinc} we give more details on our notation.


\section{\matht{SU/U} duality}
\label{sec: SU/U duality}

The first duality we consider involves Chern-Simons gauge theories with unitary and special unitary groups, as well as bosonic and fermionic matter in the fundamental representation, which is complex. We propose the following duality:
\be
\label{SU/U duality grav}
SU(N)_{k - \frac{N_f}2} \text{ with $N_s$ $\phi$, $N_f$ $\psi$} \quad\longleftrightarrow\quad U(k)_{-N + \frac{N_s}2}\text{ with $N_f$ $\phi$, $N_s$ $\psi$ }  \times U\big( k(N-N_s) \big)_1
\ee
for
\be
N \geq N_s \;,\qquad k \geq N_f \;,\qquad (N,k) \neq (N_s, N_f) \;.
\ee
We indicate scalar fields as $\phi$ and fermionic fields as $\psi$, and in this case they are both complex. Thus the theory on the LHS has $N_s$ scalars and $N_f$ fermions in the fundamental representation, while the theory on the RHS has $N_f$ scalars and $N_s$ fermions. On the RHS, $U\big(k(N - N_s)\big)_1$ is a trivial spin-TQFT \cite{Seiberg:2016gmd} not coupled to matter, which represents the gravitational coupling $-2k(N-N_s) \text{CS}_g$ (see Appendix \ref{app: level-rank dualities} for our conventions).

This proposal reproduces the boson/fermion dualities of \cite{Aharony:2015mjs, Hsin:2016blu} for $N_s=0$ or \mbox{$N_f=0$}, \ie{} when we take a single matter species, as well as the level-rank dualities when $N_s = N_f = 0$. We summarize those dualities, for reference, in Appendix \ref{app: summary}. Our proposal also agrees with \cite{Jain:2013gza} where the case $N_s = N_f =1$ was studied in the large $N,k$ limit. In the following we will assume $N_s,N_f\geq 1$.

On both sides of (\ref{SU/U duality grav}) there is a manifest global symmetry $SU(N_s) \times SU(N_f) \times U(1)^2 \rtimes \bZ_2^\cC$: each $SU$ factor acts on one matter species, one $U(1)$ acts anti-diagonally on scalars and fermions, the other $U(1)$ is baryonic on the LHS and magnetic on the RHS, while $\bZ_2^\cC$ is charge conjugation (see Appendix \ref{app: charge conjugation} for our notation). We will be more precise in Section~\ref{sec: faithful symmetry} and show that the symmetry that acts faithfully on gauge-invariant operators is in fact
\be
\label{faithful symmetry SU/U}
G = \frac{U(N_s) \times U(N_f)}{\bZ_N} \rtimes \bZ_2^\cC \;.
\ee
The case of gauge group $SU(2) \cong USp(2)$ is special and is analyzed in detail in Section~\ref{sec: case of SU(2)}, however (\ref{faithful symmetry SU/U}) is still true. On both sides of (\ref{SU/U duality grav}) we include all gauge-invariant relevant operators compatible with those symmetries. Let us list them in the $SU$ theory. First, there are the quadratic mass terms
\be
\label{mass terms}
|\phi|^2 = \phi^\dag_{\alpha I} \phi^{\alpha I} \;,\qquad\qquad \bar\psi\psi = \bar\psi_{\alpha B} \psi^{\alpha B}
\ee
where $\alpha=1,\dots,N$ is in the fundamental of $SU(N)$, $I=1,\dots, N_s$ is in the fundamental of $SU(N_s)$ and $B=1,\dots, N_f$ is in the fundamental of $SU(N_f)$. Then there are the quartic scalar couplings
\be
\label{quartic scalar terms}
\big( \phi^\dag_{\alpha I} \phi^{\alpha I} \big)^2 \;,\qquad\qquad \phi^\dag_{\alpha I} \phi^{\alpha J} \phi^\dag_{\beta J} \phi^{\beta I} \;.
\ee
The fermionic quartic couplings are irrelevant in the UV, and we will assume that they remain such in the IR.

Finally, there are mixed scalar-fermion quartic couplings:
\be
\label{def mixed couplings}
\cO_\text{d} = \big( \phi^\dag_{\alpha I} \phi^{\alpha I} \big) \big( \bar\psi_{\beta A} \psi^{\beta A} \big) \;,\qquad\qquad \cO_\text{m} = \phi^\dag_{\alpha I} \phi^{\beta I} \bar\psi_{
\beta A} \psi^{\alpha A} \;.
\ee
Both $\cO_\text{d}$ and $\cO_\text{m}$ are marginal at the UV free fixed point. In the 't~Hooft large $N$ and $k$ limit (with $N/k$ fixed), the operator $\cO_\text{d}$ gets a large IR anomalous dimension $\Delta_\text{d} = 4 + \cO\big( \frac 1N \big)$ in the critical boson theory, namely in the scalar theory deformed by the first (and possibly the second) operator in (\ref{quartic scalar terms}) and with a single tuning to set the scalar mass to zero. The presence of the fermions and of gauge interactions does not change this conclusion. Hence this operator is irrelevant in the IR. On the contrary, $\cO_\text{m}$ does not get anomalous dimension at leading order in $N$, $\Delta_\text{m} = 3 + \cO\big( \frac 1N \big)$, in the critical boson and regular fermion theory (with both scalar and fermion masses tuned to zero). In the following we will assume that $\cO_\text{m}$ is present in the IR (at least when it exists as an operator independent from $\cO_\text{d}$): as we will see, its presence in the two theories (with a very specific sign for its coefficient) is crucial for the duality to work. On the contrary, even though we expect $\cO_d$ to be marginally irrelevant in the UV, its presence in the theories would not modify our discussion and so we will not make assumptions about it.

The operators $\cO_\text{d}$ and $\cO_\text{m}$ also behave differently when $\phi$ gets a vacuum expectation value (VEV). Indeed $\cO_\text{d}$ gives a uniform mass to all fermions in the theory, while $\cO_\text{m}$ only gives mass to those components that are not charged under the unbroken gauge group (but uniformly across the flavors).

Let us mention that also sextic scalar couplings, schematically $|\phi|^6$, similarly to $\cO_\text{d}$ are marginal in the UV free theory but are expected to be irrelevant and not to modify the discussion in the IR (at least as long as their coefficients are positive) because we are not tuning the quartic couplings.%
\footnote{See \cite{Aharony:toappear} for a large $N$ analysis of CS-matter theories where the quartic scalar coupling is tuned.}

For generic values of the parameters, both theories in (\ref{SU/U duality grav}) are either completely gapped or reduce to Goldstone bosons. We can study the phase diagram as we vary the mass terms for the couplings $m_\phi^2 |\phi|^2$ and $m_\psi \bar\psi \psi$ in (\ref{mass terms}). Along lines where one function of $m_\phi^2$ and $m_\psi$ is tuned, we reproduce the Chern-Simons gauge theories with one matter species, either scalars or fermions, involved in the dualities of \cite{Aharony:2015mjs, Hsin:2016blu} and conjectured to have a non-trivial IR fixed point. Those lines correspond to tuning either the IR scalar or fermion mass to zero, respectively. Classically (or in the 't~Hooft large $N$ limit \cite{Jain:2013gza}) those lines meet at a multicritical IR fixed point, that we indicate as $m_\phi^2 = m_\psi =0$. In the full quantum theory, we do not know whether all gapless lines meet at a single point, or whether they form a more intricate net---possibly involving first-order transitions. Only in a few cases we will find indications of the first scenario. Leaving such a central region aside, we will study the phase diagrams in detail below.

\subsection{The faithful global symmetry}
\label{sec: faithful symmetry}

Let us find the symmetry group $G$ that acts faithfully on gauge-invariant operators, for both theories in (\ref{SU/U duality grav}). This analysis will be independent of the duality, and valid for all values of $N,k,N_s, N_f$.

Consider first the theory on the LHS of (\ref{SU/U duality grav}). The faithfully-acting symmetry is
\be
\label{faithful symmetry LHS}
SU(N)_{k - \frac{N_f}2} \text{ with } N_s\, \phi,\, N_f\, \psi : \qquad\qquad G = \frac{U(N_s) \times U(N_f)}{\bZ_N} \rtimes \bZ_2^\cC \;.
\ee
The $\bZ_N$ quotient is generated by $\big( e^{2\pi i /N} \unit, e^{2\pi i /N} \unit \big)$, and it corresponds to the center of $SU(N)$. Then $\bZ_2^\cC$ is charge conjugation (see Appendix \ref{app: charge conjugation} for details). When $k = \frac{N_f}2 \in \bZ$ the theory has also time-reversal invariance $\bZ_2^\cT$. The case $N=2$ deserves more attention, and is treated in Section~\ref{sec: case of SU(2)}, however the conclusion is the same.

Next consider the theory on the RHS: $U(k)_{-N + \frac{N_s}2} \text{ with } N_f\, \phi,\, N_s\, \psi$. The theory has charge conjugation symmetry $\bZ_2^\cC$ (for $N = \frac{N_s}2 \in \bZ$ it also has $\bZ_2^\cT$ time-reversal invariance), so let us write
\be
U(k)_{-N + \frac{N_s}2} \text{ with } N_f\, \phi,\, N_s\, \psi: \qquad\qquad G = \wh G \rtimes \bZ_2^\cC \;. \qquad\qquad
\ee
The theory has a magnetic symmetry $U(1)_M$ and the bare CS level for the gauge group is $N_s - N$, therefore monopole operators of magnetic charge $1$ have charge $(N_s - N)$ under the diagonal $U(1) \subset U(k)$. Since fundamentals have charge 1 under that $U(1)$, the symmetry group is
\be
\label{symmetry RHS}
\wh G = \frac{U(N_f) \times U(N_s) \times U(1)_M}{U(1)_*} \qquad\qquad U(1)_* = \big( e^{2\pi i \alpha}, e^{2\pi i \alpha}, e^{2\pi i (N_s-N)\alpha} \big)
\ee
with $\alpha \in [0,1)$. For $N_s \neq N$ we can use $U(1)_*$ to remove $U(1)_M$. Thus we can write
\be
\wh G = \frac{U(N_f) \times U(N_s)}{\bZ_{| N-N_s |}} \quad\text{for } N_s \neq N \;,\qquad \wh G = \frac{U(N_f) \times U(N_s)}{U(1)} \times U(1)_M \quad\text{for } N_s = N
\ee
where in the second expression the quotient is by the diagonal $U(1)$.

To compare with the symmetry (\ref{faithful symmetry LHS}) of the theory on the LHS, we notice that there is an isomorphism
\be
\label{symmetry isomorphism I}
\frac{U(N_f) \times U(N_s)}{\bZ_{|n|}} \cong \frac{U(N_f) \times U(N_s)}{\bZ_{|n+N_f|}} \cong \frac{U(N_f) \times U(N_s)}{\bZ_{|n+N_s|}}
\ee
for $n\in\bZ$, where each expression is valid when the order of the group in the denominator is not zero. To exhibit the isomorphism we parametrize $\big( U(N_f) \times U(N_s) \big)/\bZ_{|n|}$ as
\be
\Big( g \in SU(N_f) ,\; u \in U(1),\; h \in SU(N_s),\; w \in U(1) \Big)
\ee
with the identifications
\bea
\label{old identifications}
(g,u,h,w) & \;\sim\; \Big( e^{2\pi i/N_f} g,\; e^{-2\pi i /N_f} u,\; h,\; w \Big) \;\sim\; \Big( g,\; u,\; e^{2\pi i/N_s} h,\; e^{-2\pi i /N_s} w \Big) \\
& \;\sim\; \Big( g,\; e^{2\pi i/n} u,\; h,\; e^{2\pi i /n} w \Big) \;.
\eea
The isomorphism is given by
\be
(u,w) \;\mapsto\; \Big( \tilde u = u^{\frac{n}{n+N_f}},\; \tilde w = w \, u^{-\frac{N_f}{n+N_f}} \Big)
\ee
which is well-defined thanks to the identifications. It maps (\ref{old identifications}) to the identifications for $\big( U(N_f) \times U(N_s) \big)/\bZ_{|n+N_f|}$. If $n=N_f$, the identifications (\ref{old identifications}) can be reorganized such that $u$ describes $U(1)/\bZ_{N_f} \cong U(1)'$, while $\big( g,h,t = \frac wu \big)$ describe $\big( U(N_f) \times U(N_s) \big)/U(1)$. Thus we also have the isomorphism
\be
\label{symmetry isomorphism II}
\frac{U(N_f) \times U(N_s)}{U(1)} \times U(1)' \cong \frac{U(N_f) \times U(N_s)}{\bZ_{N_f}} \cong \frac{U(N_f) \times U(N_s)}{\bZ_{N_s}} \;.
\ee
Using the two isomorphisms, the symmetries agree on the LHS and RHS of (\ref{SU/U duality grav}).

More directly, we can start from (\ref{symmetry RHS}) and rewrite $U(N_f) \cong \big( U(1) \times SU(N_f) \big)/\bZ_{N_f}$ and similarly for $U(N_s)$. We can use $U(1)_*$ to remove the $U(1)$ inside $U(N_s)$. Then we use an $N$-fold multiple cover $U(1)_B$ of $U(1)_M$, meaning that there is a projection map $\pi:U(1)_B \to U(1)_M$ that maps $e^{i\beta} \to e^{iN\beta}$, and we can write $U(1)_M = U(1)_B/\bZ_N$. This is natural from the point of view of the duality, because the monopole of charge 1 in the RHS is mapped to a ``baryon'' of charge $N$ in the LHS. We obtain
\bea
\wh G &= \frac{U(1) \times SU(N_f) \times SU(N_s) \times U(1)_B}{\bZ_{N_f} \times \bZ_{N_s} \times \bZ_N} \qquad & \bZ_{N_f}: & \big( e^{2\pi i /N_f}, e^{-2\pi i /N_f}, 1, 1 \big) \\
&& \bZ_{N_s}: & \big( e^{-2\pi i /N_s}, 1,e^{-2\pi i / N_s}, e^{2\pi i /N_s} \big) \\
&& \bZ_N : & \big( 1,1,1,e^{2\pi i /N} \big)
\eea
where we have indicated the generators of the quotient groups. We parametrize $U(1) \times U(1)_B$ as $\big( e^{2\pi i \gamma}, e^{2\pi i \beta} \big)$ and change coordinates to $U(1)'' \times U(1)_B = \big( e^{2\pi i (\gamma+\beta)}, e^{2\pi i \beta} \big)$. We find \mbox{$\wh G = \frac{U(1)'' \times SU(N_f) \times SU(N_s) \times U(1)_B}{\bZ_{N_f} \times \bZ_{N_s} \times \bZ_N}$} and the quotient is generated by \mbox{$\bZ_{N_f}: \big( e^{2\pi i /N_f}, e^{-2\pi i /N_f}, 1, 1 \big)$}, \mbox{$\bZ_{N_s}: \big( 1, 1,e^{-2\pi i / N_s}, e^{2\pi i /N_s} \big)$} and \mbox{$\bZ_N : \big( e^{2\pi i /N} ,1,1,e^{2\pi i /N} \big)$}. Finally we use $\bZ_{N_f} \times \bZ_{N_s}$ to form $U(N_f) \times U(N_s)$ and recover
\be
\wh G = \frac{U(N_f) \times U(N_s)}{\bZ_N} \qquad\qquad \bZ_N : \big( e^{2\pi i /N}, e^{2\pi i /N} \big)
\ee
as on the LHS of (\ref{SU/U duality grav}).

For $N_s=0$ or $N_f=0$ the analysis here reproduces the result in \cite{Benini:2017dus}.

\subsubsection[The case of $SU(2)$]{The case of \matht{SU(2)}}
\label{sec: case of SU(2)}

The case of $SU(2)$ gauge group deserves more attention, because $SU(2) \cong USp(2)$. Here we neglect time-reversal symmetry, which is preserved if and only if the CS level is zero.

For $N_s=0$ there are only fermions with no potential. Thus the symmetry of $SU(2)_{k - \frac{N_f}2}$ with $N_f$ $\psi$ is $G = USp(2N_f)/\bZ_2$, as manifest in the $USp$ description (Section~\ref{sec: USp duality}).

For $N_f=0$ there are only scalars with a potential. For $N_s=1$ there is only one gauge-invariant quartic coupling we can write, $(\phi_\alpha^\dag \phi^\alpha)^2$, and it preserves $G = USp(2)/\bZ_2 \cong SO(3)$.

For $N_s>1$ we write the quadratic gauge invariant
\be
\cO = \sum_{\alpha=1}^2 \sum_{I=1}^{N_s} |\phi^{\alpha I}|^2 \;.
\ee
In the $USp$ notation, we introduce $\Phi_{\alpha i}$ with $i=1,\dots, 2N_s$ and subject to $\Phi_{\alpha i} \epsilon^{\alpha\beta} \Omega^{ij} = \Phi^*_{\beta j}$, where $\Omega^{ij}$ is the $USp(2N_s)$ invariant tensor. We can set $\phi^{\alpha I} = \Phi_{\alpha I}$ for $I=1,\dots, N_s$ and use the constraint to fix the other components of $\Phi$. Then we define $\cM_{ij} = \Phi_{\alpha i} \Phi_{\beta j} \epsilon^{\alpha\beta}$, and it follows that $\cO = - \frac12 \Tr \cM\Omega$. Since the gauge group $SU(2) \cong USp(2)$ has only rank 1, it is easy to prove that
\be
\Tr \cM \Omega \cM \Omega = \frac12 \big( \Tr \cM \Omega \big)^2
\ee
for all $N_s \geq 1$. Define the matrix $\cN_{\alpha\beta} = \Phi_{\alpha i} \Phi_{\beta j} \Omega^{ij}$. This is a $2\times 2$ antisymmetric matrix, thus it must be $\cN_{\alpha\beta} = - \frac12 (\Tr \cM\Omega) \epsilon_{\alpha\beta}$. The formula follows. Thus, if we only include the quartic potential
\be
V = \cO^2 = \frac14 \big( \Tr \cM \Omega \big)^2 = \frac12 \Tr \cM\Omega \cM \Omega
\ee
the theory preserves $G = USp(2N_s)/\bZ_2$ symmetry. There are no other quartic couplings we can write that preserve this symmetry.

However, for $N_s>1$ there is another coupling that preserves only \mbox{$G = \big( U(N_s) / \bZ_2\big) \rtimes \bZ_2^\cC$}, namely
\be
\label{extra SU2 scalar coupling}
\phi^{\alpha I} \phi^\dag_{\alpha J} \, \phi^{\beta J} \phi^\dag_{\beta I} \qquad\text{ or }\qquad \phi^{\alpha I} \phi^{\beta J} \epsilon_{\alpha\beta} \, \phi^\dag_{\gamma J} \phi^\dag_{\delta I} \epsilon^{\gamma\delta}
\ee
in $SU(2)$ notation. The two couplings above satisfy a linear relation. Define $\cP\ud{\beta}{\gamma} = \phi^{\beta J} \phi^\dag_{\gamma J}$: this is a $2\times 2$ Hermitian matrix that can be decomposed as $\cP = c_0 \unit + c_n \sigma_n$, where \mbox{$n=1,2,3$} indicates the three Pauli matrices. Rewriting the couplings in terms of $\cP$ one finds $\phi^{\alpha I} \phi^\dag_{\alpha J} \phi^{\beta J} \phi^\dag_{\beta I} - \phi^{\alpha I} \phi^{\beta J} \epsilon_{\alpha\beta} \phi^\dag_{\gamma J} \phi^\dag_{\delta I} \epsilon^{\gamma\delta} = \cO^2$. The coupling (\ref{extra SU2 scalar coupling}) is present on the LHS of (\ref{SU/U duality grav}), therefore there is no enhanced symmetry for $N_s>1$.

For $N_s,N_f\geq 1$ there is the mixed coupling
\be
\cO_\text{d} = \big(\phi^\dag_{\alpha I}  \phi^{\alpha I} \big)\big( \bar\psi_{\beta J} \psi^{\beta J} \big) = \frac14 \big( \Tr \cM\Omega\big) \, \big( \Tr \cM^\Psi \Omega \big) = \frac12 \Phi_{\alpha i} \Phi_{\beta j} \Omega^{ij} \Psi_{\gamma x} \Psi_{\delta y} \Omega^{xy} \epsilon^{\beta\gamma} \epsilon^{\delta\alpha}
\ee
that preserves $G = \big( USp(2N_s) \times USp(2N_f) \big)/\bZ_2$. Here $\cM^\Psi$ is the gauge-invariant fermion bilinear. The identity follows from the same argument as above, using $\cN$ and $\cN^\Psi$. There is another coupling that preserves only $G = \big( U(N_s) \times U(N_f) \big)/\bZ_2 \rtimes \bZ_2^\cC$, namely
\be
\cO_\text{m} = \phi^\dag_{\alpha I} \phi^{\beta I} \bar\psi_{\beta X} \psi^{\alpha X} \;.
\ee
This coupling is independent from $\cO_\text{d}$ even for $N_s = N_f = 1$ (while, as before, the coupling $\phi^\dag_{\alpha I} \phi^{\gamma I} \bar\psi_{\beta X} \psi^{\delta X} \epsilon^{\alpha\beta} \epsilon_{\gamma\delta}$ is not independent). Since $\cO_\text{m}$ is present on the LHS of (\ref{SU/U duality grav}), there is no enhanced symmetry for $N_s, N_f\geq 1$ with respect to (\ref{faithful symmetry LHS}).

\subsection{Phase diagram}

We can study relevant deformations of the two theories in (\ref{SU/U duality grav}) that preserve the full symmetry $G = \big( U(N_s) \times U(N_f) \big)/\bZ_N \rtimes \bZ_2^\cC$. They are described by the operators
\be
\label{mass deformations}
m_\phi^2 |\phi|^2 \qquad\text{ and }\qquad m_\psi \bar\psi\psi \;.
\ee
Notice that, in the absence of time-reversal symmetry, the scalar and fermion mass can mix. Our analysis will be classical, therefore valid for large values of the masses compared with the Yang-Mills regulator $g_\text{YM}^2$. As we commented above, we do not know the detailed structure of the phase diagram in the vicinity of the origin $m_\phi^2 = m_\psi = 0$. Nonetheless, we find consistent results with no need to invoke new quantum phases (possibly triggered by spontaneous symmetry breaking) around the origin. (See \eg{} \cite{Komargodski:2017keh, Gomis:2017ixy} for examples where the appearance of quantum phases has been argued, and it is crucial for the dualities to work.)

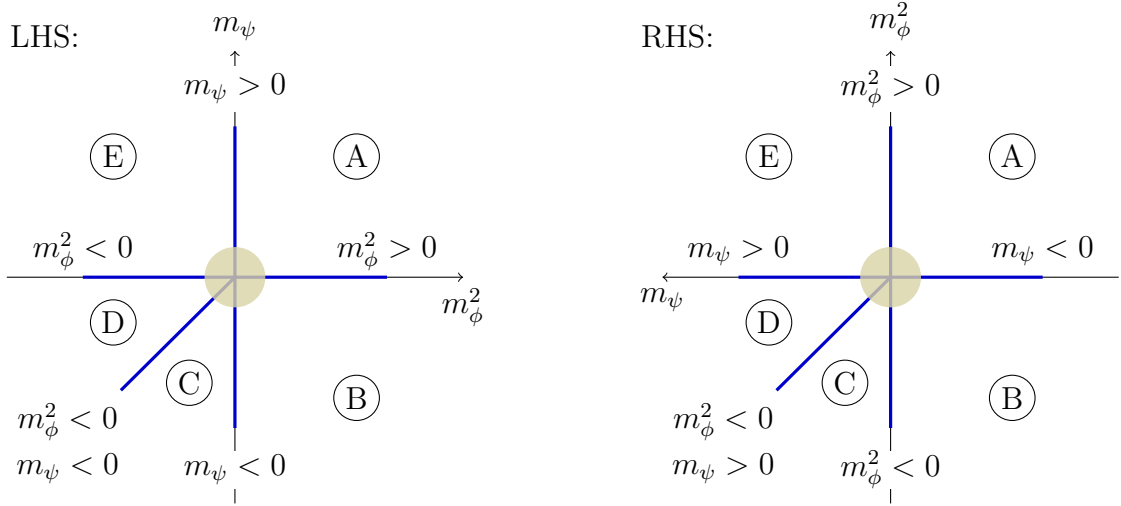
\begin{figure}[t]
\centering
\hspace{\stretch{1}}
\begin{tikzpicture}[scale=1]
\node at (-2.5,3.2) {LHS:};
\draw [->] (-3,0)--(3,0) node[below] {$m_\phi^2$}; \draw [->] (0,-3)--(0,3) node[above] {$m_\psi$};
\draw [very thick, blue!80!black] (0,0)--(2,0); \draw [very thick, blue!80!black] (0,-2)--(0,2); \draw [very thick, blue!80!black] (-1.5,-1.5)--(0,0); \draw [very thick, blue!80!black] (0,0)--(-2,0);
\fill [lightgray!70!yellow, fill opacity=.8] (0,0) circle [radius=.4];
\fill [white]  (-.5,2.2) rectangle (.5,2.8) node[midway, black] {$m_\psi>0$};
\node at (2,.35) {$m_\phi^2>0$};
\fill [white] (-.5, -2.3) rectangle (.5,-2.8) node[midway, black] {$m_\psi<0$};
\node at (-2, .35) {$m_\phi^2<0$};
\node [align=center] at (-2.2, -2.2) {$m_\phi^2<0$ \\ $m_\psi<0$};
\draw (1.6,1.6) circle [radius=.3] node {A}; \draw (1.6,-1.6) circle [radius=.3] node {B};
\draw (-.6,-1.4) circle [radius=.3] node {C}; \draw (-1.6,-.6) circle [radius=.3] node {D};
\draw (-1.6,1.6) circle [radius=.3] node {E};
\end{tikzpicture}
\hspace{\stretch{2}}
\begin{tikzpicture}[scale=1]
\node at (-2.8,3.2) {RHS:};
\draw [->] (3,0)--(-3,0) node[below] {$m_\psi$}; \draw [->] (0,-3)--(0,3) node[above] {$m_\phi^2$};
\draw [very thick, blue!80!black] (0,0)--(2,0); \draw [very thick, blue!80!black] (0,-2)--(0,2); \draw [very thick, blue!80!black] (-1.5,-1.5)--(0,0); \draw [very thick, blue!80!black] (0,0)--(-2,0);
\fill [lightgray!70!yellow, fill opacity=.8] (0,0) circle [radius=.4];
\fill [white]  (-.5,2.2) rectangle (.5,2.8) node[midway, black] {$m_\phi^2>0$};
\node at (2,.35) {$m_\psi<0$};
\fill [white] (-.5, -2.3) rectangle (.5,-2.8) node[midway, black] {$m_\phi^2<0$};
\node at (-2, .35) {$m_\psi>0$};
\node [align=center] at (-2.2, -2.2) {$m_\phi^2<0$ \\ $m_\psi>0$};
\draw (1.6,1.6) circle [radius=.3] node {A}; \draw (1.6,-1.6) circle [radius=.3] node {B};
\draw (-.6,-1.4) circle [radius=.3] node {C}; \draw (-1.6,-.6) circle [radius=.3] node {D};
\draw (-1.6,1.6) circle [radius=.3] node {E};
\end{tikzpicture}
\hspace{\stretch{1}}
\caption{Masks for the phases of the various dualities. The phases in circles are either fully gapped (possibly with topological order) or contain Goldstone bosons. The thick blue lines correspond to the tuning of one mass parameter that conjecturally yields extra massless matter. The shaded circle in the middle covers the detailed structure of the phase diagram around the origin, which we do not know precisely.
\label{fig: masks}}
\end{figure}

We propose the following map of operators across the duality:
\bea
|\phi|^2 \qquad&\longleftrightarrow\qquad - \bar\psi\psi \\
\bar\psi\psi \qquad&\longleftrightarrow\qquad \;\;\; |\phi|^2 \;.
\eea
This reproduces the proposal in \cite{Aharony:2015mjs, Karch:2016sxi, Seiberg:2016gmd, Hsin:2016blu} for the case of a single matter species ($N_s=0$ or $N_f=0$) as well as the proposal in \cite{Jain:2013gza} for the case $N_s = N_f=1$ at large $N,k$, and---as we will see---it allows to match the phase diagrams.

We draw a qualitative picture of the two phase diagrams in Figure \ref{fig: masks}. The regions A through E are fully gapped for $N_s<N$ and can contain a Goldstone mode on the LHS for $N_s = N$. The thick lines are critical lines where extra modes become massless, as explained below. For convenience, we use Figure \ref{fig: masks} as a ``mask'' and list the theories that describe the various phases and critical lines in tables, such as Table \ref{tab: SU/U dualities case 1} and \ref{tab: SU/U dualities case 2}. Let us now explain which theories live on the critical lines.

Turning on a mass $m_\psi$ for the fermions, these can be integrated out leaving a Chern-Simons gauge theory coupled to scalars, possibly with shifted CS level due to the fermions. Classically the scalars remain massless; quantum mechanically a mass term will be generated, but with a suitable tuning of the scalar mass in the UV one obtains a fixed line in the phase diagram where the scalars are massless (provided the conjecture in \cite{Hsin:2016blu} is correct). We will keep this tuning implicit.

Similarly, turning on a positive mass $m_\phi^2>0$ for the scalars, these can be integrated out leaving a CS theory coupled to fermions. The fermions are massless along a fixed line in the phase diagram.

Turning on a negative mass $m_\phi^2<0$ for the scalars, the latter condense. Their expectation value breaks the gauge group, and gives mass to all scalars (but possible Goldstone bosons) and some fermions. Let us consider the two sides of (\ref{SU/U duality grav}) separately. First consider the LHS: $SU(N)_{k- \frac{N_f}2}$ with $N_s\,\phi$, $N_f\,\psi$. Up to a gauge and flavor rotation, the scalar VEV is
\be
\phi^{\alpha I} \propto \mat{ \unit_{N_s} \\ 0}_{\alpha I}
\ee
for $N > N_s$ (the case $N=N_s$ is similar). All scalars get a mass, either by Higgs mechanism or because of the quartic potential. The gauge group is broken to $SU(N-N_s)_{k - \frac{N_f}2}$. Because of the mixed coupling $\cO_\text{m}$ in (\ref{def mixed couplings}) and since
\be
\phi^\dag_{\alpha I} \phi^{\beta I} \propto \mat{ \unit_{N_s} & 0 \\ 0 & 0}_{\alpha\beta} \;,
\ee
the $N_s N_f$ fermion components neutral under the unbroken gauge group get a mass. Thus the theory along the critical line is $SU(N-N_s)_{k - \frac{N_f}2}$ with $N_f\,\psi$ in the fundamental representation.

The presence of the mixed coupling $\cO_\text{m}$ is crucial to give mass to the fermion components that are neutral under the unbroken gauge group. Those components are not reproduced by the dual theory in the corresponding phase, and so the duality would not work without $\cO_\text{m}$. The sign in front of the coupling $\cO_\text{m}$ determines the sign of the mass of the extra fermion components, which in turn determines the shift of the gravitational coupling. Only for one sign this matches the gravitational coupling in the dual, therefore we conclude that the mixed coupling on the LHS must be
\be
\label{sign mixed term LHS}
+ \cO_\text{m}
\ee
with positive sign.%
\footnote{On the contrary, the coupling $\cO_\text{d}$---even if present---would not qualitatively change the phase diagram. It would induce an equal mass for all fermions in the Higgsed phase, which would simply mix with the implicit UV tuning of the fermion mass.}

When deforming the LHS with $m_\phi^2<0$, we can at the same time turn on a fermion mass $m_\psi<0$ such that the fermions in the fundamental of $SU(N-N_s)$ are massive while the $N_s N_f$ singlet fermions remain massless. In the IR this gives $N_s N_f$ free fermions, transforming in the bifundamental representation of $U(N_s) \times U(N_f)$, plus the spin-TQFT $SU(N-N_s)_{k-N_f}$ (with suitable gravitational coupling). This is the oblique critical line in Figure \ref{fig: masks}.

Once again, the positive sign in (\ref{sign mixed term LHS}) is crucial for the duality to work. With negative sign, the position of the critical line in the phase diagram would change (it would move in the middle of phase E) and the TQFT would change: both features would not match with the dual description.

The discussion for $U(k)_{-N + \frac{N_s}2}$ with $N_f\,\phi$, $N_s\,\psi$---on the RHS of (\ref{SU/U duality grav})---is similar. For $m_\phi^2<0$ (and $N_f \leq k$) the scalar VEV breaks the gauge group to $U(k - N_f)_{-N + \frac{N_s}2}$, all scalars become massive as well as the $N_sN_f$ fermion components that are neutral under the unbroken gauge group. The IR theory is $U(k - N_f)_{-N + \frac{N_s}2}$ with $N_s\,\psi$. The gravitational coupling matches with the dual theory only if the mixed coupling on the RHS is
\be
- \cO_\text{m}
\ee
with negative sign. Turning on both $m_\phi^2<0$ and $m_\psi>0$ one finds another critical line with $N_sN_f$ free fermions plus the spin-TQFT $U(k-N_f)_{-N+N_s}$ (with gravitational coupling).

We can rephrase the condition on the mixed coupling in the following way: The theories involved in the duality (\ref{SU/U duality grav}) have a coupling $\pm\cO_\text{m}$, where the sign is the same as that of the CS level. In fact we can apply time reversal to (\ref{SU/U duality grav}), then both the CS level and the coupling $\cO_\text{m}$ change sign.

\begin{table}[ht!]
{\renewcommand{\arraystretch}{1.46}\begin{center}
$
\begin{array}{|lll|}
\hline
\multicolumn{3}{|c|}{SU(N)_{k - \frac{N_f}2} \text{ with $N_s$ $\phi$, $N_f$ $\psi$}} \text{ (LHS)}\\[.5em]
\hline\hline
m_\psi>0: &	SU(N)_k \text{ with $N_s$ $\phi$ } \times U(0)_1 & \qquad\qquad \\
	& \multicolumn{2}{r|}{A:\; SU(N)_k \times U(0)_1} \\
m_\phi^2>0: &	SU(N)_{k-\frac{N_f}2} \text{ with $N_f$ $\psi$} & \\
	& \multicolumn{2}{r|}{B:\; SU(N)_{k-N_f} \times U(NN_f)_1} \\
m_\psi<0: &	SU(N)_{k-N_f} \text{ with $N_s$ $\phi$ } \times U(NN_f)_1 & \\
	& \multicolumn{2}{r|}{C:\; SU(N-N_s)_{k-N_f} \times U(NN_f)_1} \\
m_\phi^2<0,\, m_\psi<0: & N_s N_f\, \psi \times SU(N-N_s)_{k-N_f} \times U\big( (N-N_s) N_f \big)_1 & \\
	& \multicolumn{2}{r|}{D:\; SU(N-N_s)_{k-N_f} \times U\big( (N-N_s)N_f \big)_1} \\
m_\phi^2<0: &	SU(N-N_s)_{k - \frac{N_f}2} \text{ with $N_f$ $\psi$} & \\
	& \multicolumn{2}{r|}{E:\; SU(N-N_s)_k \times U(0)_1} \\
\hline
\end{array}
$ \\[1.5em]
$
\begin{array}{|lll|}
\hline
\multicolumn{3}{|c|}{U(k)_{-N + \frac{N_s}2} \text{ with $N_f$ $\phi$, $N_s$ $\psi$ } \times U\big( k(N-N_s) \big)_1} \text{ (RHS)}\\[.5em]
\hline\hline
m_\phi^2>0: &	U(k)_{-N + \frac{N_s}2} \text{ with $N_s$ $\psi$ } \times U\big( k(N-N_s) \big)_1 & \quad \\
	& \multicolumn{2}{r|}{A:\; U(k)_{-N} \times U(kN)_1} \\
m_\psi<0: &	U(k)_{-N} \text{ with $N_f$ $\phi$ } \times U(kN)_1 & \\
	& \multicolumn{2}{r|}{B:\; U(k-N_f)_{-N} \times U(kN)_1} \\
m_\phi^2<0: &	U(k-N_f)_{-N + \frac{N_s}2} \text{ with $N_s$ $\psi$ } \times U(kN - kN_s + N_fN_s)_1 & \\
	& \multicolumn{2}{r|}{C:\; U(k-N_f)_{-N + N_s} \times U(kN - kN_s + N_fN_s)_1} \\
m_\phi^2<0,\, m_\psi>0: &  N_s N_f\, \psi \times U(k-N_f)_{-N + N_s} \times U\big( k(N-N_s) \big)_1 & \\
	& \multicolumn{2}{r|}{D:\; U(k-N_f)_{-N + N_s} \times U\big( k(N-N_s) \big)_1} \\
m_\psi>0: &	U(k)_{-N+N_s} \text{ with $N_f$ $\phi$ } \times U\big( k(N-N_s) \big)_1 & \\
	& \multicolumn{2}{r|}{E:\; U(k)_{-N+N_s} \times U\big( k(N-N_s)\big)_1} \\
\hline
\end{array}
$
\end{center}}
\caption{Phase diagram of the $SU/U$ dualities, for $N> N_s$ and $k \geq N_f$.
\label{tab: SU/U dualities case 1}}
\end{table}

\begin{table}[ht!]
{\renewcommand{\arraystretch}{1.46}\begin{center}
$
\begin{array}{|lll|}
\hline
\multicolumn{3}{|c|}{SU(N)_{k - \frac{N_f}2} \text{ with $N$ $\phi$, $N_f$ $\psi$}} \text{ (LHS)}\\[.5em]
\hline\hline
m_\psi>0: &	SU(N)_k \text{ with $N$ $\phi$ } \times U(0)_1 & \qquad\qquad\qquad\qquad\;\;\; \\
	& \multicolumn{2}{r|}{A:\; SU(N)_k \times U(0)_1} \\
m_\phi^2>0: &	SU(N)_{k-\frac{N_f}2} \text{ with $N_f$ $\psi$} & \\
	& \multicolumn{2}{r|}{B:\; SU(N)_{k-N_f} \times U(NN_f)_1} \\
m_\psi<0: &	SU(N)_{k-N_f} \text{ with $N$ $\phi$ } \times U(NN_f)_1 & \\
	& \multicolumn{2}{r|}{C:\; S^1 \times U(NN_f)_1} \\
m_\phi^2<0,\, m_\psi<0: & N N_f\, \psi \times S^1 & \\
	& \multicolumn{2}{r|}{D:\; S^1 \times U(0)_1} \\
m_\phi^2<0: &	S^1 \times U(0)_1 & \\
	& \multicolumn{2}{r|}{E:\; S^1 \times U(0)_1} \\
\hline
\end{array}
$ \\[1.5em]
$
\begin{array}{|lll|}
\hline
\multicolumn{3}{|c|}{U(k)_{- \frac N2} \text{ with $N_f$ $\phi$, $N$ $\psi$}} \text{ (RHS)}\\[.5em]
\hline\hline
m_\phi^2>0: &	U(k)_{-\frac N2} \text{ with $N$ $\psi$} & \qquad\qquad\qquad\qquad \\
	& \multicolumn{2}{r|}{A:\; U(k)_{-N} \times U(kN)_1} \\
m_\psi<0: &	U(k)_{-N} \text{ with $N_f$ $\phi$ } \times U(kN)_1 & \\
	& \multicolumn{2}{r|}{B:\; U(k-N_f)_{-N} \times U(kN)_1} \\
m_\phi^2<0: &	U(k-N_f)_{- \frac N2} \text{ with $N$ $\psi$ } \times U(NN_f)_1 & \\
	& \multicolumn{2}{r|}{C:\; U(k-N_f)_0 \times U(NN_f)_1} \\
m_\phi^2<0,\, m_\psi>0: & N N_f\, \psi \times U(k-N_f)_0 & \\
	& \multicolumn{2}{r|}{D:\; U(k-N_f)_0 \times U(0)_1} \\
m_\psi>0: &	U(k)_0 \text{ with $N_f$ $\phi$ } \times U(0)_1 & \\
	& \multicolumn{2}{r|}{E:\; U(k)_0 \times U(0)_1} \\
\hline
\end{array}
$
\end{center}}
\caption{Phase diagram of the $SU/U$ dualities, for $N = N_s$ and $k \geq N_f$.
\label{tab: SU/U dualities case 2}}
\end{table}

The various phases and critical lines for the SU/U dualities, in the case $N>N_s$ and $k \geq N_f$, are reported in Table \ref{tab: SU/U dualities case 1}. We recall that we assume $N_s, N_f \geq 1$. In the range $N>N_s$ and $k \geq N_f$ there is no (classical) symmetry breaking. The analysis is valid for the two theories in (\ref{SU/U duality grav}) independently of the dualities. In the tables we also indicate the trivial spin-TQFTs $U(n)_1$ that appear in the various phases, both to keep track of the gravitational couplings and to remind ourselves that the claimed dualities involve \emph{spin} theories. Extra observables in the various phases (which provide extra checks of the dualities and help distinguishing massive phases) are the couplings to background fields for global symmetries and the corresponding counterterms: these will be considered in Section~\ref{sec: SU/U coupling to background}.

Comparing the various phases (see Appendix \ref{app: summary}), we find that they are dual for
\be
N > N_s \geq 1 \;,\qquad\qquad k \geq N_f \geq 1 \;.
\ee
Notice that for $k = N_f$ (and $N > N_s$) the vertical line in the lower half plane (corresponding to $m_\psi<0$ on the LHS and $m_\phi^2<0$ on the RHS) disappears since it is gapped. We reduce to the duality $SU(N)_0$ with $N_s$ $\phi$ $\leftrightarrow$ $\emptyset$, expressing confinement. Moreover phases $B$ and $C$ are identical.

The phases and critical lines for $N = N_s$ and $k \geq N_f$ are in Table \ref{tab: SU/U dualities case 2}. In that table, $S^1$ refers to a compact Goldstone boson. Comparing the various phases, we find that they are dual for
\be
N = N_s \geq 1 \;,\qquad\qquad k > N_f \geq 1 \;.
\ee
Notice that the horizontal line on the left half plane (corresponding to $m_\phi^2<0$ on the LHS and $m_\psi>0$ on the RHS) disappears since it is identical to phases $D$ and $E$ which are described by the $S^1$ Goldstone mode.

Putting together the two cases we find:
\be
\label{range SU/U}
\text{Range of dualities:}\qquad\quad N \geq N_s \;, \quad k \geq N_f \;,\quad (N,k) \neq (N_s,N_f) \;.
\ee
As we explained before, the cases $N=1$ or $k=1$ are somehow special because the interaction $(\phi\bar\psi)(\psi\bar\phi)$ is not independent from $|\phi|^2 \bar\psi\psi$ and we might expect the latter to be marginally irrelevant in the UV. Moreover, our classical analysis of the phase diagrams kept the tuning of mass terms implicit, and so it should be regarded as a qualitative picture. Finally, it appears to be possible to make sense of the dualities also for larger values of $N_s$ and $N_f$, invoking quantum phases with spontaneous symmetry breaking along the lines of \cite{Komargodski:2017keh} (see also \cite{Gaiotto:2017tne}); we leave the analysis of this possibility for future work.

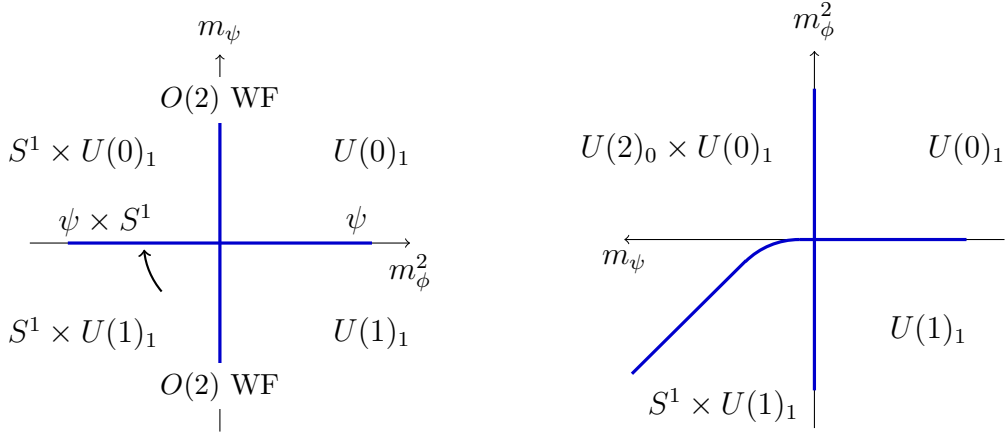
\begin{figure}[t!]
\centering
\hspace{\stretch{1}}
\begin{tikzpicture}[scale=1]
\draw [->] (-2.5,0)--(2.5,0) node[below] {$m_\phi^2$}; \draw [->] (0,-2.5)--(0,2.5) node[above] {$m_\psi$};
\draw [very thick, blue!80!black] (-2,0)--(2,0); \draw [very thick, blue!80!black] (0,-2)--(0,2);
\draw [->, thick] (0,0) ++(220:1.0) arc (220:185:1.0);
\node at (1.8,.3) {$\psi$}; \node at (-1.5,.3) {$\psi \times S^1$};
\fill [white]  (-.5,1.6) rectangle (.5,2.2) node[midway, black] {\small $O(2)$ WF};
\fill [white]  (-.5,-1.6) rectangle (.5,-2.2) node[midway, black] {\small $O(2)$ WF};
\node at (2,1.2) {$U(0)_1$}; \node at (-1.8,1.2) {$S^1 \times U(0)_1$};
\node at (2, -1.2) {$U(1)_1$}; \node at (-1.8, -1.2) {$S^1 \times U(1)_1$};
\end{tikzpicture}
\hspace{\stretch{1}}
\begin{tikzpicture}[scale=1]
\draw [->] (2.5,0)--(-2.5,0) node[below] {$m_\psi$}; \draw [->] (0,-2.5)--(0,2.5) node[above] {$m_\phi^2$};
\draw [very thick, blue!80!black] (0,0)--(2,0); \draw [very thick, blue!80!black] (0,-2)--(0,2);
\draw [very thick, blue!80!black] (0,0)--(-.2,0); \draw [very thick, blue!80!black] (-.2,0) arc [radius = 1, start angle = 90, end angle = 135]; \draw [very thick, blue!80!black] (-2.40,-1.78)--(-.90,-.28);
\node at (2,1.2) {$U(0)_1$}; \node at (-1.8,1.2) {$U(2)_0 \times U(0)_1$};
\node at (1.5, -1.2) {$U(1)_1$}; \node at (-1.2, -2.2) {$S^1 \times U(1)_1$};
\end{tikzpicture}
\hspace{\stretch{1}}
\caption{$\qquad O(2)$ WF $\times$ $\psi$ $\qquad\longleftrightarrow\qquad$ $U(2)_{-1/2}$ with $\phi,\psi$. $\quad$ Phase diagram. 
\label{fig: O2 WF}}
\end{figure}

\subsection{Some simple examples}

One of the simplest examples is $N=N_s = N_f = 1$, $k=2$:
\be
O(2) \text{ WF } \times\; \psi \qquad\longleftrightarrow\qquad U(2)_{-\frac12} \text{ with $\phi$, $\psi$} \;.
\ee
The gravitational coupling is $U(0)_1$ on both sides.
The theory on the LHS is decoupled in two parts (we know that $(\phi\bar\psi)(\psi\bar\phi) \equiv |\phi|^2 \bar\psi\psi$ is irrelevant): the $O(2)$ Wilson-Fisher fixed point and a free Dirac fermion; such a theory is time-reversal invariant. We summarize the phase diagram in Figure~\ref{fig: O2 WF}. On the left we took into account that the coupling $(\phi\bar\psi)(\psi\bar\phi)$ is not present and moved a gapless line accordingly; hence, on the right we implemented the fact that around the origin the lines should cross perpendicularly, as implied by the duality. This example generalizes to
\be
O(2) \text{ WF } \times\; N_f\, \psi \qquad\longleftrightarrow\qquad U(k)_{-\frac12} \text{ with $N_f$ $\phi$, 1 $\psi$}
\ee
with $k>N_f$. Again the gravitational coupling is $U(0)_1$ on both sides.

In these examples the duality predicts that the theory on the RHS, namely $U(k)_{-\frac12}$ with $N_f$ $\phi$, 1 $\psi$ (and $k>N_f$) has a multicritical fixed point where the four lines meet at a single point. At such a multicritical fixed point the IR dynamics factorizes into two critical fixed points (and develops time-reversal invariance quantum mechanically), explaining why four lines meet at a single point.

\begin{figure}[t!]
\centering
\hspace{\stretch{1}}
\begin{tikzpicture}[scale=1]
\draw [->] (-2.5,0)--(2.5,0) node[below] {$m_\phi^2$}; \draw [->] (0,-2.5)--(0,2.5) node[above] {$m_\psi$};
\draw [very thick, blue!80!black] (-2,0)--(2,0); \draw [very thick, blue!80!black] (0,0)--(0,2);
\draw [very thick, blue!80!black] (0,0)--(0,-.2); \draw [very thick, blue!80!black] (0,-.2) arc [radius = 1, start angle = 0, end angle = -45]; \draw [very thick, blue!80!black] (-1.78,-2.40)--(-.28,-.90);
\fill [lightgray!70!yellow, fill opacity=.6] (0,0) circle [radius=.4];
\draw [densely dashed] (-.6,-.6)--(.6,.6);
\node at (1.6,.3) {$SU(2)_{1/2} \; \psi$}; \node at (-1.8,.3) {$\psi$}; \node at (-1.9,-2) {$\psi$};
\fill [white]  (-.5,1.6) rectangle (.5,2.2) node[midway, black] {$SU(2)_1 \; \phi$};
\node at (2,1.2) {$SU(2)_1 \times U(0)_1$}; \node at (-1.8,1.2) {$U(0)_1$};
\node at (2, -1.2) {$U(2)_1$}; \node at (-1.8, -.8) {$U(1)_1$};
\end{tikzpicture}
\hspace{\stretch{1}}
\begin{tikzpicture}[scale=1]
\draw [->] (2.5,0)--(-2.5,0) node[below] {$m_\psi$}; \draw [->] (0,-2.5)--(0,2.5) node[above] {$m_\phi^2$};
\draw [very thick, blue!80!black] (-2,0)--(2,0); \draw [very thick, blue!80!black] (0,-2)--(0,2);
\draw [->, thick] (0,0) ++(230:1.4) arc (230:265:1.4);
\fill [lightgray!70!yellow, fill opacity=.6] (0,0) circle [radius=.4];
\draw [densely dashed] (-.6,-.6)--(.6,.6);
\node at (2,1.2) {$U(1)_{-2} \times U(2)_1$}; \node at (-1.8,1.2) {$U(0)_1$};
\node at (1.8, -1.2) {$U(2)_1$}; \node at (-1.8, -1.2) {$U(1)_1$};
\end{tikzpicture}
\hspace{\stretch{1}}
\caption{$\qquad SU(2)_{1/2}$ with $\phi, \psi$ $\qquad\longleftrightarrow\qquad$ $U(1)_{-3/2}$ with $\phi,\psi$ $\times\, U(1)_1$. $\;$ Phase diagram. \newline
On both sides we emphasized an emergent time-reversal symmetry (with an anomaly) with respect to the dashed line.
\label{fig: SU2}}
\end{figure}
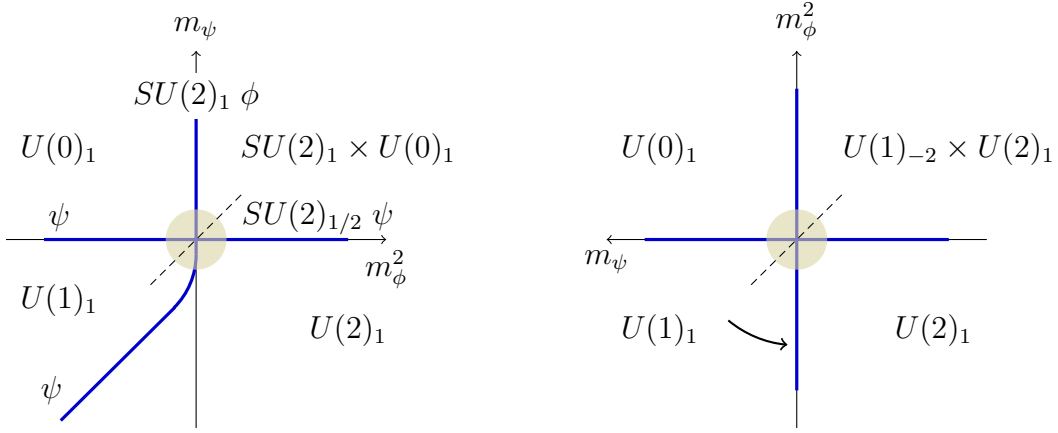

Another simple example is $k=N_s = N_f = 1$, $N=2$:
\be
\label{SU(2) U(1)3/2 duality}
SU(2)_{\frac12} \text{ with $\phi$, $\psi$} \qquad\longleftrightarrow\qquad U(1)_{-\frac32} \text{ with $\phi$, $\psi$} \;.
\ee
There is a gravitational coupling $U(1)_1$ on the RHS. The phase diagram is summarized in Figure~\ref{fig: SU2}. On the left we drew a bent line to match the diagram on the right around the origin; on the right we took into account that we expect the coupling $(\phi\bar\psi)(\psi\bar\phi) \equiv |\phi|^2\bar\psi\psi$ not to be present in the IR, and moved a gapless line accordingly. Two of the thick lines in the phase diagram correspond to a free Dirac fermion, while the other two correspond to a CFT (and its time reversal) with $SO(3)$ global symmetry, discussed in \cite{Aharony:2016jvv}.

The theory on the RHS also appears in a $U/U$ duality (see Figure~\ref{fig: U(1)3/2} and the discussion in Section~\ref{sec: U/U duality}) which is part of a family but can also be found by combining the Abelian dualities of \cite{Seiberg:2016gmd} (see Section~\ref{sec: U(1) 3/2 time rev}). The $U/U$ duality implies that the theory develops time-reversal invariance quantum mechanically in the IR, along the line $m_\phi^2 = - m_\psi$. On the other hand, the theory on the LHS can be obtained as a relevant deformation of $USp(2)_\frac12$ with a scalar and a fermion, which in turn appears in a $USp$ duality (see Section~\ref{sec: USp duality}). The duality for the $USp(2)$ gauge theory implies a duality for the $SU(2)$ gauge theory, and the latter implies that the theory on the LHS develops time-reversal invariance around the origin along the line $m_\phi^2 = m_\psi$. As we see here, the two conclusions are compatible with the SU/U duality (\ref{SU(2) U(1)3/2 duality}) that relates the two theories. The predicted time-reversal invariance (with an anomaly) implies a symmetry of the phase diagram around the origin with respect to the dashed line at $45^\circ$. This however is not enough to guarantee that the four lines meet at a single point.

\subsection{RG flows}
\label{sec: RG flows}

We can start from the duality (\ref{SU/U duality grav}) with parameters $(N,k,N_s,N_f)$ and give mass to a single flavor, either a scalar or a fermion. We accompany this deformation with a tuning of the symmetry-preserving mass deformations (\ref{mass terms}) such that the remaining scalars and fermions stay massless. By choosing positive or negative mass, we end up with the same duality as in (\ref{SU/U duality grav}) but with reduced parameters:
\be
\label{RG flow pattern}
(N,k,N_s,N_f) \to \begin{cases} (N,k, N_s, N_f-1) \qquad & \wt m_\psi>0 \\ (N,k-1,N_s, N_f-1) & \wt m_\psi<0 \\ (N,k,N_s-1,N_f) & \wt m_\phi^2>0 \\ (N-1, k, N_s-1,N_f) & \wt m_\phi^2<0 \;. \end{cases}
\ee
We have indicated with a tilde the mass of the single field. The constraint (\ref{range SU/U}) is preserved along the RG flow.

Therefore, the proposed list of dualities is consistent with massive RG flows.

\subsection{Coupling to a background}
\label{sec: SU/U coupling to background}

We are interested in what counterterms for background fields coupled to the global symmetries of the theory should we put on the RHS of the duality, if we set them to zero on the LHS (given that, the counterterms can be shifted by the same amount on both sides). For continuous symmetries, such counterterms modify the contact terms in three-point functions of the currents, which obviously should match across the duality.

In order to understand those counterterms, we simply give mass to the matter fields and compare the phases that we obtain. Coupling to an $SU(N_s) \times SU(N_f)$ background is simple, and the duality with counterterms for those groups takes the form
\begin{multline}
\label{SU/U duality with SU background}
SU(N)_{k - \frac{N_f}2} \times SU(N_s)_{L_s} \times SU(N_f)_{L_f - \frac N2} \text{ with $\phi$ in $(\rep{N}, \rep{N_s}, \rep1)$, $\psi$ in $(\rep{N}, \rep1, \rep{N_f})$} \quad\longleftrightarrow \\
U(k)_{-N + \frac{N_s}2} \times SU(N_s)_{L_s + \frac k2} \times SU(N_f)_{L_f} \text{ with $\phi$ in $(\rep{k}, \rep1, \rep{N_f})$, $\psi$ in $(\rep{k}, \rep{N_s}, \rep1)$} \;.
\end{multline}
Here the first group is dynamical while the other two are global symmetries coupled to a classical background, and we have indicated their CS counterterms. One can check that both sides give equal counterterms%
\footnote{Level-rank dualities can be used on dynamical fields, but not on background fields.}
in all phases in Figure \ref{fig: masks}.

The coupling to the two $U(1)$ factors, their mapping through the duality and the corresponding counterterms are a bit more involved. To express them in a precise way, we write the duality in a Lagrangian form and explicitly couple the two sides to $U(1) \times U(1)$ background fields $A$ and $B$. The duality reads
\begin{align}
& |D_{b+A}\phi|^2 + i \bar\psi \Dslash_{b-A} \psi - \phi^4 - \phi^2\psi^2 + \frac{k}{4\pi} \Tr_N \Big( bdb- \frac{2i}3 b^3 \Big) + \frac1{2\pi} cd\big( B -  \Tr_N b \big) \quad\longleftrightarrow \nn\\
& |D_{f-A}\phi|^2  + i \bar\psi \Dslash_{f+A} \psi - \phi^4 + \phi^2\psi^2 - \frac{N-N_s}{4\pi} \Tr_k \Big( fdf - \frac{2i}3 f^3 \Big) \nn\\
&\qquad + \frac1{2\pi} (\Tr_k f) d(B + N_s A) + \frac{N_s k}{4\pi} AdA -2 (N- N_s)k \, \text{CS}_g \;.
\label{SU/U duality with U(1)2 background}
\end{align}
Here $b$, $f$ and $c$ are dynamical $U(N)$, $U(k)$ and $U(1)$ gauge fields, respectively, while $A$, $B$ are background $U(1)$ gauge fields.%
\footnote{The duality as written in (\ref{SU/U duality with U(1)2 background}) is well-defined on spin manifolds. Since the theories involved in the $SU/U$ and $U/U$ dualities satisfy the spin/charge relation, they can be placed on more general non-spin manifolds with the help of a spin$_c$ connection \cite{Seiberg:2016gmd}. Indeed one could generalize (\ref{SU/U duality with U(1)2 background}) such that it makes sense on non-spin manifolds, along the lines of \cite{Hsin:2016blu}, but we will not do so here.}
The quartic couplings are schematically indicated as $\phi^4$ and $\phi^2\psi^2$, and recall that the potential appears in the Lagrangian as $-V$. The theory on the RHS is a $U(k)$ CS gauge theory at level $-N + \frac{N_s}2$, and the magnetic current couples to $B + N_sA$. On the LHS, instead, we can integrate out $c$ to fix $\Tr b = B$ and thus the theory is an $SU(N)$ CS gauge theory at level $k-\frac{N_f}2$. Notice that when $B=0$ the dynamical gauge field is a standard $SU(N)$ gauge field, but when $B\neq 0$ the dynamical field describes non-trivial $PSU(N)$ bundles with (generalized) second Stiefel-Whitney class equal to $B \mod N$. Substituting back in the Lagrangian, $B$ couples to the ``baryonic'' current giving charge $1$ to the baryons.

It is instructive to check that, upon mass deformations, (\ref{SU/U duality with U(1)2 background}) reproduces the dualities with a single matter species with the correct coupling to a $U(1)$ background and the correct counterterms, that we have summarized in Appendix \ref{app: summary}. For instance, take the LHS of (\ref{SU/U duality with U(1)2 background}) and deform it with \mbox{$m_\psi>0$}.{} Shifting the dynamical gauge fields as $b \to b - A \unit_N$ and $c \to c - kA$ we get
\be
\cL_\text{LHS} = |D_b\phi|^2 - \phi^4 + \frac{k}{4\pi} \Tr_N \Big( bdb- \frac{2i}3 b^3 \Big) + \frac1{2\pi} cd\big( B + NA - \Tr_N b \big) - \frac{k}{2\pi} BdA - \frac{Nk}{4\pi} AdA \;.
\ee
Then take the RHS of (\ref{SU/U duality with U(1)2 background}) and deform it with $m_\phi^2>0$. Shifting the dynamical gauge field as $f \to f - A\unit_k$ we get
\begin{multline}
\cL_\text{RHS} = i \bar\psi \Dslash_f \psi - \frac{N-N_s}{4\pi} \Tr_k \Big( fdf - \frac{2i}3 f^3 \Big) + \frac1{2\pi} (\Tr_k f) d(B + N A) - 2(N-N_s) k \, \text{CS}_g \\
- \frac{k}{2\pi} BdA - \frac{Nk}{4\pi} AdA \;.
\end{multline}
The duality between the Lagrangians $\cL_\text{LHS}$ and $\cL_\text{RHS}$ is precisely the duality in \cite{Hsin:2016blu},%
\footnote{The gauge field $A$ in \cite{Hsin:2016blu} should not be confused with the one here. $A_\text{there}$ is a spin$_c$ connection, that should be set to zero to compare with our formul\ae. On the other hand, $A_\text{here}$ is a regular gauge field which, together with $B$, describes the $U(1) \times U(1)$ background.}
that we reported in (\ref{SU/U duality 1 species background}), up to the fact that the two theories are coupled to a linear combination of the two $U(1)$'s given by $B + NA$ and there are equal extra counterterms on both sides. The case of $m_\phi^2>0$ on the LHS and $m_\psi<0$ on the RHS is similar.

Alternatively, take the LHS of (\ref{SU/U duality with U(1)2 background}) and deform it with $m_\psi<0$. Shifting the dynamical gauge fields as $b \to b - A \unit_N$ and $c \to c + (2N_f-k)A$ we get
\begin{multline}
\cL_\text{LHS} = |D_b\phi|^2 - \phi^4 + \frac{k-N_f}{4\pi} \Tr_N \Big( bdb- \frac{2i}3 b^3 \Big) + \frac1{2\pi} cd\big( B+NA - \Tr_N b \big) \\
-2NN_f\, \text{CS}_g + \frac{2N_f - k}{2\pi} BdA - \frac{Nk}{4\pi} AdA \;.
\end{multline}
Then take the RHS of (\ref{SU/U duality with U(1)2 background}) and deform it with $m_\phi^2<0$. In this case $N_f$ scalars get a VEV, fixing $(f - A \unit_k)\phi = 0$. This means that $f$ breaks into a block $A \unit_{N_f}$ and a block $\tilde f$ of dimension $k-N_f$. Moreover $N_s N_f$ fermions get a negative mass, and they are coupled to $2A$. After shifting the dynamical gauge field as $\tilde f \to \tilde f - A\unit_{k-N_f}$ we get
\begin{multline}
\cL_\text{RHS} = i \bar\psi \Dslash_{\tilde f} \psi - \frac{N-N_s}{4\pi} \Tr_{k-N_f} \Big( \tilde fd\tilde f - \frac{2i}3 \tilde f^3 \Big) + \frac1{2\pi} (\Tr_{k-N_f} \tilde f) d(B + N A) \\
-2 \big( N_sN_f+Nk-N_sk\big) \text{CS}_g + \frac{2N_f - k}{2\pi} BdA - \frac{Nk}{4\pi} AdA \;.
\end{multline}
Once again, the duality between $\cL_\text{LHS}$ and $\cL_\text{RHS}$ is precisely the one in \cite{Hsin:2016blu}, that we reported in (\ref{SU/U duality 1 species background}), up to the fact that the coupling is to $B+NA$ and there are equal extra CS counterterms on both sides. The case of $m_\phi^2<0$ on the LHS and $m_\psi>0$ on the RHS is similar.

Given the duality in (\ref{SU/U duality with U(1)2 background}) with coupling to the $U(1) \times U(1)$ background, we can produce new dualities by adding CS counterterms on both sides and then making $A$, $B$ or a linear combination of them dynamical. For instance, we can add $\frac1{2\pi} BdC$ on both sides---where $C$ is a new $U(1)$ background field---and then make $B$ dynamical. Integrating out $B$ on one of the two sides, we are left with a duality which is precisely the parity transformed of (\ref{SU/U duality with U(1)2 background}). This is a consistency check.

More interestingly, we can start with (\ref{SU/U duality with U(1)2 background}), add $\frac1{2\pi} BdC \pm \frac1{4\pi} BdB$ on both sides and make $B$ dynamical. The LHS becomes
\begin{multline}
\cL_\text{LHS} = |D_{b+A}\phi|^2 + i \bar\psi \Dslash_{b-A} \psi - \phi^4 - \phi^2\psi^2 + \frac{k}{4\pi} \Tr_N \Big( bdb- \frac{2i}3 b^3 \Big) \\
\pm \frac1{4\pi} (\Tr_N b) d (\Tr_N b) + \frac1{2\pi} (\Tr_N b)dC
\end{multline}
while the RHS becomes
\begin{multline}
\cL_\text{RHS} = |D_{f-A}\phi|^2 + i \bar\psi \Dslash_{f+A} \psi - \phi^4 + \phi^2\psi^2 - \frac{N-N_s}{4\pi} \Tr_k \Big( fdf - \frac{2i}3 f^3 \Big) \mp \frac1{4\pi} (\Tr_k f)d(\Tr_k f) \\
+ \frac1{2\pi} (\Tr_k f)d (\mp C + N_s A) \mp \frac1{4\pi} CdC + \frac{N_s k}{4\pi} AdA - 2 \big( k (N-N_s) \pm 1 \big) \text{CS}_g \;.
\end{multline}
These are two $U/U$ dualities that will be analyzed in more detail in Section~\ref{sec: U/U duality}.

\subsection{Baryonic and monopole operators}
\label{sec: SU/U monopoles}

The $SU(N)$ theory on the LHS of (\ref{SU/U duality grav}) has baryonic operators, which are mapped to monopole operators in the $U(k)$ theory on the RHS. We would like to specify the operator map precisely.

Let us start reviewing how baryonic operators are mapped to monopole operators in the theories with a single matter species \cite{Radicevic:2015yla, Aharony:2015mjs, Aharony:2016jvv}. In $SU(N)_{k-\frac{N_f}2}$ with $N_f$ $\psi$ the simplest baryonic operators are
\be
\label{baryons SU(N) psi}
\epsilon_{\alpha_1\ldots\alpha_N} \; \psi^{\alpha_1B_1} \ldots \psi^{\alpha_N B_N} \;.
\ee
The fermions are antisymmetric in the gauge indices $\alpha_i$ and have antisymmetric statistics, therefore they are totally symmetric in the pairs $\big( B_i, \text{spin}_i \frac12 \big)$ where the first entry $B_i$ is a flavor index of $SU(N_f)$ while the second entry is an index for the spacetime spin, that we have always kept implicit in this paper. For instance, if $N_f=1$ then the baryonic operators have spacetime spin $\frac N2$. If $N_f>1$ then there is a baryonic operator with spin $\frac N2$ that transforms in the totally symmetric $N$-index representation of $SU(N_f)$, as well as other baryonic operators whose spin is correlated with the representation under the global $SU(N_f)$ symmetry.

In $U(k)_{-N}$ with $N_f$ $\phi$ the corresponding monopole operators are
\be
\cM_{\alpha_1 \ldots \alpha_N} \; \phi^{\alpha_1B_1} \ldots \phi^{\alpha_N B_N} \;.
\ee
Here $\cM$ is a bare monopole operator with monopole charge $1$. Because of Chern-Simons interactions, it transforms in the $N^\text{th}$ symmetric power of the antifundamental of $U(k)$, and to form a gauge invariant it should be multiplied by $N$ scalar fields $\phi$. In the monopole background the ground state of the scalar field $\phi$ has spacetime spin $\frac12$ \cite{Wu:1976ge}: in terms of spin (or monopole) spherical harmonics%
\footnote{The spin spherical harmonics $Y^s_{j,j_3}$ have $j \in \frac12 \bZ_{\geq 0}$, $j=s=j_3 \mod 1$, and $|s|, |j_3| \leq j$. They are sections of the line bundle on $S^2$ with first Chern class $2s$, and are eigenfunctions of the covariant Laplacian with eigenvalue $j(j+1)-s^2$ and orbital angular momentum $j$.}
$Y^s_{j,j_3}$, the wavefunctions are $Y^{1/2}_{1/2, \pm1/2}$. The scalars are symmetric in the gauge indices and have symmetric statistics, therefore they are totally symmetric in the pairs $\big( B_i, \text{spin}_i \frac12 \big)$. We see that the quantum numbers of these monopole operators precisely match those of the baryons in (\ref{baryons SU(N) psi}).

In $SU(N)_k$ with $N_s$ $\phi$ the simplest baryonic operators are
\be
\label{baryons SU(N) phi}
\epsilon_{\alpha_1 \ldots \alpha_N} \; \phi^{\alpha_1 I_1} \ldots \phi^{\alpha_{N_s} I_{N_s}} \; \partial_\bullet \phi^{\alpha_{N_s+1} I_{N_s+1}} \ldots \partial_\bullet\phi^{\alpha_N I_N} \;.
\ee
In this expression we have assumed $N \geq N_s$. Since the gauge indices are antisymmetrized and the scalars have symmetric statistics, we cannot simply take a product of the fields $\phi$. Instead, in order to get a non-vanishing operator, (at least) $N-N_s$ of them should be acted upon by various numbers of derivatives that we have indicated concisely by $\partial_\bullet \equiv \partial_{\mu_1} \dots \partial_{\mu_\ell}$ (see \cite{Shenker:2011zf} for a counting at large $N$). We should remember that the scalars obey (in the free theory) $\partial^2 \phi = 0$. The first flavor indices $I_1, \dots, I_{N_s}$ are totally antisymmetrized and form a singlet of $SU(N_s)$, while the symmetry pattern for the remaining $N-N_s$ is correlated with the spacetime spin in such a way that the pairs $(I_i, \text{spin}_i)$ are antisymmetric.

In $U(k)_{-N + \frac{N_s}2}$ with $N_s$ $\psi$ the corresponding monopole operators are
\be
\cM_{\alpha_1 \ldots \alpha_{N-N_s}} \; \tilde\partial_\bullet \psi^{\alpha_1 I_{N_s+1}} \ldots \tilde\partial_\bullet \psi^{\alpha_{N-N_s} I_N} \;.
\ee
The bare CS term is $-(N-N_s)$, therefore the bare monopole $\cM$ transforms in the $(N-N_s)^\text{th}$ symmetric power of the antifundamental representation of $U(k)$, and should be dressed by $N-N_s$ fermion fields $\psi$ to form a gauge invariant. In the monopole background the fermion field $\psi$ has a state of spin zero and a state of spin $1$; we use here the ground states of spin $1$. The fields $\psi$ are symmetric in the gauge indices and have antisymmetric statistics, therefore they are antisymmetrized in the pairs $(I_i, \text{spin}_i)$. The notation $\tilde\partial_\bullet$ indicates some number of derivatives acting on $\psi$. This number can be zero, however each insertion of $\psi$ already carries spin $1$. Therefore we can identify $\partial_\bullet = \tilde\partial_\bullet \partial_\mu$. A more precise statement is that $\psi$ is in a state $Y^1_{j,j_3}$ where $j$ equals the spacetime spin of $\partial_\bullet \phi$. Notice that the harmonic $Y^1_{0,0}$ does not exist, consistently with the fact that we should not take $\partial^2\phi$ in (\ref{baryons SU(N) phi}). We see that the quantum numbers of these monopole operators precisely match those of the baryons in (\ref{baryons SU(N) phi}).

Let us now move to the general case of the duality (\ref{SU/U duality grav}) with $N_s, N_f\geq1$. We can read off the precise mapping of symmetries from (\ref{SU/U duality with SU background})-(\ref{SU/U duality with U(1)2 background}). In $SU(N)_{k-\frac{N_f}2}$ with $N_s$ $\phi$, \mbox{$N_f$ $\psi$} the simplest baryonic operators are
\be
\label{baryons I}
\cB^{(r)} = \epsilon_{\alpha_1 \ldots \alpha_N} \; \psi^{\alpha_1B_1} \ldots \psi^{\alpha_r B_r} \; \phi^{\alpha_{r+1} I_1} \ldots \phi^{\alpha_{r+N_s} I_{N_s}} \; \partial_\bullet \phi^{\alpha_{r+N_s+1} I_{N_s+1}} \ldots \partial_\bullet \phi^{\alpha_N I_{N-r}}
\ee
for $0 \leq r \leq N-N_s$, and
\be
\label{baryons II}
\cB^{(r)} = \epsilon_{\alpha_1 \ldots \alpha_N} \; \psi^{\alpha_1B_1} \ldots \psi^{\alpha_r B_r} \; \phi^{\alpha_{r+1} I_1} \ldots \phi^{\alpha_N I_{N-r}}
\ee
for $N-N_s \leq r \leq N$. In the first class of baryons the number of fields $\phi$ exceeds $N_s$ and since the flavor indices are antisymmetrized, we should include derivatives to form non-vanishing operators. In the second class the number of $\phi$'s is smaller than $N_s$ and the derivatives are not necessary. The charges of those operators are:
\be
\label{quantum numbers baryons/monopoles}
\begin{array}{c|ccc}
\cB^{(r)} & U(1)_B & U(1)_A & SU(N_s) \times SU(N_f) \times \text{spin} \\[.2em]
\hline \rule{0pt}{1.4em}
0 \leq r \leq N-N_s & 1 & N-2r & \quad \big(\rep{N_s}, \text{spin}\big)^{\otimes_A(N-N_s-r)} \otimes \big(\rep{N_f}, \text{spin}\frac12 \big)^{\otimes_S \, r} \quad \\[.3em]
N-N_s \leq r \leq N & 1 & N-2r & \rep{\overline N_s}^{\otimes_A(r-N+N_s)} \otimes \big( \rep{N_f}, \text{spin}\frac12 \big)^{\otimes_S \, r}
\end{array}
\ee
Here $\rep{N_s}$ and $\rep{N_f}$ refer to the fundamentals of $SU(N_s)$ and $SU(N_f)$, respectively, while $\otimes_S$ and $\otimes_A$ are the symmetric and antisymmetric products. In the first line, ``spin'' refers to the particular spin representation of each term in the product, which depends on the number of derivatives in $\partial_\bullet$ as explained above. In the second line we used $\rep{N_s}^{\otimes_A(N-r)} \cong \rep{\overline N_s}^{\otimes_A(r-N+N_s)}$.

In $U(k)_{-N+\frac{N_s}2}$ with $N_f$ $\phi$, $N_s$ $\psi$ the bare CS level is $-(N-N_s)$ and therefore the basic bare monopole $\cM$ transforms in the $(N-N_s)^\text{th}$ symmetric power of the antifundamental representation under the gauge group. The gauge-invariant monopole operators corresponding to the baryonic operators (\ref{baryons I})-(\ref{baryons II}) are thus
\be
\label{monopoles I}
\cB^{(r)} = \cM_{\alpha_1 \ldots \alpha_{N-N_s}} \; \phi^{\alpha_1 B_1} \dots \phi^{\alpha_r B_r} \; \tilde\partial_\bullet \psi^{\alpha_{r+1} I_{N_s+1}} \dots \tilde\partial_\bullet \psi^{\alpha_{N-N_s} I_{N-r}}
\ee
for $0 \leq r \leq N-N_s$, and
\be
\label{monopoles II}
\cB^{(r)} = \cM_{\alpha_1 \ldots \alpha_{N-N_s}} \; \phi^{\alpha_1 B_1} \dots \phi^{\alpha_r B_r} \; \overline\psi_{\alpha_{N-N_s+1} I_1} \ldots \overline\psi_{\alpha_r I_{r-N+N_s}}
\ee
for $N-N_s \leq r \leq N$. In the first class (\ref{monopoles I}) we recall that the fields $\phi$ in the monopole background carry spin $\frac12$, while for the fields $\psi$ we take the ground states with spin $1$ and identify $\partial_\bullet = \tilde\partial_\bullet \partial_\mu$ as before---more precisely each $\psi$ is in a state $Y^1_{j,j_3}$. In the second class (\ref{monopoles II}), instead, for the fields $\overline\psi$ we take the ground state $Y^0_{0,0}$ with spin $0$. In this way we precisely reproduce the quantum numbers in (\ref{quantum numbers baryons/monopoles}).


\section{\matht{U/U} duality}
\label{sec: U/U duality}

The second duality we consider involves Chern-Simons gauge theories with unitary groups, but with a different level for the $SU$ and the $U(1)$ parts, as well as bosonic and fermionic matter in the fundamental representation (which is complex). We propose the following duality:
\begin{multline}
\label{U/U duality grav}
U(N)_{k - \frac{N_f}2, k - \frac{N_f}2 \pm N} \text{ with $N_s$ $\phi$, $N_f$ $\psi$} \qquad\longleftrightarrow \\
U(k)_{-N + \frac{N_s}2, -N + \frac{N_s}2 \mp k } \text{ with $N_f$ $\phi$, $N_s$ $\psi$ } \times U\big( k(N-N_s) \pm 1 \big)_1
\end{multline}
for
\be
\label{U/U duality range}
N \geq N_s \;,\qquad k \geq N_f \;,\qquad (N,k) \neq (N_s, N_f) \;.
\ee
Without matter, the notation $U(N)_{k_1, k_1 + Nk_2}$ represents the Chern-Simons theory with Lagrangian
\be
\label{U_k1,k2 Lagrangian}
\cL_\text{CS} = \frac{k_1}{4\pi} \Tr_N \Big( bdb - \frac{2i}3 b^3 \Big) + \frac{k_2}{4\pi} (\Tr_N b) d (\Tr_N b)
\ee
while $U(N)_k \equiv U(N)_{k,k}$. In (\ref{U/U duality grav}) the bare CS level in the Lagrangian are $k, k\pm N$ on the LHS and $N_s-N, N_s-N\mp k$ on the RHS. The theory on the LHS of (\ref{U/U duality grav}) includes a mixed coupling $+\cO_\text{m}$ (\ref{def mixed couplings}) in the potential, while the theory on the RHS includes $-\cO_\text{m}$. As in Section~\ref{sec: SU/U duality}, those couplings are crucial to reproduce the same phase diagram. On the RHS, the trivial spin-TQFT $U\big(k(N - N_s) \pm1 \big)_1$ represents the gravitational coupling $-2\big( k(N-N_s) \pm1\big) \text{CS}_g$.

\begin{table}[ht!]
{\renewcommand{\arraystretch}{1.42}\begin{center}
$
\begin{array}{|lll|}
\hline
\multicolumn{3}{|c|}{U(N)_{k - \frac{N_f}2, k - \frac{N_f}2 \pm N} \text{ with $N_s$ $\phi$, $N_f$ $\psi$}} \text{ (LHS)}\\[.5em]
\hline\hline
m_\psi>0: &	U(N)_{k,k\pm N} \text{ with $N_s$ $\phi$ } \times U(0)_1 & \qquad\; \\
	& \multicolumn{2}{r|}{A:\; U(N)_{k,k \pm N} \times U(0)_1} \\
m_\phi^2>0: &	U(N)_{k-\frac{N_f}2, k - \frac{N_f}2 \pm N} \text{ with $N_f$ $\psi$ } & \\
	& \multicolumn{2}{r|}{B:\; U(N)_{k-N_f,k - N_f \pm N} \times U(NN_f)_1} \\
m_\psi<0: &	U(N)_{k-N_f,k - N_f \pm N} \text{ with $N_s$ $\phi$ } \times U(NN_f)_1 & \\
	& \multicolumn{2}{r|}{C:\; U(N-N_s)_{k-N_f,k-N_f \pm(N-N_s)} \times U(NN_f)_1} \\
m_\phi^2<0,\, m_\psi<0: & N_s N_f\, \psi \times U(N-N_s)_{k-N_f,k-N_f\pm(N-N_s)} \times U\big(N_f (N-N_s) \big)_1 & \\
	& \multicolumn{2}{r|}{D:\; U(N-N_s)_{k-N_f,k-N_f\pm (N-N_s)} \times U\big( N_f(N-N_s) \big)_1} \\
m_\phi^2<0: &	U(N-N_s)_{k - \frac{N_f}2, k - \frac{N_f}2 \pm (N-N_s)} \text{ with $N_f$ $\psi$ } & \\
	& \multicolumn{2}{r|}{E:\; U(N-N_s)_{k,k\pm (N-N_s)} \times U(0)_1} \\
\hline
\end{array}
$ \\[1.5em]
$
\begin{array}{|lll|}
\hline
\multicolumn{3}{|c|}{U(k)_{-N + \frac{N_s}2, - N + \frac{N_s}2 \mp k} \text{ with $N_f$ $\phi$, $N_s$ $\psi$ } \times U\big( k(N-N_s) \pm1 \big)_1 } \text{ (RHS)}\\[.5em]
\hline\hline
m_\phi^2>0: &	U(k)_{-N + \frac{N_s}2,-N + \frac{N_s}2 \mp k} \text{ with $N_s$ $\psi$ } \times U\big( k(N-N_s) \pm1 \big)_1 & \\
	& \multicolumn{2}{r|}{A:\; U(k)_{-N,-N\mp k} \times U(kN\pm1)_1} \\
m_\psi<0: &	U(k)_{-N,-N\mp k} \text{ with $N_f$ $\phi$ } \times U(kN \pm1)_1 & \\
	& \multicolumn{2}{r|}{B:\; U(k-N_f)_{-N,-N \mp(k-N_f)} \times U(kN\pm1)_1} \\
m_\phi^2<0: &	U(k-N_f)_{-N + \frac{N_s}2, -N + \frac{N_s}2 \mp (k-N_f)} \text{ with $N_s$ $\psi$ } \times U(kN - kN_s + N_fN_s \pm1)_1 & \\
	& \multicolumn{2}{r|}{C:\; U(k-N_f)_{-N + N_s, -N + N_s \mp(k-N_f)} \times U(kN - kN_s + N_fN_s \pm 1)_1} \\
\makebox[0pt][l]{\raisebox{.7em}[0pt]{$m_\phi^2<0$,}}\raisebox{-.7em}[0pt][0pt]{$m_\psi>0$:} & N_s N_f\, \psi \times U(k-N_f)_{-N + N_s, -N + N_s \mp(k-N_f)} \times U\big( k(N-N_s) \pm1 \big)_1 & \\
	& \multicolumn{2}{r|}{D:\; U(k-N_f)_{-N + N_s, -N + N_s \mp (k-N_f)} \times U\big( k(N-N_s) \pm1\big)_1} \\
m_\psi>0: &	U(k)_{-N+N_s, -N + N_s \mp k} \text{ with $N_f$ $\phi$ } \times U\big( k(N-N_s) \pm1 \big)_1 & \\
	& \multicolumn{2}{r|}{E:\; U(k)_{-N+N_s, -N + N_s \mp k} \times U\big( k(N-N_s) \pm 1\big)_1} \\
\hline
\end{array}
$
\end{center}}
\caption{Phase diagram of the $U/U$ dualities. These tables are valid for $N_s \leq N$ and $N_f \leq k$.
\label{tab: U/U dualities}}
\end{table}

As noted at the end of Section~\ref{sec: SU/U coupling to background}, this duality can be derived from the $SU/U$ duality. One couples a $U(1)$ global symmetry---the one that is a baryonic symmetry on one side and a magnetic symmetry on the other side---to a gauge field $c$, adds a suitable CS conterterm, and makes $c$ dynamical. Repeating the process, one can conversely derive the $SU/U$ duality from the $U/U$ duality.

The various phases and critical lines for the $U/U$ dualities, in the case $N\geq N_s$ and $k \geq N_f$, are reported in Table \ref{tab: U/U dualities}. Using the dualities in \cite{Hsin:2016blu}, the two phase diagrams match (including the gravitational couplings) in the claimed range of parameters. Notice that for $N = N_s$ (and $k \geq N_f$ on the LHS, or $k > N_f$ on the RHS) the horizontal line in the left half plane (corresponding to $m_\phi^2<0$ on the LHS and $m_\psi>0$ on the RHS) disappears since it is gapped---moreover phases $D$ and $E$ are identical. We reduce for $k>N_f$ to the duality $U(k)_{0,\mp k}$ with $N_f$ $\phi$ $\times \, U(\pm 1)_1 \leftrightarrow U(0)_1$, expressing confinement. The same happens for $k = N_f$ (and $N>N_s$ on the LHS, or $N\geq N_s$ on the RHS): the vertical line in the lower half plane (corresponding to $m_\psi<0$ on the LHS and $m_\phi^2<0$ on the RHS) disappears because it is gapped, and phases $B$ and $C$ coincide.

\begin{figure}[t!]
\centering
\hspace{\stretch{1}}
\begin{tikzpicture}[scale=1]
\draw [->] (-2.5,0)--(2.5,0) node[below] {$m_\phi^2$}; \draw [->] (0,-2.5)--(0,2.5) node[above] {$m_\psi$};
\draw [very thick, blue!80!black] (-2,0)--(2,0); \draw [very thick, blue!80!black] (0,-2)--(0,2);
\draw [->, thick] (0,0) ++(220:1.8) arc (220:185:1.8);
\fill [lightgray!70!yellow, fill opacity=.6] (0,0) circle [radius=.4];
\draw [densely dashed] (-.6,-.6)--(.6,.6);
\node at (1.8,1.2) {$U(1)_2 \times U(0)_1$};
\node at (-1.5,1.2) {$U(0)_1$};
\node at (1.5, -1.2) {$U(2)_1$};
\node at (-1.0, -1.8) {$U(1)_1$};
\end{tikzpicture}
\hspace{\stretch{1}}
\begin{tikzpicture}[scale=1]
\draw [->] (2.5,0)--(-2.5,0) node[below] {$m_\psi$}; \draw [->] (0,-2.5)--(0,2.5) node[above] {$m_\phi^2$};
\draw [very thick, blue!80!black] (-2,0)--(2,0); \draw [very thick, blue!80!black] (0,-2)--(0,2);
\draw [->, thick] (0,0) ++(230:1.8) arc (230:265:1.8);
\fill [lightgray!70!yellow, fill opacity=.6] (0,0) circle [radius=.4];
\draw [densely dashed] (-.6,-.6)--(.6,.6);
\node at (1.8,1.2) {$U(1)_{-2} \times U(2)_1$};
\node at (-1.5,1.2) {$U(0)_1$};
\node at (1.5, -1.2) {$U(2)_1$};
\node at (-1.9, -1.1) {$U(1)_1$};
\end{tikzpicture}
\hspace{\stretch{1}}
\caption{$\qquad U(1)_{3/2}$ with $\phi,\psi$ $\qquad\longleftrightarrow\qquad$ $U(1)_{-3/2}$ with $\phi,\psi$ $\times\, U(1)_1$. $\quad$ Phase diagram. \newline
We emphasized a quantum time-reversal symmetry (with an anomaly) with respect to the dashed line.
\label{fig: U(1)3/2}}
\end{figure}

The cases $N=N_s$, $k = N_f$ should be studied separately, since the phases in Table \ref{tab: U/U dualities} do not match directly. Consider first the two Abelian cases with $N=N_s = k = N_f = 1$. The case with upper sign is
\be
\label{U(1)3/2 duality}
U(1)_{\frac32} \text{ with $\phi$, $\psi$} \qquad\longleftrightarrow\qquad U(1)_{-\frac32} \text{ with $\phi$, $\psi$ } \times U(1)_1 \;.
\ee
The phase diagram is summarized in Figure~\ref{fig: U(1)3/2}, taking into account that there is no independent $\cO_\text{m}$ coupling, while we expect $\cO_\text{d}$ to be irrelevant. The corresponding shift of the critical lines is indicated by arrows. Comparing the gapless lines after such a shift we find:
\be
\label{scheme deformations U(1)3/2}
{\renewcommand{\arraystretch}{1.5}
\begin{array}{ll|ll}
\multicolumn{2}{c|}{U(1)_{3/2} \text{ with $\phi$, $\psi$}} & \multicolumn{2}{c}{U(1)_{-3/2} \text{ with $\phi$, $\psi$ } \times U(1)_1} \\
\hline
m_\psi>0: & U(1)_2 \text{ with $\phi$ } \times U(0)_1 & m_\phi^2>0: & U(1)_{-\frac32} \text{ with $\psi$ } \times U(1)_1 \\
m_\phi^2>0: & U(1)_\frac32 \text{ with $\psi$ } & m_\psi<0: & U(1)_{-2} \text{ with $\phi$ } \times U(2)_1 \\
m_\psi<0: & U(1)_1 \text{ with $\phi$ } \times U(1)_1 & m_\phi^2<0: & \psi \times U(1)_1 \\
m_\phi^2<0: & \psi & m_\psi>0: & U(1)_{-1} \text{ with $\phi$ } \times U(1)_1
\end{array}}
\ee
We find a perfect match, making use of the dualities in \cite{Seiberg:2016gmd, Hsin:2016blu}. We thus conjecture that this duality is correct. In fact in Section~\ref{sec: U(1) 3/2 time rev} we derive this duality from the Abelian dualities of \cite{Seiberg:2016gmd}. This duality expresses the fact that the theory has a time-reversal invariant line in its phase diagram, with an anomaly. Applying a time-reversal transformation to the LHS of (\ref{U(1)3/2 duality}) (see Appendix \ref{app: spinc}) and then using the duality we can write
\be
U(1)_\frac32 \text{ with $\phi$, $\psi$} \quad\stackrel{T}{\longrightarrow}\quad U(1)_{-\frac32} \text{ with $\phi$, $\psi$ } \times U(1)_{-1} \;\stackrel{\text{duality}}{\cong}\; U(1)_\frac32 \text{ with $\phi$, $\psi$ } \times U(2)_{-1} \;.
\ee
Therefore time reversal is a quantum symmetry of the theory, up to the anomalous shift of the gravitational coupling (the counterterms for global symmetries also suffer from anomalous shifts). The action of this time-reversal symmetry on the mass operators is $|\phi|^2 \;\stackrel{T}{\longleftrightarrow}\; \bar\psi\psi$, hence the theory is time-reversal invariant along the line $m_\phi^2 = m_\psi$ (dashed in Figure~\ref{fig: U(1)3/2}) while phases at opposite points with respect to the line are related by time reversal.

\begin{figure}[t!]
\centering
\hspace{\stretch{1}}
\begin{tikzpicture}[scale=1]
\draw [->] (-2.5,0)--(2.5,0) node[below] {$m_\phi^2$}; \draw [->] (0,-2.5)--(0,2.5) node[above] {$m_\psi$};
\draw [very thick, blue!80!black] (-2,0)--(2,0); \draw [very thick, blue!80!black] (0,-2)--(0,2);
\draw [->, thick] (0,0) ++(220:1.8) arc (220:185:1.8);
\fill [lightgray!70!yellow, fill opacity=.6] (0,0) circle [radius=.4];
\draw [densely dashed] (-.6,-.6)--(.6,.6);
\node at (1.8,1.2) {$S^1 \times U(0)_1$};
\node at (-1.5,1.2) {$U(0)_1$};
\node at (1.5, -1.2) {$U(0)_1$};
\node at (-1.0, -1.8) {$U(1)_1$};
\node at (1.8,.3) {$\phi$}; \node at (-1.5,.3) {$\psi$};
\fill [white]  (-.5,1.6) rectangle (.5,2.2) node[midway, black] {$\phi$};
\fill [white]  (-.5,-1.6) rectangle (.5,-2.2) node[midway, black] {$\psi$};
\end{tikzpicture}
\hspace{\stretch{1}}
\begin{tikzpicture}[scale=1]
\draw [->] (2.5,0)--(-2.5,0) node[below] {$m_\psi$}; \draw [->] (0,-2.5)--(0,2.5) node[above] {$m_\phi^2$};
\draw [very thick, blue!80!black] (-2,0)--(2,0); \draw [very thick, blue!80!black] (0,-2)--(0,2);
\draw [->, thick] (0,0) ++(230:1.3) arc (230:265:1.3);
\fill [lightgray!70!yellow, fill opacity=.6] (0,0) circle [radius=.4];
\draw [densely dashed] (-.6,-.6)--(.6,.6);
\node at (1.8,1.2) {$S^1 \times U(0)_1$};
\node at (-1.5,1.2) {$U(0)_1$};
\node at (1.5, -1.2) {$U(0)_1$};
\node at (-1.9, -1.1) {$U(1)_{-1}$};
\node at (1.8,.3) {$\phi$}; \node at (-1.5,.3) {$\psi$};
\fill [white]  (-.5,1.6) rectangle (.5,2.2) node[midway, black] {$\phi$};
\fill [white]  (-.2,-1.6) rectangle (.2,-2.2) node[midway, black] {$\psi$};
\end{tikzpicture}
\hspace{\stretch{1}}
\caption{$\qquad U(1)_{-1/2}$ with $\phi,\psi$ $\qquad\text{vs.}\qquad$ $U(1)_{1/2}$ with $\phi,\psi$ $\times\,U(-1)_1$. $\;$ Phase diagrams. \newline
The two theories are not dual as the phases do not match (the two theories are mapped into each other by time reversal). However each diagram is symmetric with respect to the dashed line, due to a self-duality.
\label{fig: U(1)1/2}}
\end{figure}
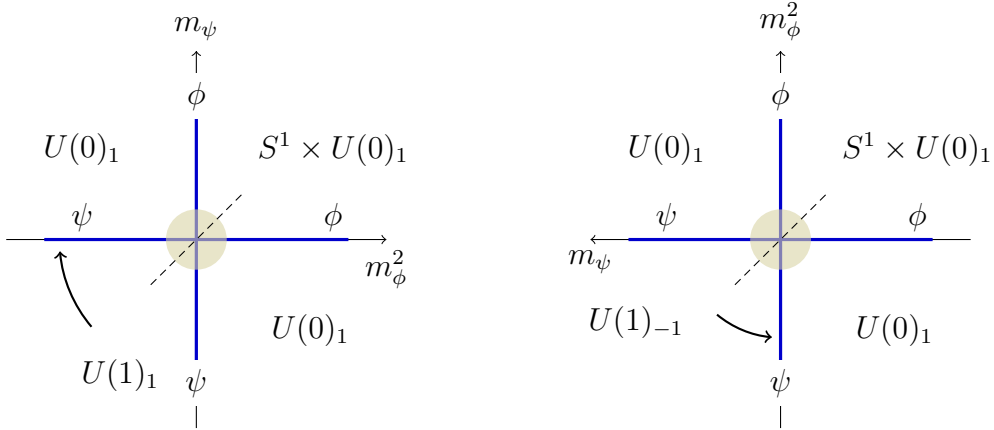

The case with lower sign is
\be
\label{non duality U(1) 1/2}
U(1)_{-\frac12} \text{ with } \phi,\psi \qquad\text{vs.}\qquad U(1)_{\frac12} \text{ with } \phi,\psi \times U(-1)_1 \;.
\ee
The two phase diagrams are schematically summarized in Figure~\ref{fig: U(1)1/2}. Also in this case, some of the phases (gapless and gapped) do not match. However, as opposed to the previous case, they still do not match even after the shift of a gapless line due to the facts that $\cO_\text{m}$ is not an independent operator and we expect $\cO_\text{d}$ to be irrelevant. Comparing the gapless lines we find:
\be
\label{phase diagram U(1) 1/2 with phi psi}
{\renewcommand{\arraystretch}{1.5}
\begin{array}{ll|c|ll}
\multicolumn{2}{c|}{U(1)_{-1/2} \text{ with } \phi,\psi} & & \multicolumn{2}{c}{U(1)_{1/2} \text{ with $\phi,\psi$ } \times U(-1)_1} \\
\hline
m_\psi>0: & U(1)_0 \text{ with $\phi$ } \times U(0)_1 & = & m_\phi^2>0: & U(1)_{\frac12} \text{ with $\psi$ } \times U(-1)_1 \\
m_\phi^2>0: & U(1)_{-\frac12} \text{ with $\psi$} & = & m_\psi<0: & U(1)_0 \text{ with $\phi$ } \times U(0)_1 \\
m_\psi<0: & U(1)_{-1} \text{ with $\phi$ } \times U(1)_1 & \raisebox{-.1em}{\text{\Large$\times$}} & m_\phi^2<0: & \psi \times U(-1)_1 \\
m_\phi^2<0: & \psi & \raisebox{-.1em}{\text{\Large$\times$}} & m_\psi>0: & U(1)_1 \text{ with $\phi$ } \times U(-1)_1
\end{array}}
\ee
In the first two lines there is a precise match, including the gravitational couplings. In the last two lines, instead, there is a match of degrees of freedom---a free fermion in all cases---but the gravitational couplings do not match on the two sides. Thus, we will not regard this as a good duality. (Notice that the two theories in (\ref{non duality U(1) 1/2}) are mapped into each other by time reversal).

From the phase diagram in Figure~\ref{fig: U(1)1/2} and from (\ref{phase diagram U(1) 1/2 with phi psi}), looking at the LHS for concreteness, one might suspect that the two gapped phases $U(0)_1$, the two gapless lines $\phi$ (which represent the $O(2)$ Wilson-Fisher fixed point) and the two gapless lines $\psi$, respectively, are identical. In that case it would be natural to expect that the line $\phi$ and the line $\psi$ do not touch, and $U(0)_1$ is one connected phase with no phase transitions in the middle. However, a closer inspection of the counterterms for global symmetries reveals that they are different in the two phases $U(0)_1$---see Section~\ref{sec: U(1)-1/2 phi psi phases with background}. This implies that the two phases are different, that they must be separated by a phase transition and thus that the gapless lines must meet.

Although the two theories in (\ref{non duality U(1) 1/2}) do not seem to be dual---at least in the simple way discussed in this paper---at thus we do not see an emergent time-reversal symmetry, yet each of the phase diagrams in Figure~\ref{fig: U(1)1/2} appears to be symmetric with respect to the dashed line. In fact, as we discuss in Section~\ref{sec: U(1)-1/2 phi psi}, each of the two theories has a self-duality (with an anomaly) that exchanges the scalar with the fermion, thus explaining the specularity of its phase diagram. The self-duality maps $|\phi|^2 \leftrightarrow \bar\psi\psi$.

The other cases with $N=N_s$, $k=N_f$ and $Nk>1$ can be studied in a similar way. Some of the phases, denoted in Figure~\ref{fig: masks} as C and D, do not match. We notice that on the LHS the phases D and E are the same phase (because there is no gapless line between them), and similarly on the RHS the phases B and C are the same phase. We should then identify phase C on the LHS with phase D on the RHS, however for $Nk>1$ they are different. Therefore, even taking into account possible shifts of the gapless lines, we do not find evidence of a duality and discard this case. (We also do not find evidence of a self-duality.)

As discussed in Section~\ref{sec: RG flows} for the $SU/U$ dualities, also the proposed $U/U$ dualities are consistent under RG flows triggered by a mass term---either positive or negative---for a single scalar or fermion. Starting with a $U/U$ duality and integrating out a single matter field, possibly taking into account a partial breaking of the gauge group, one obtains another $U/U$ duality with smaller values of $N,k, N_s, N_f$ as in (\ref{RG flow pattern}) that remain within the range (\ref{U/U duality range}).

\subsection{Global symmetry, background fields and monopole operators}

Let us first determine the global symmetry that acts faithfully on gauge-invariant operators in the theory
\be
U(N)_{k - \frac{N_f}2, k - \frac{N_f}2 + j N} \text{ with $N_s$ $\phi$, $N_f$ $\psi$}
\ee
for generic integer values of $N, k, N_s, N_f, j$ and $N_s, N_f \geq 1$, independently of the dualities. The case $j=0$ is analyzed in Section~\ref{sec: faithful symmetry}. First of all there is $\bZ_2^\cC$ charge conjugation symmetry (and time-reversal symmetry for $k = \frac{N_f}2 \in \bZ$ and $j=0$). We write
\be
G = \wh G \times \bZ_2^\cC \;.
\ee
To determine $\wh G$ we use the same argument as in Section~\ref{sec: faithful symmetry}. There is a $U(1)_M$ magnetic symmetry. The bare CS levels correspond to the Lagrangian terms
\be
\cL_\text{CS} = \frac k{4\pi} \Tr_N \Big( bdb - \frac{2i}3 b^3 \Big) + \frac j{4\pi} (\Tr_N b) d (\Tr_N b)
\ee
where $b$ is the dynamical $U(N)$ gauge field, therefore a monopole operator of magnetic charge $1$ has charge $k + jN$ under the gauged diagonal $U(1) \subset U(N)$. Since fundamentals have charge $1$ under that $U(1)$, the symmetry group is
\be
\wh G = \frac{U(N_s) \times U(N_f) \times U(1)_M}{U(1)_*} \qquad\qquad U(1)_* = \big( e^{2\pi i \alpha}, e^{2\pi i \alpha}, e^{2\pi i(k + jN)\alpha} \big)
\ee
with $\alpha \in [0,1)$. For $k + jN \neq 0$ we can use $U(1)_*$ to remove $U(1)_M$. Thus we can write
\be
\wh G = \frac{U(N_s) \times U(N_f)}{\bZ_{|k + jN|}} \qquad\qquad\text{for $k+jN \neq 0$}
\ee
and
\be
\wh G = \frac{U(N_s) \times U(N_f)}{U(1)} \times U(1)_M \qquad\qquad\text{for $k+jN=0$}
\ee
where the quotient is by the diagonal $U(1)$.

In the dualities (\ref{U/U duality grav}) we have $j= \pm1$. It is easy to check that the faithful global symmetry agrees on the two sides of the duality, exploiting the isomorphisms (\ref{symmetry isomorphism I}) and (\ref{symmetry isomorphism II}).

Next, we can identify the relative counterterms on the two sides of the duality, for background fields coupled to the global symmetry. The counterterms for the $SU(N_s) \times SU(N_f)$ factor of the global symmetry are exactly the same as in the $SU/U$ dualities, written in (\ref{SU/U duality with SU background}). The counterterms for the $U(1)^2$ factor of the global symmetry, as well as the precise map of the two Abelian global symmetry factors across the duality, are conveniently captured by the Lagrangian form of the duality, as derived at the end of Section~\ref{sec: SU/U coupling to background} from the $SU/U$ duality:
\begin{align}
\label{U/U duality with U(1)2 background}
\cL_\text{LHS} &= |D_{b+A}\phi|^2 + i \bar\psi \Dslash_{b-A} \psi - \phi^4 - \phi^2\psi^2 + \frac{k}{4\pi} \Tr_N \Big( bdb- \frac{2i}3 b^3 \Big) \\
&\quad \pm \frac1{4\pi} (\Tr_N b) d (\Tr_N b) + \frac1{2\pi} (\Tr_N b)dB \nn\\
\cL_\text{RHS} &= |D_{f-A}\phi|^2 + i \bar\psi \Dslash_{f+A} \psi - \phi^4 + \phi^2\psi^2 - \frac{N-N_s}{4\pi} \Tr_k \Big( fdf - \frac{2i}3 f^3 \Big) \mp \frac1{4\pi} (\Tr_k f)d(\Tr_k f) \nn\\
&\quad + \frac1{2\pi} (\Tr_k f)d (\mp B + N_s A) \mp \frac1{4\pi} BdB + \frac{N_s k}{4\pi} AdA - 2 \big( k (N-N_s) \pm 1 \big) \text{CS}_g \;. \nn
\end{align}
Here $b,f$ are dynamical $U(N)$ and $U(k)$ gauge fields, respectively, while $A,B$ are background $U(1)$ gauge fields.

Finally, we can verify that the map of Abelian global symmetry factors implied by (\ref{U/U duality with U(1)2 background}) is consistent with the map of basic monopole operators between the dual theories in (\ref{U/U duality grav}). Consider first the theory $U(N)_{k- \frac{N_f}2, k - \frac{N_f}2 + N}$ with $N_s$ $\phi$, $N_f$ $\psi$ (duality with upper sign). The bare CS levels for the $SU(N)$ and $U(1)$ gauge factors are $k$ and $k+N$, respectively. The simplest bare monopole $\cM$ has magnetic gauge fluxes $(1, 0, \dots ,0)$ under the maximal torus $U(1)^N$ (up to Weyl transformations), breaking the gauge group to $U(1) \times U(N-1)$. Because of CS interactions, $\cM$ has charge $k+N$ under $U(1)$ and it transforms as the highest weight of the symmetric $k^\text{th}$ power of the fundamental representation of $SU(N)$. To form a gauge-invariant operator, we should dress it with $k+N$ fields transforming in the antifundamental representation of $U(N)$, $N$ of which are contracted into an $SU(N)$ singlet---an ``anti-baryon''. The simplest gauge invariants constructed with the highest weight can be schematically written as
\be
\label{U monopoles 1}
\cB^{(r_1, r_2)} = \cM \;\; \underbrace{ \underbrace{\wb\phi_{1I}}_{r_1} \;\; \underbrace{\tilde\partial_\bullet \, \wb\psi_{1B}}_{k-r_1} }_{k} \;\; \underbrace{ \underbrace{\wb\psi_{\dots B}}_{r_2} \;\; \underbrace{ \wb\phi_{\dots I}}_{N_s} \;\; \underbrace{ \partial_\bullet\, \wb\phi_{\dots I}}_{N-N_s - r_2} }_N
\ee
for $0 \leq r_2 \leq N-N_s$ and
\be
\label{U monopoles 2}
\cB^{(r_1, r_2)} = \cM \;\; \underbrace{ \underbrace{\wb\phi_{1I}}_{r_1} \;\; \underbrace{\tilde\partial_\bullet \, \wb\psi_{1B}}_{k-r_1} }_{k} \;\; \underbrace{ \underbrace{\wb\psi_{\dots B}}_{r_2} \;\; \underbrace{ \wb\phi_{\dots I}}_{N - r_2} }_N
\ee
for $N-N_s \leq r_2 \leq N$, with $0\leq r_1 \leq k$ in both cases. We assumed $N_s \leq N$. A gauge index ``$1$'' corresponds to the lowest weight of the antifundamental representation, a gauge index ``$\dots$'' is antisymmetrized, while $I$ and $B$ are antifundamental (because lower) indices of $SU(N_s)$ and $SU(N_f)$, respectively. The notations $\partial_\bullet$ and $\tilde\partial_\bullet$ are explained in Section~\ref{sec: SU/U monopoles}: they indicate the smallest number of different derivatives that make the operator non-vanishing after antisymmetrization, such a number can be zero for $\tilde\partial_\bullet$ but not for $\partial_\bullet$, and we should not use $\partial^2$ in neither of the two expressions. All fields with a gauge index ``$1$'' feel the monopole background and have modes with spin shifted by $\frac12$ (among the antisymmetrized indices there is only one ``$1$''). In particular the first group of modes have spin $\frac12$ (harmonics $Y^{1/2}_{1/2,\pm1/2}$), while in the second group we take the mode of spin $0$ before taking derivatives (harmonics $Y^0_{j,j_3}$). From the groups with antisymmetrized gauge indices we get one extra $\text{spin}\frac12$ representation. The quantum numbers of these operators are
\be
\label{quantum numbers U monopoles}
\begin{array}{c|c|c}
\quad \cB^{(r_1, r_2)} \quad & \quad U(1)_B:\; 1 \quad & U(1)_A:\; k - N + 2(r_2 - r_1) \\[.5em]
\hline
\multicolumn{3}{l}{\rule{0pt}{1.6em}
\big(\rep{\wb N_s}, \text{spin}\frac12 \big)^{\otimes_S \, r_1} \otimes \big( \rep{\wb N_f}, \text{spin}\frac12 \big)^{\otimes_S\, r_2} \otimes \big( \rep{\wb N_f}, \text{spin}_i \big)^{\otimes_A(k-r_1)} \otimes \big( \rep{\wb N_s}, \text{spin}_j \big)^{\otimes_A (N - N_s  - r_2)} \otimes \text{spin}\tfrac12
}\\[.6em]
\hline
\multicolumn{3}{l}{\rule{0pt}{1.6em}
\big(\rep{\wb N_s}, \text{spin}\frac12 \big)^{\otimes_S \, r_1} \otimes \big( \rep{\wb N_f}, \text{spin}\frac12 \big)^{\otimes_S\, r_2} \otimes \big( \rep{\wb N_f}, \text{spin}_i \big)^{\otimes_A(k-r_1)} \otimes \rep{\wb N_s}^{\otimes_A (N - r_2)} \otimes \text{spin}\tfrac12
}
\end{array}
\ee
The second and third row refer to (\ref{U monopoles 1}) and (\ref{U monopoles 2}), respectively. We used that the fourth group of fields in (\ref{U monopoles 1}) is a total singlet. 

In the dual theory $U(k)_{-N + \frac{N_s}2, - N + \frac{N_s}2 - k}$ with $N_f$ $\phi$, $N_s$ $\psi$, the corresponding monopole operators  are constructed in a similar way. From (\ref{U/U duality with U(1)2 background}) the basic bare monopole with charge $1$ under $U(1)_B$ has magnetic fluxes $(-1, 0, \dots, 0)$ under the maximal torus $U(1)^N$, and we indicate it as $\wb\cM$. Such a bare monopole has charge $N-N_s+k$ under the diagonal gauge $U(1)$, and it transforms as the highest weight of the symmetric $(N-N_s)^\text{th}$ power of the fundamental representation of $SU(k)$. The basic gauge-invariant operators are then
\be
\label{U monopoles 3}
\cB^{(r_1, r_2)} = \wb\cM \;\; \underbrace{ \underbrace{\wb\phi_{1B}}_{r_2} \;\; \underbrace{\tilde\partial_\bullet \, \wb\psi_{1I}}_{N-N_s-r_2} }_{N-N_s} \;\; \underbrace{ \underbrace{\wb\psi_{\dots I}}_{r_1} \;\; \underbrace{ \tilde\partial_\bullet \, \wb\phi_{\dots B}}_{k-r_1} }_k
\ee
for $0 \leq r_2 \leq N-N_s$ and
\be
\label{U monopoles 4}
\cB^{(r_1, r_2)} = \wb\cM \;\; \underbrace{\wb\phi_{1B}}_{r_2} \;\; \underbrace{\psi^{1I}}_{r_2-N+N_s} \;\; \underbrace{ \underbrace{\wb\psi_{\dots I}}_{r_1} \;\; \underbrace{ \tilde\partial_\bullet \, \wb\phi_{\dots B}}_{k-r_1} }_k
\ee
for $N-N_s \leq r_2 \leq N$. In the second group of fields in (\ref{U monopoles 3}) we take the modes of spin $1$ and identify $\partial_\bullet = \tilde\partial_\bullet \partial_\mu$ with the last group in (\ref{U monopoles 1}) (precisely, we use the harmonics $Y^1_{j,j_3}$ where $j$ equals the spacetime spin of $\partial_\bullet \wb\phi$), while in the second group in (\ref{U monopoles 4}) we take the mode of spin $0$ (harmonic $Y^0_{0,0}$). The quantum numbers of these operators are exactly the same as in (\ref{quantum numbers U monopoles}).

The basic monopole operators in the theories involved in the $U/U$ dualities (\ref{U/U duality grav}) with lower sign are constructed in a similar way. The only difference is that on the LHS we use a ``baryon'' (as opposed to an anti-baryon) to dress the bare monopole, while on the RHS we use a bare monopole $\cM$ (as opposed to the anti-monopole $\wb\cM$) dressed by fields in the fundamental times an anti-baryon. The quantum numbers match in that case too.


\section{\matht{USp} duality}
\label{sec: USp duality}

\begin{table}[t!]
{\renewcommand{\arraystretch}{1.43}
$$
\begin{array}{|lll|}
\hline
\multicolumn{3}{|c|}{USp(2N)_{k - \frac{N_f}2} \text{ with $N_s$ $\phi$, $N_f$ $\psi$}} \text{ (LHS)}\\[.5em]
\hline\hline
m_\psi>0: &	USp(2N)_k \text{ with $N_s$ $\phi$ } \times U(0)_1 & \qquad\qquad \\
	& \multicolumn{2}{r|}{A:\; USp(2N)_k \times U(0)_1} \\
m_\phi^2>0: &	USp(2N)_{k-\frac{N_f}2} \text{ with $N_f$ $\psi$} & \\
	& \multicolumn{2}{r|}{B:\; USp(2N)_{k-N_f} \times U(2NN_f)_1} \\
m_\psi<0: &	USp(2N)_{k-N_f} \text{ with $N_s$ $\phi$ } \times U(2NN_f)_1 & \\
	& \multicolumn{2}{r|}{C:\; USp\big(2(N-N_s)\big)_{k-N_f} \times U(2NN_f)_1} \\
m_\phi^2<0,\, m_\psi<0: & 2 N_s N_f \, \psi \times USp\big(2(N-N_s)\big)_{k-N_f} \times U\big(2N_f (N-N_s) \big)_1 & \\
	& \multicolumn{2}{r|}{D:\; USp\big(2(N-N_s)\big)_{k-N_f} \times U\big( 2N_f(N-N_s) \big)_1} \\
m_\phi^2<0: &	USp\big(2(N-N_s)\big)_{k - \frac{N_f}2} \text{ with $N_f$ $\psi$} & \\
	& \multicolumn{2}{r|}{E:\; USp\big(2(N-N_s)\big)_k \times U(0)_1} \\
\hline
\end{array}
$$
$$
\begin{array}{|lll|}
\hline
\multicolumn{3}{|c|}{USp(2k)_{-N + \frac{N_s}2} \text{ with $N_f$ $\phi$, $N_s$ $\psi$ } \times U\big( 2k(N-N_s) \big)_1} \text{ (RHS)}\\[.5em]
\hline\hline
m_\phi^2>0: &	USp(2k)_{-N + \frac{N_s}2} \text{ with $N_s$ $\psi$ } \times U\big( 2k(N-N_s) \big)_1 & \\
	& \multicolumn{2}{r|}{A:\; USp(2k)_{-N} \times U(2kN)_1} \\
m_\psi<0: &	USp(2k)_{-N} \text{ with $N_f$ $\phi$ } \times U(2kN)_1 & \\
	& \multicolumn{2}{r|}{B:\; USp\big(2(k-N_f)\big)_{-N} \times U(2kN)_1} \\
m_\phi^2<0: &	USp\big(2(k-N_f)\big)_{-N + \frac{N_s}2} \text{ with $N_s$ $\psi$ } \times U(2kN - 2kN_s + 2N_fN_s)_1 & \\
	& \multicolumn{2}{r|}{C:\; USp\big(2(k-N_f)\big)_{-N + N_s} \times U(2kN - 2kN_s + 2N_fN_s)_1} \\
m_\phi^2<0,\, m_\psi>0: & 2 N_s N_f \, \psi \times USp\big(2(k-N_f)\big)_{-N + N_s} \times U\big( 2k(N-N_s) \big)_1 & \\
	& \multicolumn{2}{r|}{D:\; USp\big(2(k-N_f)\big)_{-N + N_s} \times U\big( 2k(N-N_s) \big)_1} \\
m_\psi>0: &	USp(2k)_{-N+N_s} \text{ with $N_f$ $\phi$ } \times U\big( 2k(N-N_s) \big)_1 & \\
	& \multicolumn{2}{r|}{E:\; USp(2k)_{-N+N_s} \times U\big( 2k(N-N_s)\big)_1} \\
\hline
\end{array}
$$
}
\caption{Phase diagram for $USp$ dualities. These tables are valid for $N_s \leq N$ and $N_f \leq k$.
\label{tab: USp dualities}}
\end{table}

The third duality we consider involves Chern-Simons theories with (unitary) symplectic groups as well as bosonic and fermionic matter in the fundamental representation, which is pseudo-real. We propose the following duality:%
\footnote{In our notation $USp(2N)$ is the compact unitary symplectic group of rank $N$. In particular one identifies $USp(2) \cong SU(2)$. Elsewhere the notation $Sp(N)$ is used sometimes.}
\begin{multline}
\label{USp duality grav}
USp(2N)_{k - \frac{N_f}2} \text{ with } N_s\, \phi,\, N_f\, \psi \qquad\longleftrightarrow \\
USp(2k)_{-N + \frac{N_s}2} \text{ with $N_f$ $\phi$, $N_s$ $\psi$ } \times U\big( 2k(N-N_s) \big)_1 \;.
\end{multline}
We recall that in the symplectic case it is convenient to double the number of fields and impose a reality constraint. We use $\Phi_{\alpha I}$ and $\Psi_{\alpha A}$, where $\alpha=1,\dots,2N$ is for $USp(2N)$, $I = 1, \dots, 2N_s$ is for $USp(2N_s)$ and $A =1,\dots, 2N_f$ is for $USp(2N_f)$. We impose
\be
\Phi_{\alpha I}^* = \Omega^{\alpha\beta} \, \Omega^{IJ} \, \Phi_{\beta J} \;,\qquad\qquad \Psi_{\alpha A}^c = \Omega^{\alpha\beta} \, \Omega^{AB} \, \Psi_{\beta B}
\ee
where, with some abuse of notation, we have indicated by the same symbol $\Omega$ the three invariant symplectic forms of $USp(2N)$, $USp(2N_s)$ and $USp(2N_f)$, while $^c$ is the charge conjugate. Then, even before turning on any potential, the two theories in (\ref{USp duality grav}) are invariant under the faithfully-acting symmetry
\be
G = \frac{USp(2N_s) \times USp(2N_f)}{\bZ_2} \;,
\ee
where $\bZ_2$ is generated by $(-\unit, -\unit)$ that is part of the gauge group.

In both theories in (\ref{USp duality grav}) we include the following quartic interactions, that preserve the full symmetry $G$:
\bea
\label{USp relevant couplings}
& \big( \Phi_{\alpha I} \Phi_{\beta J} \Omega^{\alpha\beta} \Omega^{IJ} \big)^2 \\
& \big( \Phi_{\alpha I} \Phi_{\beta J} \Omega^{\alpha\beta} \big) \Omega^{JK} \big( \Phi_{\gamma K} \Phi_{\delta L} \Omega^{\gamma\delta} \Big) \Omega^{LI} \\
\cO_\text{m} = \; & \big( \Phi_{\alpha I} \Phi_{\beta J} \Omega^{IJ} \big) \Omega^{\beta\gamma} \big( \Psi_{\gamma A} \Psi_{\delta B} \Omega^{AB} \Big) \Omega^{\delta\alpha} \;.
\eea
The first two are classically relevant. The third one is classically marginal and we conjecture that it is present in the IR. As in Section~\ref{sec: SU/U duality}, we add $+\cO_\text{m}$ to the potential on the LHS, and $-\cO_\text{m}$ on the RHS. This is crucial for the duality to work. Instead we do not include
\be
\label{Phi^2 Psi^2 USp(2N) diagonal}
\cO_\text{d} = \big( \Phi_{\alpha I} \Phi_{\beta J} \Omega^{\alpha\beta} \Omega^{IJ} \big) \big( \Psi_{\gamma A} \Psi_{\delta B} \Omega^{\gamma\delta} \Omega^{AB} \Big)
\ee
which is a ``double trace operator''. This is also classically marginal, but it is marginally irrelevant at large $N$ and so we expect that it is marginally irrelevant also at finite $N$. In any case the presence of this operator would not change our discussion, once its effect is absorbed in the tuning of the IR masses. Some care should be used when one of $N$, $N_s$, $N_f$ is $1$: in that case some of the interactions above will be identified.

The phase diagrams of the two theories are summarized in Table~\ref{tab: USp dualities}, following the masks in Figure~\ref{fig: masks}. The duality is consistent in the following domain.
\be
\text{Range of dualities:} \qquad N \geq N_s\;,\qquad k \geq N_f \;.
\ee
For $N = N_s$ the horizontal gapless line in the left half space disappears, while for $k = N_f$ the vertical gapless line in the lower half plane disappears, and when both conditions are met both lines disappear.

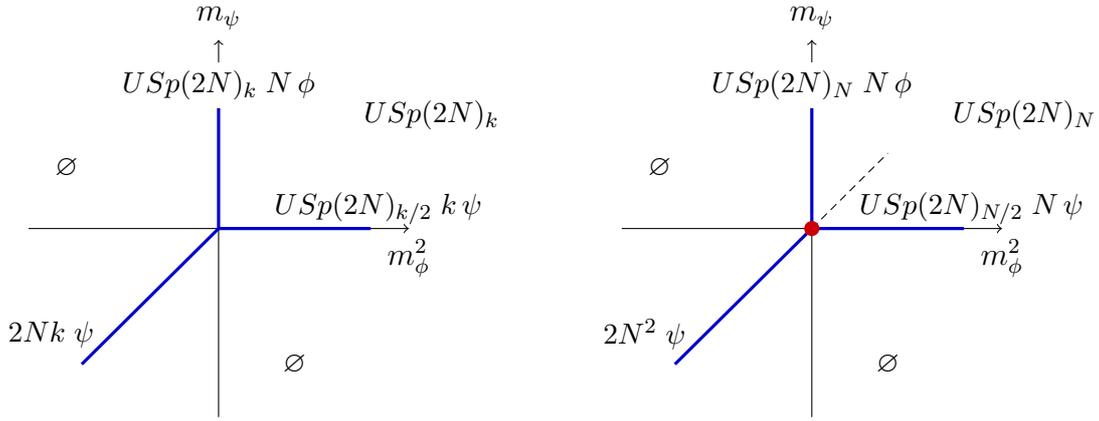
\begin{figure}[t!]
\centering
\hspace{\stretch{1}}
\begin{tikzpicture}[scale=1]
\draw [->] (-2.5,0)--(2.5,0) node[below] {$m_\phi^2$}; \draw [->] (0,-2.5)--(0,2.5) node[above] {$m_\psi$};
\draw [very thick, blue!80!black] (0,0)--(2,0); \draw [very thick, blue!80!black] (0,0)--(0,2); \draw [very thick, blue!80!black] (0,0)--(-1.8, -1.8);
\fill [white]  (-.5,1.6) rectangle (.5,2.2) node[midway, black] {\small $USp(2N)_k \; N\, \phi$};
\node at (2.1,.27) {\small $USp(2N)_{k/2} \; k\,\psi$};
\node at (-2.2,-1.4) {\small $2Nk\;\psi$};
\node at (2.8,1.5) {\small $USp(2N)_k$};
\node at (-2.0,.8) {\small $\varnothing$};
\node at (1.0, -1.8) {\small $\varnothing$};
\end{tikzpicture}
\hspace{\stretch{1}}
\begin{tikzpicture}[scale=1]
\draw [->] (-2.5,0)--(2.5,0) node[below] {$m_\phi^2$}; \draw [->] (0,-2.5)--(0,2.5) node[above] {$m_\psi$};
\draw [very thick, blue!80!black] (0,0)--(2,0); \draw [very thick, blue!80!black] (0,0)--(0,2); \draw [very thick, blue!80!black] (0,0)--(-1.8, -1.8);
\draw [densely dashed] (0,0)--(1,1);
\fill [red!80!black] (0,0) circle [radius=.1];
\fill [white]  (-.5,1.6) rectangle (.5,2.2) node[midway, black] {\small $USp(2N)_N \; N\, \phi$};
\node at (2.1,.27) {\small $USp(2N)_{N/2} \; N\,\psi$};
\node at (-2.2,-1.4) {\small $2N^2\;\psi$};
\node at (2.8,1.5) {\small $USp(2N)_N$};
\node at (-2.0,.8) {\small $\varnothing$};
\node at (1.0, -1.8) {\small $\varnothing$};
\end{tikzpicture}
\hspace{\stretch{1}}
\caption{(Left) Phase diagram of $USp(2N)_{\frac k2}$ with $N$ $\phi$, $k$ $\psi$. We have not indicated gravitational couplings for simplicity. The symbol $\varnothing$ indicates a gapped state with no topological order. (Right) Phase diagram specialized to the case $N=k$. In this case there is emergent time-reversal symmetry (with an anomaly) along the dashed line. We conjecture that there exists a time-reversal invariant tri-critical fixed point at the origin.
\label{fig: USp 3 lines}}
\end{figure}

There are two interesting subclasses of dualities. The first subclass corresponds to the special case just mentioned, namely $N = N_s$ and $k = N_f$:
\be
USp(2N)_\frac k2 \text{ with $N$ $\phi$, $k$ $\psi$} \qquad\longleftrightarrow\qquad USp(2k)_{-\frac N2} \text{ with $k$ $\phi$, $N$ $\psi$} \;.
\ee
The phase diagram for the theory on the LHS is depicted in Figure~\ref{fig: USp 3 lines} (left). In this case there are only three gapless lines in the phase diagram, and we might expect that they simply meet at a multi-critial fixed point.

The second subclass corresponds to the special case $N=k$ and $N_s = N_f$:
\begin{multline}
USp(2N)_{N - \frac{N_s}2} \text{ with } N_s\, (\phi \text{ and } \psi) \qquad\longleftrightarrow \\
USp(2N)_{-N + \frac{N_s}2} \text{ with } N_s\, (\phi \text{ and } \psi) \; \times U\big( 2N(N-N_s) \big)_1 \;.
\end{multline}
These theories have (in general) five gapless lines, and the duality implies that there is emergent time-reversal symmetry (with an anomaly) in the IR along the line $m_\phi^2 = m_\psi$. The intersection of the two subclasses corresponds to the special case $N=N_s = k = N_f$. In this case the phase diagram is as in Figure~\ref{fig: USp 3 lines} (right): there are only three gapless lines that conjecturally meet at a multi-critical fixed point with emergent time-reversal symmetry.

The simplest example is the case $N=N_s = k = N_f = 1$:
\be
\label{USp(2) duality}
USp(2)_{\frac12} \text{ with $\phi$, $\psi$} \qquad\longleftrightarrow\qquad USp(2)_{-\frac12} \text{ with $\phi$, $\psi$} \;.
\ee
Notice that in $USp(2N)_{k - \frac{N_f}2}$ with $N_s$ $\phi$, $N_f$ $\psi$, for $N=1$ and/or $N_s=1$ there is a unique $\Phi^4$ coupling, in the sense that the first two couplings in (\ref{USp relevant couplings}) are proportional. Moreover for $N=1$ there is a unique $\Phi^2\Psi^2$ coupling, in the sense that $\cO_\text{m} = \frac12 \cO_\text{d}$.%
\footnote{For $N=1$, $\Phi_{\alpha I} \Phi_{\beta J} \Omega^{IJ}$ is proportional to $\Omega_{\alpha\beta}$. Contracting with $\Omega^{\alpha\beta}$ we find $\Phi_{\alpha I} \Phi_{\beta J} \Omega^{IJ} = \frac12 C_1\, \Omega_{\alpha\beta}$ with $C_1 = \Phi_{\alpha I} \Phi_{\beta J} \Omega^{\alpha\beta} \Omega^{IJ}$. Similarly $\Psi_{\gamma A} \Psi_{\delta B} \Omega^{AB} = \frac12 C_2\, \Omega_{\gamma\delta}$ with $C_2 = \Psi_{\gamma A} \Psi_{\delta B} \Omega^{\gamma\delta} \Omega^{AB}$. Therefore the first coupling in (\ref{USp relevant couplings}) is equal to $C_1^2$ while the second one is equal to $\frac12 C_1^2$. Similarly $\cO_\text{d} = C_1 C_2$ while $\cO_\text{m} = \frac12 C_1 C_2$. For $N_s=1$ we can repeat the argument on the scalar coupling.}
We have represented the two phase diagrams in Figure~\ref{fig: USp(2) duality} (including the gravitational couplings in gapped phases but not along gapless lines). Along the oblique gapless line we find two Dirac fermions, transforming in the bifundamental representation%
\footnote{We write them in terms of four Dirac fermions $\Psi_{IA}$ with a reality constraint $\Psi_{IA}^c = \Omega^{IJ} \Omega^{AB} \Psi_{JB}$.}
of the global symmetry $\big( USp(2) \times USp(2)\big) / \bZ_2$.
Along the other two gapless lines we find $SU(2)_1$ with $\phi$ (and its time reversal): a CFT with $SO(3)$ global symmetry, studied to some extent in \cite{Aharony:2016jvv}.

\begin{figure}[t!]
\centering
\hspace{\stretch{1}}
\begin{tikzpicture}[scale=1]
\draw [->] (-2.5,0)--(2.5,0) node[below] {$m_\phi^2$}; \draw [->] (0,-2.5)--(0,2.5) node[above] {$m_\psi$};
\draw [very thick, blue!80!black] (0,0)--(2,0); \draw [very thick, blue!80!black] (0,0)--(0,2);
\draw [densely dashed] (-.6,-.6)--(.6,.6);
\draw [very thick, blue!80!black] (0,0)--(-1.6,-1.6);
\fill [red!80!black] (0,0) circle [radius=.1];
\node at (1.8,.3) {\small $SU(2)_{1/2} \; \psi$};
\node at (-1.6,-1.0) {\small $2\psi$};
\fill [white]  (-.5,1.6) rectangle (.5,2.2) node[midway, black] {\small $SU(2)_1 \; \phi$};
\node at (2.4,1.2) {\small $SU(2)_1 \times U(0)_1$};
\node at (-1.8,1.0) {\small $U(0)_1$}; \node at (1.6, -1.4) {\small $U(2)_1$};
\end{tikzpicture}
\hspace{\stretch{1}}
\begin{tikzpicture}[scale=1]
\draw [->] (2.5,0)--(-2.5,0) node[below] {$m_\psi$}; \draw [->] (0,-2.5)--(0,2.5) node[above] {$m_\phi^2$};
\draw [very thick, blue!80!black] (0,0)--(2,0); \draw [very thick, blue!80!black] (0,0)--(0,2);
\draw [densely dashed] (-.6,-.6)--(.6,.6);
\draw [very thick, blue!80!black] (0,0)--(-1.6,-1.6);
\fill [red!80!black] (0,0) circle [radius=.1];
\node at (2.2,.3) {\small $SU(2)_{-1} \; \phi$};
\node at (-1.6,-1.0) {\small $2\psi$};
\fill [white]  (-.5,1.6) rectangle (.5,2.2) node[midway, black] {\small $SU(2)_{-\frac12} \; \psi$};
\node at (2.4,1.2) {\small $SU(2)_{-1}\times U(2)_1$};
\node at (-1.8,1.0) {\small $U(0)_1$}; \node at (1.6, -1.4) {\small $U(2)_1$};
\end{tikzpicture}
\hspace{\stretch{.5}}
\caption{$\qquad USp(2)_\frac12$ with $\phi, \psi$ $\qquad\quad\longleftrightarrow\quad\qquad$ $USp(2)_{-\frac12}$ with $\phi,\psi$. $\quad$ Phase diagram. \newline
We have not indicated the gravitational couplings along gapless lines, for simplicity. The duality predicts emergent IR time-reversal invariance at the origin and along the dashed line.
\label{fig: USp(2) duality}}
\end{figure}

It is interesting to compare the $USp(2)$ theory in (\ref{USp(2) duality}) with the $SU(2)$ theory in (\ref{SU(2) U(1)3/2 duality}). In $USp(2)$ with $N_s=1$ scalars there is a unique gauge-invariant $\Phi^4$ quartic coupling that preserves $USp(2)$ global symmetry. Similarly, in $SU(2)$ with $N_s=1$ there is a unique gauge-invariant $\phi^4$ quartic coupling that preserves $U(1)$ global symmetry: it is the very same coupling, it preserves a larger $USp(2) \cong SU(2)$ global symmetry, and in fact the two theories are the same (see also Section~\ref{sec: case of SU(2)}). We have used this fact in Figure~\ref{fig: USp(2) duality} to write the gapless lines in terms of $SU(2)$ gauge theories. In $USp(2)$ with $N_s = N_f =1$ scalars and fermions there is a unique gauge-invariant $\Phi^2\Psi^2$ quartic coupling $\cO_\text{d}$ (\ref{Phi^2 Psi^2 USp(2N) diagonal}) that preserves $USp(2) \times USp(2)$ global symmetry. On the contrary, in $SU(2)$ with $N_s = N_f=1$ there are two gauge-invariant $\phi^2\psi^2$ couplings that preserve $U(1) \times U(1)$ global symmetry:
\be
(\phi_\alpha^* \phi^\alpha)(\bar\psi_\alpha \psi^\alpha) \qquad\text{ and }\qquad (\phi^\alpha \bar\psi_\alpha)(\psi^\alpha \phi_\alpha^*) \;.
\ee
The first one preserves $USp(2) \times USp(2)$ global symmetry, while the second one preserves $U(1) \times U(1)$ (and we expect it to be marginally relevant in the UV). Thus the $USp(2)$ and $SU(2)$ theories with a scalar and a fermion are different, and the latter is expected to be a relevant deformation of the former. In fact, the relevant deformation splits the gapless line $2\psi$ in Figure~\ref{fig: USp(2) duality} into two lines, by giving mass to one of the two fermions.

\begin{figure}[t!]
\centering
\hspace{\stretch{1}}
\begin{tikzpicture}[scale=1]
\draw [->] (-2.5,0)--(2.5,0) node[below] {$m_\phi^2$}; \draw [->] (0,-2.5)--(0,2.5) node[above] {$m_\psi$};
\draw [very thick, blue!80!black] (-2,0)--(2,0); \draw [very thick, blue!80!black] (0,0)--(0,2);
\draw [very thick, blue!80!black] (0,0)--(0,-.2); \draw [very thick, blue!80!black] (0,-.2) arc [radius = 1, start angle = 0, end angle = -45]; \draw [very thick, blue!80!black] (-1.78,-2.40)--(-.28,-.90);
\fill [lightgray!70!yellow, fill opacity=.6] (0,0) circle [radius=.4];
\draw [densely dashed] (-.6,-.6)--(.6,.6);
\node at (1.6,.3) {\small $SU(2)_{1/2} \; \psi$}; \node at (-1.8,.3) {\small $\psi$}; \node at (-2,-2.2) {\small $\psi$};
\fill [white]  (-.5,1.6) rectangle (.5,2.2) node[midway, black] {\small $SU(2)_1 \; \phi$};
\node at (2.4,1.3) {\small $SU(2)_1 \times U(0)_1$}; \node at (-1.8,1.2) {\small $U(0)_1$};
\node at (2, -1.2) {\small $U(2)_1$}; \node at (-1.8, -.8) {\small $U(1)_1$};
\end{tikzpicture}
\hspace{\stretch{1}}
\begin{tikzpicture}[scale=1]
\draw [->] (2.5,0)--(-2.5,0) node[below] {$m_\psi$}; \draw [->] (0,-2.5)--(0,2.5) node[above] {$m_\phi^2$};
\draw [very thick, blue!80!black] (0,0)--(2,0); \draw [very thick, blue!80!black] (0,-2)--(0,2);
\draw [very thick, blue!80!black] (0,0)--(-.2,0); \draw [very thick, blue!80!black] (-.2,0) arc [radius = 1, start angle = 90, end angle = 135]; \draw [very thick, blue!80!black] (-2.40,-1.78)--(-.90,-.28);
\fill [lightgray!70!yellow, fill opacity=.6] (0,0) circle [radius=.4];
\draw [densely dashed] (-.6,-.6)--(.6,.6);
\node at (2.2,.3) {\small $SU(2)_{-1} \times U(2)_1 \; \phi$}; \node at (-2.6,-1.6) {\small $\psi$}; \node [right] at (0,-1.8) {\small $\psi$};
\fill [white]  (-.5,1.6) rectangle (.5,2.2) node[midway, black] {\small $SU(2)_{-\frac12} \; \psi$};
\node at (2.4,1.3) {\small $SU(2)_{-1} \times U(2)_1$}; \node at (-1.8,1.2) {\small $U(0)_1$};
\node at (2, -1.2) {\small $U(2)_1$}; \node at (-1,-2) {\small $U(1)_1$};
\end{tikzpicture}
\hspace{\stretch{.5}}
\caption{$\qquad SU(2)_{1/2}$ with $\phi$, $\psi$ $\qquad\;\;\longleftrightarrow\;\;\;\qquad$ $SU(2)_{-1/2}$ with $\phi$, $\psi$. $\quad$ Phase diagram. \newline
We have not indicated the gravitational couplings along gapless lines, for simplicity. The duality predicts emergent time-reversal invariance around the origin, along the dashed line.
\label{fig: SU(2) duality}}
\end{figure}
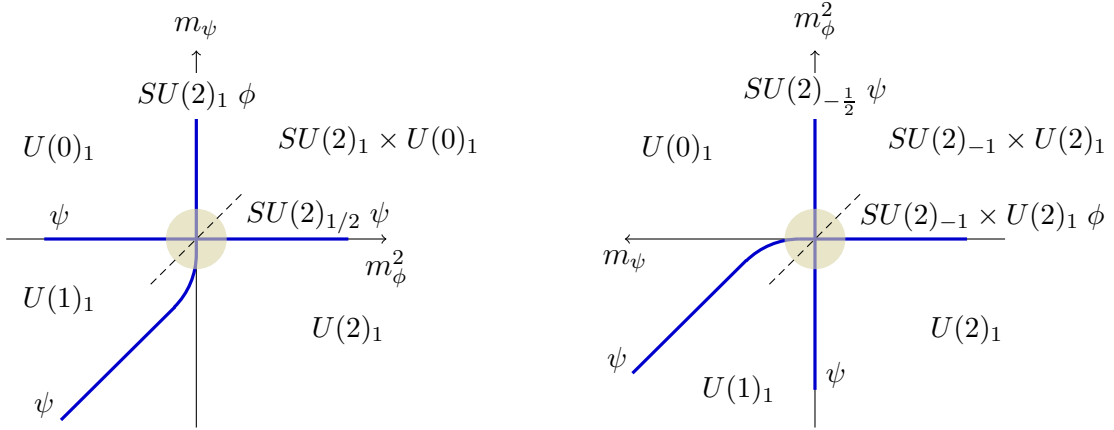

We propose that deforming the $USp$ duality in (\ref{USp(2) duality}) by $(\phi^\alpha \bar\psi_\alpha)(\psi^\alpha \phi_\alpha^*) \leftrightarrow - (\phi^\alpha \bar\psi_\alpha)(\psi^\alpha \phi_\alpha^*)$ we obtain the following $SU$ duality:
\be
SU(2)_\frac12 \text{ with $\phi$, $\psi$} \qquad\longleftrightarrow\qquad SU(2)_{-\frac12} \text{ with $\phi$, $\psi$} \;.
\ee
The phase diagram is summarized in Figure~\ref{fig: SU(2) duality}. This duality implies that the theory has emergent time-reversal invariance in the IR. In fact, this theory is precisely the one in the duality (\ref{SU(2) U(1)3/2 duality}) and it is dual to the two theories in (\ref{U(1)3/2 duality}): the three dualities are compatible.


\section{\matht{SO} duality}
\label{sec: SO duality}

The fourth duality we consider involves Chern-Simons theories with special orthogonal groups as well as bosonic and fermionic matter fields in the fundamental representation, which is real. For simplicity, we use the same symbols $\phi$ and $\psi$ as before, but we should keep in mind that for $SO$ theories $\phi$ is a real scalar and $\psi$ is a Majorana fermion (when a field has no gauge interactions, to avoid confusion we write $\phi_\bR$ or $\psi_\bR$). We propose the following duality:
\begin{multline}
\label{SO duality grav}
SO(N)_{k - \frac{N_f}2} \text{ with $N_s$ $\phi$, $N_f$ $\psi$} \qquad\longleftrightarrow \\
SO(k)_{-N + \frac{N_s}2} \text{ with $N_f$ $\phi$, $N_s$ $\psi$ } \times SO\big( k(N-N_s) \big)_1 \;.
\end{multline}
The last factor in the second line represents a gravitational coupling $-k(N-N_s) \text{CS}_g$. We propose this duality in the range
\be
\label{SO parameter range}
N \geq N_s \;,\qquad k \geq N_f \;,\qquad N+k \geq N_s + N_f + 3 \;,
\ee
as explained below. It might be possible to extend this range along the lines of \cite{Komargodski:2017keh} or with the observation in (\ref{special SO(2) duality}). Notice that the range (\ref{SO parameter range}) reproduces the range of the $SO$ dualities with a single matter species \cite{Aharony:2016jvv} after setting $N_f =0$.

The proposal (\ref{SO duality grav}) reproduces the boson/fermion dualities with a single matter species of \cite{Aharony:2016jvv, Metlitski:2016dht} for $N_s=0$ or \mbox{$N_f=0$}, as well as the level-rank dualities when $N_s = N_f = 0$ (see the summary in Appendix \ref{app: summary}). The proposal is also consistent under RG flows triggered by a mass term, either positive or negative, for a single scalar or fermion.

The theories in (\ref{SO duality grav}) have global symmetry $G_0 = O(N_s) \times O(N_f) \times \bZ_2^\cC \times \bZ_2^M$, not necessarily acting faithfully. The first two factors act on the matter fields $\phi_{\alpha I}$ and $\psi_{\alpha B}$ in the fundamental representation, respectively, through the indices $I=1,\dots,N_s$ and $B =1,\dots, N_f$, while $\alpha=1,\dots, N$ is a gauge index. The generator of ``charge conjugation'' $\bZ_2^\cC$ maps $\phi_{1I} \mapsto -\phi_{1I}$ and $\psi_{1A} \mapsto - \psi_{1A}$ while leaving all other components invariant.%
\footnote{For $N$ odd, $O(N) \cong \bZ_2 \times SO(N)$ where $\bZ_2 = \{\unit, - \unit\}$. Therefore for $N$ odd, $\bZ_2^\cC$ is already contained into $O(N_s) \times O(N_f)$ and is not independent. For $N$ even, $-\unit_N \in SO(N)$ therefore the diagonal $\bZ_2$ in $O(N_s) \times O(N_f)$ generated by $\{-\unit_{N_s}, - \unit_{N_f}\}$ is gauged.}
Finally, $\bZ_2^M$ is a magnetic symmetry giving charge to monopole operators. As in the dualities with a single matter species \cite{Aharony:2016jvv}, the duality exchanges $\bZ_2^\cC$ with $\bZ_2^M$.

The quadratic operators invariant under $G_0$ are%
\footnote{Fermions are contracted in a Lorentz-invariant way, keeping the $\epsilon$-tensor implicit.}
\be
m_\phi^2 \; \phi_{\alpha I} \phi_{\alpha I} \qquad\qquad\text{and}\qquad\qquad m_\psi\; \psi^\sT_{\alpha B} \psi_{\alpha B} \;,
\ee
whose coefficients we tune to find phase transitions. The quartic operators invariant under $G_0$, classically relevant or marginal in the UV, are
\be
\label{SO quartic couplings}
(\phi_{\alpha I} \phi_{\alpha I})^2 \;,\qquad \phi_{\alpha I} \phi_{\alpha J} \phi_{\beta J} \phi_{\beta I} \;,\qquad \cO_\text{d} = (\phi_{\alpha I} \phi_{\alpha I})(\psi^\sT_{\beta A} \psi_{\beta A}) \;,\qquad \cO_\text{m} = \phi_{\alpha I} \phi_{\beta I} \psi^\sT_{\beta C} \psi_{\alpha C} \;.
\ee
Paralleling the discussion in Section~\ref{sec: SU/U duality}, the quartic scalar couplings are present in the theories, the mixed coupling $\cO_\text{m}$ is assumed to be present in the IR potential with positive sign on the LHS and negative sign on the RHS, while the mixed coupling $\cO_\text{d}$, even if present in the IR, does not affect the discussion here.

The case $N=2$ is special because $SO(2) \cong U(1)$, in particular the magnetic symmetry is enhanced from $\bZ_2^M$ to $U(1)_M$. Also the flavor symmetry can be enhanced. For instance, since four-Fermi interactions are irrelevant, $SO(2)_{k-\frac{N_f}2}$ with $N_f$ $\psi$ is the same as $U(1)_{k - \frac{N_f}2}$ with $N_f$ $\psi$ (notice that the fermions are Majorana in the $SO$ theory and Dirac in the $U$ theory) and the flavor symmetry is enhanced from $O(N_f)$ to $SU(N_f)$. With a single scalar there is a unique quartic scalar coupling, $(\phi_\alpha \phi_\alpha)^2$, therefore $SO(2)_k$ with $1$ $\phi$ is the same as $U(1)_k$ with $1$ $\phi$. For $N_s\geq 2$ the two quartic scalar couplings are independent, therefore the $SO(2)$ theory is different from the $U(1)$ theory. Similarly, for $N_s = N_f=1$ the mixed couplings $\cO_\text{d}$ and $\cO_\text{m}$ are different: while $\cO_\text{d}$ is present in the $U(1)$ theory, $\cO_\text{m}$ is not and so its presence distinguishes the $SO(2)$ theory. The same is true for all other cases with $N_s, N_f \geq 1$.

\begin{table}[t!]
{\renewcommand{\arraystretch}{1.39}
$$
\begin{array}{|lll|}
\hline
\multicolumn{3}{|c|}{SO(N)_{k - \frac{N_f}2} \text{ with $N_s$ $\phi$, $N_f$ $\psi$}} \text{ (LHS)}\\[.5em]
\hline\hline
m_\psi>0: &	SO(N)_k \text{ with $N_s$ $\phi$ } \times SO(0)_1 & \qquad\qquad\qquad \\
	& \multicolumn{2}{r|}{A:\; SO(N)_k \times SO(0)_1} \\
m_\phi^2>0: &	SO(N)_{k-\frac{N_f}2} \text{ with $N_f$ $\psi$} & \\
	& \multicolumn{2}{r|}{B:\; SO(N)_{k-N_f} \times SO(NN_f)_1} \\
m_\psi<0: &	SO(N)_{k-N_f} \text{ with $N_s$ $\phi$ } \times SO(NN_f)_1 & \\
	& \multicolumn{2}{r|}{C:\; SO(N-N_s)_{k-N_f} \times SO(NN_f)_1} \\
m_\phi^2<0,\, m_\psi<0: & N_s N_f\, \psi_\bR \times SO(N-N_s)_{k-N_f} \times SO\big((N-N_s)N_f \big)_1 & \\
	& \multicolumn{2}{r|}{D:\; SO(N-N_s)_{k-N_f} \times SO\big( (N-N_s) N_f \big)_1} \\
m_\phi^2<0: &	SO(N-N_s)_{k - \frac{N_f}2} \text{ with $N_f$ $\psi$} & \\
	& \multicolumn{2}{r|}{E:\; SO(N-N_s)_k \times SO(0)_1} \\
\hline
\end{array}
$$
$$
\begin{array}{|lll|}
\hline
\multicolumn{3}{|c|}{SO(k)_{-N + \frac{N_s}2} \text{ with $N_f$ $\phi$, $N_s$ $\psi$ } \times SO\big( k(N-N_s) \big)_1} \text{ (RHS)}\\[.5em]
\hline\hline
m_\phi^2>0: & SO(k)_{-N + \frac{N_s}2} \text{ with $N_s$ $\psi$ } \times SO\big( k(N-N_s) \big)_1 & \;\;\qquad \\
	& \multicolumn{2}{r|}{A:\; SO(k)_{-N} \times SO(kN)_1} \\
m_\psi<0: &	SO(k)_{-N} \text{ with $N_f$ $\phi$ } \times SO(kN)_1 & \\
	& \multicolumn{2}{r|}{B:\; SO(k-N_f)_{-N} \times SO(kN)_1} \\
m_\phi^2<0: &	SO(k-N_f)_{-N + \frac{N_s}2} \text{ with $N_s$ $\psi$ } \times SO(kN - kN_s + N_fN_s)_1 & \\
	& \multicolumn{2}{r|}{C:\; SO(k-N_f)_{-N + N_s} \times SO(kN - kN_s + N_fN_s)_1} \\
m_\phi^2<0,\, m_\psi>0: & N_s N_f\, \psi_\bR \times SO(k-N_f)_{-N + N_s} \times SO\big( k(N-N_s) \big)_1 & \\
	& \multicolumn{2}{r|}{D:\; SO(k-N_f)_{-N + N_s} \times SO\big( k(N-N_s) \big)_1} \\
m_\psi>0: &	SO(k)_{-N+N_s} \text{ with $N_f$ $\phi$ } \times SO\big( k(N-N_s) \big)_1 & \\
	& \multicolumn{2}{r|}{E:\; SO(k)_{-N+N_s} \times SO\big( k(N-N_s)\big)_1} \\
\hline
\end{array}
$$
}
\caption{Phase diagram of the $SO$ dualities. Here $\phi$ are real scalars and $\psi$ are Majorana fermions. The validity range of the table is explained in the main text.
\label{tab: SO dualities}}
\end{table}

The phase diagrams for the two theories in (\ref{SO duality grav}), assuming $N_s, N_f \geq 1$, are reported in Table~\ref{tab: SO dualities}, following the masks in Figure~\ref{fig: masks}. The table should be read with some care. If $N_s \leq N-2$ and $N_f \leq k-2$, the table is valid without subtleties. Then, using the dualities in \cite{Aharony:2016jvv}, all gapless lines and gapped phases match across the duality. If $N_s = N-1$ then on the LHS the gauge group is completely broken for $m_\phi^2<0$, and thus the factors $SO(N-N_s)_\#$ in Table~\ref{tab: SO dualities} LHS should be dropped (but the gravitational contributions should be kept). Similarly, if $N_f = k-1$ then on the RHS the gauge group is completely broken for $m_\phi^2<0$ and the factors $SO(k-N_f)_\#$ should be dropped. We find that for $N_s \leq N-1$ and $N_f \leq k-1$ there still is a match of phases between the two sides, with the exception of the case $(N_s, N_f) = (N-1,k-1)$.

If $N_s = N$ then on the LHS the gauge group is completely broken for $m_\phi^2<0$ and, moreover, the $\bZ_2^\cC$ charge conjugation symmetry is spontaneously broken%
\footnote{Here we are assuming that the IR relative strength of the two quartic scalar couplings is in a certain range, elaborated in Section \ref{sec: potential}. In this case, up to gauge and flavor rotations, $\phi_{\alpha I}$ is proportional to $\unit_N$ which is not invariant under $\bZ_2^\cC$ but does not break $SO(N_s)$. There exists another regime, though, for which the induced VEV of $\phi_{\alpha I}$ has a unique non-zero entry along the diagonal; such a VEV breaks $SO(N_s) \to SO(N_s-1)$ (and preserves $\bZ_2^\cC$) leaving the Goldstone bosons of an $S^{N-1}$ NLSM.}
giving rise to two vacua. Then we should substitute the factors $SO(N-N_s)_\#$ in Table~\ref{tab: SO dualities} LHS with a ``$\bZ_2$'' that represents those two gapped states. Notice that the horizontal gapless line in the left half plane disappears, and phases D and E are identical. In this case we find a match of phases between the two sides, provided $N_f \leq k-3$. To verify the match we use that for $0 \leq N_f \leq k-3$ the theory
\be
\label{SO confinement}
SO(k)_0 \text{ with $N_f$ $\phi$}
\ee
confines, with a spontaneous breaking of the $\bZ_2^M$ magnetic symmetry (we typically think of a phase with broken magnetic symmetry as confining). As a partial check of this assumption, the claim is consistent under mass deformations of the theory.

Similar comments apply to the case $N_f = k$: on the RHS the gauge group is completely broken for $m_\phi^2<0$, the vertical gapless line in the lower half plane disappears  and phases B and C are identical, the $\bZ_2^\cC$ charge conjugation symmetry is spontaneously broken and there are two vacua. We should substitute the factors $SO(k-N_f)_\#$ in Table~\ref{tab: SO dualities} LHS with a ``$\bZ_2$''. The phases match between the two sides, provided $N_s \leq N-3$. Finally, for $N_s > N$ or $N_f > k$ we find phases with a more severe spontaneous symmetry breaking, which is not classically observed on the other side of the duality. Collecting the various cases, we end up with the range in (\ref{SO parameter range}).

\subsection{Simple examples}

\begin{figure}[t!]
\centering
\hspace{\stretch{1}}
\begin{tikzpicture}[scale=1]
\draw [->] (-2.5,0)--(2.5,0) node[below] {$m_\phi^2$}; \draw [->] (0,-2.5)--(0,2.5) node[above] {$m_\psi$};
\draw [very thick, blue!80!black] (-2,0)--(2,0); \draw [very thick, blue!80!black] (0,-2)--(0,2);
\draw [->, thick] (0,0) ++(220:1.0) arc (220:185:1.0);
\node at (1.8,.3) {\small $\psi_\bR$};
\fill [white]  (-.5,1.6) rectangle (.5,2.2) node[midway, black] {\small $O(1)$ WF};
\node at (2,1.2) {\small $SO(0)_1$}; \node at (-1.8,1.0) {\small $\bZ_2 \times SO(0)_1$};
\node at (2, -1.2) {\small $SO(1)_1$}; \node at (-1.8, -1.2) {\small $\bZ_2 \times SO(1)_1$};
\node at (0,-2.8) {}; 
\end{tikzpicture}
\hspace{\stretch{1}}
\begin{tikzpicture}[scale=1]
\draw [->] (2.5,0)--(-2.5,0) node[below] {$m_\psi$}; \draw [->] (0,-2.5)--(0,2.5) node[above] {$m_\phi^2$};
\draw [very thick, blue!80!black] (0,0)--(2,0); \draw [very thick, blue!80!black] (0,-2)--(0,2);
\draw [very thick, blue!80!black] (0,0)--(-.2,0); \draw [very thick, blue!80!black] (-.2,0) arc [radius = 1, start angle = 90, end angle = 135]; \draw [very thick, blue!80!black] (-2.40,-1.78)--(-.90,-.28);
\fill [white]  (-.5,1.6) rectangle (.5,2.2) node[midway, black] {\small $SO(k)_{-\frac12}$ $\psi$};
\fill [white]  (-.5,-1.2) rectangle (.5,-1.8) node[midway, black] {\small $SO(k-1)_{-\frac12}$ $\psi$};
\node at (1.8,.3) {\small $SO(k)_{-1}$ $\phi$}; \node at (-2.6,-1.3) {\small $\psi_\bR$};
\node at (2.4,1.2) {\small $SO(0)_1$}; \node at (-2.2,1.0) {\small $SO(k)_0 \times SO(0)_1$};
\node at (2.4, -0.8) {\small $SO(1)_1$}; \node [align=center] at (-1.3, -2.5) { \footnotesize $SO(k-1)_0$ \\[-.5em] \footnotesize $\times\; SO(1)_1$};
\end{tikzpicture}
\hspace{\stretch{1}}
\caption{$\qquad O(1)$ WF $\times$ $\psi_\bR$ $\qquad\longleftrightarrow\qquad$ $SO(k)_{-1/2}$ with $\phi,\psi$. $\quad$ Phase diagram.
\label{fig: O1 WF}}
\end{figure}

A simple but interesting example is for $N=N_s = N_f = 1$ and $k\geq 4$:
\be
O(1) \text{ WF } \times\; \psi_\bR \qquad\longleftrightarrow\qquad SO(k)_{-\frac12} \text{ with $\phi$, $\psi$} \;.
\ee
On both sides the gravitational coupling is $SO(0)_1$. The theory on the LHS is the decoupled product of the $O(1)$ Wilson-Fisher fixed point (denoted as $\phi_\bR$ in our notation), also known as the 3D Ising CFT, and a free Majorana fermion. These two theories are time-reversal invariant, and are decoupled because $\phi^2 \psi^2$ is irrelevant. Therefore the duality predicts that the theory on the RHS has a multi-critical fixed point where the four gapless lines meet, the theory factorizes and develops time-reversal invariance in the IR. The two phase diagrams are in Figure~\ref{fig: O1 WF}: on the LHS we took into account that $\phi^2\psi^2$ is irrelevant and moved the $\psi_\bR$ gapless line accordingly; on the RHS we implemented such an input from the duality and crossed the gapless lines perpendicularly. The gapless lines agree using the bosonization/fermionization dualities of \cite{Aharony:2016jvv} and the gapped phases agree using (\ref{SO confinement}).

The previous example generalizes to the dualities
\be
O(1) \text{ WF } \times\; N_f\; \psi_\bR \qquad\longleftrightarrow\qquad SO(k)_{-\frac12} \text{ with $N_f$ $\phi$, 1 $\psi$}
\ee
for $k \geq N_f+3$. On both sides the gravitational coupling is $SO(0)_1$. Once again, the theory on the LHS is factorized into two decoupled sectors: the $O(1)$ WF fixed point and $N_f$ free Majorana fermions. Both sectors are time-reversal invariant. The duality predicts that the theory on the RHS has a multi-critical fixed point with the same IR properties.


\section{More Abelian dualities}
\label{sec: Abelian dualities}

In this Section---that could be read independently from the previous ones---we propose and discuss some other Abelian dualities involving scalars and/or fermions, that can be derived using the dualities in \cite{Seiberg:2016gmd}. Let us first summarize our findings. In Section~\ref{sec: U(1)-1/2 phi psi} we discuss
\be
\makebox[0pt][l]{\hspace{3em}\raisebox{1.4em}{\begin{tikzpicture} \draw [->,line width=.6pt] (0,0) arc (220:-40:.3 and .4); \end{tikzpicture}}}
U(1)_{-\frac12} \text{ with $\phi$, $\psi$} \quad\longleftrightarrow\quad U(1)_1 \text{ with 2 $\phi$ and $V_\text{EP}$} \quad\longleftrightarrow\quad U(1)^2_K \text{ with 2 $\psi$} \;.
\ee
The theory on the left has a self-duality that acts on the manifest part of the global symmetry. The theory in the middle has an extra ``easy plane'' quartic potential $V_\text{EP}$ that breaks the global symmetry to $O(2)^2$. The theory on the right has a $2\times2$ CS matrix $K = \smat{1/2 & 1 \\ 1 & 1/2}$. In Section~\ref{sec: U(1) 3/2 time rev} we discuss
\be
U(1)_\frac32 \text{ with $\phi$, $\psi$} \qquad\longleftrightarrow\qquad U(1)_{-\frac32} \text{ with $\phi$, $\psi$} \;.
\ee
This duality, already presented in Section~\ref{sec: U/U duality}, acts as time-reversal. In Section~\ref{sec: SO excluded} we discuss
\be
U(1)_2 \text{ with 2 $\phi$ and $V_\text{EP}$} \qquad\longleftrightarrow\qquad U(1)_{-1} \text{ with 2 $\psi$} \;.
\ee
The theory on the left has manifest $O(2)^2$ global symmetry in the UV, while the theory on the right has $O(2)\times SO(3)$ global symmetry. The duality then predicts IR symmetry enhancement. Finally, in Section~\ref{sec: scalar QED} we discuss dualities of QED with two matter fields:
\be
\makebox[0pt][l]{\hspace{2em}\raisebox{1.4em}{\begin{tikzpicture} \draw [->,line width=.6pt] (0,0) arc (220:-40:.3 and .4); \end{tikzpicture}}}
U(1)_0 \text{ with 2 $\phi$ and $V_\text{EP}$} \qquad\longleftrightarrow\qquad
\makebox[0pt][l]{\hspace{2em}\raisebox{1.4em}{\begin{tikzpicture} \draw [->,line width=.6pt] (0,0) arc (220:-40:.3 and .4); \end{tikzpicture}}}
U(1)_0 \text{ with 2 $\psi$} \;,
\ee
where both theories have a self-duality. The horizontal duality was already reported in \cite{Motrunich:2003fz, Karch:2016sxi}, the self-duality on the LHS in \cite{Wang:2017txt} and the self-duality on the RHS in \cite{Xu:2015lxa, Hsin:2016blu, Benini:2017dus}. We give here some more details.

As discussed in Section \ref{sec: SO duality} and in \cite{Aharony:2016jvv}, the theories $U(1)_\#$ with 2 $\psi$ coincide with the theories $SO(2)_\#$ with 2 $\psi$. The theories $U(1)_\#$ with 2 $\phi$ and $V_\text{EP}$ almost coincide with the theories $SO(2)_\#$ with 2 $\phi$: the extra quartic scalar potential $V_\text{EP}$ is precisely the one that distinguishes the $U(1)$ theory from the $SO(2)$ theory (for $N_s=2$). However, the relative strengths of the two scalar quartic couplings assumed in Section \ref{sec: SO duality} and the corresponding symmetry-breaking pattern when $N_s=N$, are different from the ones associated to $V_\text{EP}$. This point is elaborated upon in Section \ref{sec: potential} below.

To derive new Abelian dualities, we employ the following ones \cite{Seiberg:2016gmd} that include background fields:
\begin{subequations}
\label{Abelian dualities with background}
\begin{align}
\label{Abdu-a}
|D_B\phi|^2 - |\phi|^4 \quad&\longleftrightarrow\quad |D_b\sigma|^2 - |\sigma|^4 + \frac1{2\pi} bdB \\
\label{Abdu-b}
|D_B\phi|^2 - |\phi|^4 \quad&\longleftrightarrow\quad i \bar\psi \Dslash_a \psi - \frac1{2\pi} adB - \frac1{4\pi} BdB \\
\label{Abdu-c}
i\bar\psi \Dslash_A \psi \quad&\longleftrightarrow\quad |D_b\phi|^2 - |\phi|^4 + \frac1{4\pi} bdb + \frac1{2\pi}bdA \\
\label{Abdu-d}
i\bar\psi \Dslash_A \psi \quad&\longleftrightarrow\quad i \bar\zeta \Dslash_a \zeta + \frac1{2\pi} adu - \frac2{4\pi} udu + \frac1{2\pi} udA - \frac1{4\pi} AdA - 2\text{CS}_\text{g} \;.
\end{align}
\end{subequations}
Here $\phi,\sigma$ are complex bosons, $\psi, \zeta$ are Dirac fermions, $b,u,B$ are gauge fields (small caps indicate dynamical fields while capitals are background fields) and $a,A$ are spin$_c$ connections. We can treat $a,A$ as standard gauge fields in the last three lines (where the dualities are between spin theories) if we consider the theories on spin manifolds. In all cases a positive mass, or mass squared, on one side is mapped to a negative one on the other side. The parity-inverted versions are:
\begin{subequations}
\label{Abelian dualities with background t-rev}
\begin{align}
\label{Abdurev-a}
|D_B\phi|^2 - |\phi|^4 \quad&\longleftrightarrow\quad |D_b\sigma|^2 - |\sigma|^4 - \frac1{2\pi} bdB \\
\label{Abdurev-b}
|D_B\phi|^2 - |\phi|^4 \quad&\longleftrightarrow\quad i \bar\psi \Dslash_a \psi + \frac1{4\pi} ada + \frac1{2\pi} adB + \frac1{4\pi} BdB +2 \text{CS}_\text{g} \\
\label{Abdurev-c}
i\bar\psi \Dslash_A \psi \quad&\longleftrightarrow\quad |D_b\phi|^2 - |\phi|^4 - \frac1{4\pi} bdb - \frac1{2\pi}bdA - \frac1{4\pi} AdA - 2\text{CS}_\text{g} \\
\label{Abdurev-d}
i\bar\psi \Dslash_A \psi \quad&\longleftrightarrow\quad i \bar\zeta \Dslash_a \zeta + \frac1{4\pi} ada - \frac1{2\pi} adu + \frac2{4\pi} udu - \frac1{2\pi} udA + 2\text{CS}_\text{g} \;.
\end{align}
\end{subequations}

\subsection[The potential $V_\text{EP}$]{The potential \matht{V_\text{EP}}}
\label{sec: potential}

Consider an $SO(N)_\#$ theory with $N$ scalars (and possibly fermions). Up to an overall rescaling, the quartic scalar potential can be written as
\be
\label{potential with lambda}
V = \Big( 1 - \frac\lambda N \Big) \, \big( \Tr \phi^\sT \phi \big)^2 + \lambda \Tr \phi^\sT \phi \phi^\sT \phi \;,
\ee
where $\phi_{\alpha I}$ is an $N\times N$ real matrix. Here $\lambda$ is a real parameter and $\lambda > - \frac N{N-1}$ guarantees that the potential is positive definite. For generic values of $\lambda$, the potential preserves $O(N) \times O(N)$ symmetry (acting on $\phi$ from the left and the right), of which an $SO(N)$ is gauged. If we deform the potential with a negative mass squared,
\be
V = \Big( 1 - \frac\lambda N \Big) \, \big( \Tr \phi^\sT \phi \big)^2 + \lambda \Tr \phi^\sT \phi \phi^\sT \phi - m^2 \Tr \phi^\sT \phi \;,
\ee
the minima depend on the value of $\lambda$. For $\lambda>0$, up to gauge and flavor rotations, the minima are at $\phi = \phi_0 \, \unit_N$. This VEV breaks the $SO(N)$ gauge group completely as well as $\bZ_2^\cC$, while it preserves $SO(N)$ flavor rotations (up to gauge transformations). This is precisely the symmetry breaking pattern assumed in Section \ref{sec: SO duality}, therefore in that Section we assumed that $\lambda>0$ in the IR.

For $\lambda<0$ the minima are at $\phi = \text{diag}(0,\dots,0, \phi_0)$: they preserve $\bZ_2^\cC$ but break the global symmetry $SO(N) \to SO(N-1)$, resulting in Goldstone bosons that parametrize $S^{N-1}$. We might ask if we expect theories with $\lambda<0$ in the IR. At least in the case $N=2$, we can make the following observation. For $\lambda = 0$ the flavor symmetry is enhanced to $SU(2)$ (and the magnetic symmetry is enhanced as well).%
\footnote{For $\lambda=0$ the potential $V$ is invariant under $O(N^2)$ acting on the entries of $\phi$, however in the full $SO(N)$ gauge theory with $N>2$ there is no symmetry enhancement.}
This means that there are two domains $\lambda \gtrless 0$ in the RG flow, separated by the more symmetric theories at $\lambda=0$, and no RG trajectories cross from one domain to the other (at least as long as the symmetry is not spontaneously broken).

In fact, the potential $V_\text{EP}$ appearing in the dualities proposed in this Section has $\lambda<0$ (contrary to the cases in Section \ref{sec: SO duality} where $\lambda>0$)---and the theories have indeed gauge group $SO(2)$. It is convenient to regard the gauge group as $U(1)$ and use the complex notation. Let us then add some details about this particular case.

We write the ``easy plane'' potential $V_\text{EP}$, function of two complex scalars $\phi_{1,2}$, as
\be
\label{V_EP}
V_\text{EP} = |\phi_1|^4 + |\phi_2|^4 + 2(\lambda+1)\, |\phi_1|^2 |\phi_2|^2 \qquad\text{with}\qquad -2 < \lambda < 0 \;.
\ee
It is invariant under separate rotations of $\phi_1$ and $\phi_2$, under charge conjugation $\bZ_2^\cC$ and under exchange $\bZ_2^\cX$: $\phi_1 \leftrightarrow \phi_2$. We take $\lambda>-2$ in order for the potential to be positive definite. With $\lambda=0$ the symmetry would be enhanced and $(\phi_1, \phi_2)$ would transform as an $SU(2)$ doublet, besides having charge 1 under the $U(1)$ that is gauged.

There is a unique quadratic term invariant under the symmetries, $2v \big( |\phi_1|^2 + |\phi_2|^2\big)$. With positive mass squared, $v>0$, the only minimum of the deformed potential is at the origin and $V=0$ there. With negative mass squared, $v<0$, the minima depend on $\lambda$:
$$
\begin{array}{rlll}
-2<\lambda<0: & |\phi_1|^2 = |\phi_2|^2 = \frac{|v|}{2+\lambda} & V = - \frac{2v^2}{2+\lambda} & S^1 \\[.3em]
\lambda =0: & |\phi_1|^2 + |\phi_2|^2 = |v| & V = - v^2 & S^2 \\[.3em]
\lambda > 0: & \{ \phi_1=0, |\phi_2|^2 = |v| \} \cup \{ |\phi_1|^2 = |v|, \phi_2 = 0\} \quad & V = - v^2 \qquad & \text{2 vacua}
\end{array}
$$
In the last column we have indicated the set of ground states of the gauge theory. In the examples of this Section, the phase diagrams match if we choose $-2< \lambda<0$.

We can also consider deformations that keep one of the two fields massless. If we deform (\ref{V_EP}) with $2v|\phi_1|^2$ and positive $v$, the minimum is at the origin and $\phi_2$ remain massless. If we deform with negative mass squared for $\phi_1$, we should tune the mass of $\phi_2$ in such a way that the latter remains massless at the minimum. In the range $-2<\lambda \leq 0$ this can be done: the correct tuning is
\be
V = |\phi_1|^4 + |\phi_2|^4 + 2(\lambda+1)\, |\phi_1|^2 |\phi_2|^2 + 2 v \big( |\phi_1|^2 + (\lambda+1) |\phi_2|^2 \big)
\ee
with $v<0$. The minimum of this potential is at $|\phi_1|^2 = |v|$, $\phi_2=0$ (where $V = -v^2$) and at that point $\phi_2$ is massless. On the other hand, for $\lambda>0$ we encounter a subtlety. The point $\{ |\phi_1|^2 = |v|, \phi_2=0\}$ is still a local minimum of the potential, but the global minimum is at $\{\phi_1 = 0, |\phi_2|=(\lambda+1) |v| \}$ where $V = - (\lambda+1)^2 v^2$ and both $\phi_1$ and $\phi_2$ are massive. Hence---already at the classical level---there is no second-order phase transition as we tune the mass of $\phi_2$.

\subsection[Duality $U(1)_{-\frac12}$ with $\phi$, $\psi$ $\;\longleftrightarrow\;$ $U(1)_1$ with 2 $\phi$ and $V_\text{EP}$]{Duality \matht{U(1)_{-\frac12}} with \matht{\phi}, \matht{\psi} \matht{\;\longleftrightarrow} \matht{U(1)_1} with 2 \matht{\phi} and \matht{V_\text{EP}}}
\label{sec: U(1)-1/2 phi psi}

We start with the duality (\ref{Abdurev-b}) and shift the background field $B \to B + X$. Then we add a free fermion $i\bar\psi \Dslash_B \psi$ coupled to $B$, as well as a counterterm $\frac1{2\pi} BdY$, on both sides. Finally we make $B$ dynamical and rename it $b$. We obtain the duality of Lagrangians
\begin{multline}
\label{Abelian duality 1}
|D_{b+X}\phi|^2 - |\phi|^4 + i \bar\psi \Dslash_b \psi + \frac1{2\pi} bdY \quad\longleftrightarrow \\
i \bar\psi_1 \Dslash_a \psi_1 + i \bar\psi_2 \Dslash_b \psi_2 + \frac{(a+b)d(a+b)}{4\pi} + \frac1{2\pi} adX + \frac1{2\pi} bd(X+Y) + \frac1{4\pi} XdX + 2\text{CS}_\text{g} \;.
\end{multline}
Here $X,Y$ are background gauge fields. Notice that the two theories respect the spin/charge relation \cite{Seiberg:2016rsg, Seiberg:2016gmd} and can thus be defined on non-spin manifolds, provided we promote $a, b, X$ to spin$_c$ connections and add the counterterm $\frac1{4\pi} YdY$ on both sides (see Appendix \ref{app: spinc}). Here we will content ourselves with working on spin manifolds. On the LHS we have not included a term $|\phi|^2 \bar\psi\psi$, which is compatible with the symmetries, since we expect that such a term is not present in the IR. By the same reasoning we have not included four-Fermi interactions on the RHS. What we have obtained is the duality
\be
\label{Abelian duality 1 summary}
U(1)_{-\frac12} \text{ with $\phi$, $\psi$} \qquad\longleftrightarrow\qquad U(1)^2_K \text{ with 2 $\psi$}
\ee
(with gravitational coupling $U(1)_{-1}$ on the RHS) where the theory on the RHS has a CS matrix $K = \smat{1/2 & 1 \\ 1 & 1/2}$. In the following discussion we will need the phase diagram of these theories. We have already presented it (using the description on the LHS) in Figure \ref{fig: U(1)1/2} on the left, as well as in (\ref{phase diagram U(1) 1/2 with phi psi}).

Both theories have a manifest global symmetry $\wt{U(1)}_X \times \wt{U(1)}_{X+Y} \rtimes \bZ_2^\cC$ that acts faithfully. Here the two $U(1)$ factors are the ones natural on the RHS of (\ref{Abelian duality 1}), that couple to $X$ and $X+Y$ respectively. They are related to $U(1)_X \times U(1)_Y$, natural on the LHS, by an obvious transformation. On the other hand $\bZ_2^\cC$ is charge conjugation that inverts all gauge fields, in particular
\be
\bZ_2^\cC:\qquad X \to - X \;,\qquad Y \to -Y \;.
\ee
Turning off background fields, we see that the theory on the RHS also has a $\bZ_2^\cX$ symmetry that exchanges $\psi_1 \leftrightarrow\psi_2$ and $a \leftrightarrow b$. With background fields the symmetry acts as
\be
\label{Z2X symmetry}
\bZ_2^\cX\;:\qquad\quad X \leftrightarrow X+Y \;,\qquad\quad Y \leftrightarrow - Y \;,
\ee
in other words $\bZ_2^\cX$ exchanges $\wt{U(1)}_X \leftrightarrow \wt{U(1)}_{X+Y}$, and there is an anomaly given by
\be
\label{Z2X anomaly}
\cL[X,Y] \;\stackrel{\bZ_2^\cX}{\longleftrightarrow}\; \cL[X,Y] + \frac1{4\pi} YdY + \frac1{2\pi} XdY
\ee
where $\cL$ is the effective Lagrangian. The full global symmetry group is thus%
\footnote{The symmetry group can also be written as $\big( O(2)_{X + Y/2} \times O(2)_{Y/2} \big)/\bZ_2$ where the quotient is by the $-\unit$ element on both sides.}
\be
\wt{U(1)}_X \times \wt{U(1)}_{X+Y} \rtimes \big( \bZ_2^\cC \times \bZ_2^\cX \big) \;.
\ee
If the theory flows to a fixed point, possibly with a tuning of the relevant fermion-mass deformations invariant under $\bZ_2^\cX$, then we conclude that also the theory on the LHS develops the $\bZ_2^\cX$ symmetry in the IR. Such a symmetry is not manifest on the LHS---although it is manifest in its phase diagram in Figure \ref{fig: U(1)1/2} and in (\ref{phase diagram U(1) 1/2 with phi psi}).

So, let us show how $\bZ_2^\cX$ appears on the LHS of (\ref{Abelian duality 1 summary}). Combining the dualities in (\ref{Abdurev-b}) and (\ref{Abdurev-c}) in a way similar to what we did before, we obtain the duality of Lagrangians
\begin{multline}
|D_{b+X}\phi|^2 - |\phi|^4 + i \bar\psi \Dslash_b \psi + \frac1{2\pi} bdY \quad\longleftrightarrow \\
|D_{c+X+Y}\phi|^2 - |\phi|^4 + i \bar\psi \Dslash_c \psi - \frac1{2\pi} cdY - \frac1{4\pi} YdY - \frac1{2\pi}XdY \;.
\end{multline}
In the second line we integrated out a dynamical gauge field that appeared linearly. Once again, the duality is well-defined on non-spin manifolds provided we promote $b, c, X$ to spin$_c$ connections and add $\frac1{4\pi} YdY$ on both sides. This duality is a self-duality of $U(1)_{-\frac12}$ with $\phi$, $\psi$, that acts on the background fields $X,Y$ as in (\ref{Z2X symmetry}) and has the anomaly (\ref{Z2X anomaly}). We thus identify this self-duality with $\bZ_2^\cX$.

In terms of the basis $U(1)_X \times U(1)_Y$ for the continuous global symmetry, with charges $Q_X$, $Q_Y$, the self-duality leaves $Q_X$ invariant and maps $Q_Y \leftrightarrow Q_X - Q_Y$. Moreover it exchanges the two relevant deformations:
\be
\bZ_2^\cX:\qquad |\phi|^2 \;\leftrightarrow\; \bar\psi\psi \;.
\ee
This can be inferred by comparing the low-energy theories after deforming with the two operators. It is also apparent from the phase diagram in Figure \ref{fig: U(1)1/2}.

Next, combining the dualities in (\ref{Abdu-a}) and (\ref{Abdu-c}) we obtain the duality of Lagrangians
\begin{multline}
|D_{b+X}\phi|^2 - |\phi|^4 + i \bar\psi \Dslash_b \psi + \frac1{2\pi} bdY \quad\longleftrightarrow \\
|D_c \phi_1|^2 + \frac1{2\pi} cd(b+X) + |D_f\phi_2|^2 + \frac1{4\pi} fdf + \frac1{2\pi}fdb - V\big( |\phi_1|^2, |\phi_2|^2 \big) + \frac1{2\pi}bdY \;.
\end{multline}
On the RHS we have included a generic quartic potential in the scalars $\phi_{1,2}$, compatible with the global $\wt{U(1)}_X \times \wt{U(1)}_{X+Y} \rtimes \bZ_2^\cC$ and gauge symmetries, as such a potential is generated along the RG flow when we make the background gauge field $B$ dynamical (and rename it $b$). The gauge field $b$ on the RHS can be integrated out, and redefining $\phi_2 \to \phi_2^*$ we obtain
\begin{multline}
\label{Abelian duality 3}
|D_{b+X}\phi|^2 - |\phi|^4 + i \bar\psi \Dslash_b \psi + \frac1{2\pi} bdY \quad\longleftrightarrow \\
|D_c\phi_1|^2 + |D_{c+Y}\phi_2|^2 - V\big( |\phi_1|^2, |\phi_2|^2 \big) + \frac1{4\pi} cdc + \frac1{2\pi}cd(X+Y) + \frac1{4\pi}YdY \;.
\end{multline}
(On non-spin manifolds we should promote $b,X$ to spin$_c$ connections and add $\frac1{4\pi} YdY$ on both sides.)
This is the duality
\be
U(1)_{-\frac12} \text{ with $\phi$, $\psi$} \qquad\longleftrightarrow\qquad U(1)_1 \text{ with 2 $\phi$ and $V_\text{EP}$}
\ee
(with no extra gravitational counterterms). On the RHS the quartic potential $V_\text{EP}$ reduces the global symmetry.%
\footnote{With $V = \big( |\phi_1|^2 + |\phi_2|^2 \big)^2$ (or without $V$), the theory on the RHS would have $U(2) \rtimes \bZ_2^\cC$ faithfully-acting global symmetry \cite{Benini:2017dus}.}
Let us show that $V$ is precisely the ``easy plane'' potential (\ref{V_EP}).

We could entertain the possibility of a $\bZ_2^\cX$ symmetry that exchanges $\phi_1 \leftrightarrow \phi_2$ (in the absence of background fields): whether this is a symmetry of the theory depends on the potential $V$. With background fields the full action is
\be
\bZ_2^\cX\,:\qquad c \leftrightarrow c + Y \;,\qquad\quad X \leftrightarrow X+Y \;,\qquad\quad Y \leftrightarrow -Y \;,
\ee
with exactly the same anomaly as in (\ref{Z2X anomaly}). We recognize that this is the same $\bZ_2^\cX$ action discussed before, and---if the theory flows to a fixed point---we argued in the fermionic description on the RHS of (\ref{Abelian duality 1 summary}) that it is indeed a symmetry in the IR. This means that we should impose the full $\wt{U(1)}_X \times \wt{U(1)}_{X+Y} \rtimes\big( \bZ_2^\cC \times \bZ_2^\cX\big)$ symmetry on the potential $V$. We then claim that the potential has $\lambda<0$ as in (\ref{V_EP}).
This is dictated by the requirement that the theory reproduces the same phase diagram as $U(1)_{-\frac12}$ with $\phi$, $\psi$. For instance, turning on a negative mass for both $\phi_{1,2}$, the minima of the potential are at $|\phi_1|, |\phi_2| \neq 0$: the gauge symmetry is broken, as well as the $U(1)$ global symmetry that couples to $Y$. We are left with an $S^1$ NLSM coupled to $Y$ without extra counterterms. This reproduces the phase $m_\phi^2>0$, $m_\psi>0$ on the LHS of (\ref{Abelian duality 3}) as reported in (\ref{U(1)-1/2 phi psi phases with background}). The other phases and critical lines are reproduces in a similar way. We conclude that $V = V_\text{EP}$.

As noted before, the theory $U(1)_1$ with 2 $\phi$ and $V_\text{EP}$ coincides with
$$
SO(2)_1 \text{ with 2 $\phi$} \;,
$$
except that the relative strength of the two quartic scalar couplings is not the same as in the series of $SO$ dualities discussed in Section \ref{sec: SO duality}.

\subsubsection{The gapped phases}
\label{sec: U(1)-1/2 phi psi phases with background}

To conclude the discussion of $U(1)_{-\frac12}$ with $\phi$, $\psi$, let us list the gapped phases:
\bea
\label{U(1)-1/2 phi psi phases with background}
m_\phi^2>0,\, m_\psi>0: \qquad & \frac1{2\pi} bdY && \text{($S^1$ NLSM coupled to $Y$)} \\
m_\phi^2>0,\, m_\psi<0: \qquad & \frac1{4\pi} YdY \\
m_\phi^2<0,\, m_\psi<0: \qquad & - \frac1{4\pi} Xd(X+2Y) - 2\text{CS}_\text{g} \\
m_\phi^2<0,\, m_\psi>0: \qquad & - \frac1{2\pi} XdY \;.
\eea
This is a refined version of Figure \ref{fig: U(1)1/2} in which we have included the counterterms for background fields. This clearly shows that the four phases are all different.%
\footnote{Without background fields, the second and fourth phases might have looked equal. One could have then suspected that they were connected, with no transition in the middle. We can instead exclude such a scenario.}

\subsection[Time-reversal symmetry of $U(1)_\frac32$ with $\phi$, $\psi$]{Time-reversal symmetry of \matht{U(1)_\frac32} with \matht{\phi}, \matht{\psi}}
\label{sec: U(1) 3/2 time rev}

Combining the dualities in (\ref{Abdurev-b}) and (\ref{Abdu-c}) we obtain
\begin{multline}
\label{duality U(1) 3/2}
|D_{b+X}\phi|^2 - |\phi|^4 + i \bar\psi \Dslash_b \psi - \frac1{4\pi} bdb + \frac1{2\pi} bdY \quad\longleftrightarrow \\
|D_{c+X+Y}\phi|^2 - |\phi|^4 + i \bar\psi \Dslash_c\psi + \frac2{4\pi} cdc + \frac1{2\pi} cd(Y+2X) + \frac{2Xd(X+Y) + YdY}{4\pi} + 2\text{CS}_\text{g} \;.
\end{multline}
In the second line we integrated out a gauge field that appeared linearly and redefined \mbox{$\phi \to \phi^*$}. The two theories respect the spin/charge relation, thus the duality is well-defined on non-spin manifolds provided we promote $b,c,X$ to spin$_c$ connections and add the counterterms $\frac1{4\pi} YdY - 2\text{CS}_\text{g}$ on both sides. This is the duality
\be
U(1)_{-\frac32} \text{ with $\phi$, $\psi$} \qquad\longleftrightarrow\qquad U(1)_\frac32 \text{ with $\phi$, $\psi$}
\ee
(with a gravitational counterterm $U(1)_1$ on the LHS) that we already presented in Section~\ref{sec: U/U duality}. Turning off background fields, this is precisely the action of time reversal. We conclude that, if the theory flows to a fixed point, the latter is time-reversal invariant (with an anomaly that we are going to discuss).

In the presence of background fields, we should define an action of time reversal $\cT$ on the background as well:
\be
\cT:\qquad X \;\to\; X+Y \;,\qquad Y \;\to\; - Y - 2X \;.
\ee
From here we see that $\cT$ squares to $\cC$, the generator of charge conjugation $\bZ_2^\cC$, namely time reversal forms a group $\bZ_4^\cT$ in which $\bZ_2^\cC$ is the non-trivial subgroup.%
\footnote{See the recent work \cite{Barkeshli:2017rzd} for other examples in Chern-Simons TQFTs.}
The action of $\cT$ on the effective action is $\cT\cL[X,Y] = \cL_\text{dual}[X,Y] - \frac1{4\pi}\big( 2XdX + 2XdY + YdY \big)$, where $\cL$ and $\cL_\text{dual}$ correspond to the LHS and RHS of (\ref{duality U(1) 3/2}) respectively. The duality asserts that $\cL[X,Y] = \cL_\text{dual}[X,Y]$ as effective Lagrangians. We conclude that
\be
\cT\cL[X,Y] = \cL[X,Y] - \frac1{2\pi} Xd(X+Y) - \frac1{4\pi} YdY \;.
\ee
It is easy to check that there is no anomaly for $\cT^2 = \cC$.

The map of mass operators is
\be
\text{LHS} \qquad \begin{aligned}
|\phi|^2 &\quad\longleftrightarrow\quad \bar\psi\psi \\
\bar\psi\psi &\quad\longleftrightarrow\quad - |\phi|^2
\end{aligned} \qquad \text{RHS} \;.
\ee
This can be checked by deforming the two Lagrangians in (\ref{duality U(1) 3/2}) and comparing the resulting theories, making use of (\ref{Abelian dualities with background}) and (\ref{Abelian dualities with background t-rev}). This is essentially a refined version of the phase diagram (\ref{scheme deformations U(1)3/2}) in which we keep background fields under consideration. If we completely gap the theory we find:
\bea
m_\phi^2>0,\, m_\psi>0:\qquad& \frac1{4\pi} YdY + 2 \text{CS}_\text{g} \\
m_\phi^2>0,\, m_\psi<0:\qquad& \frac2{4\pi} bdb + \frac1{2\pi}bdY + \frac1{4\pi}YdY + 2 \text{CS}_\text{g} \\
m_\phi^2<0,\, m_\psi<0:\qquad& - \frac2{4\pi}XdX - \frac1{2\pi}XdY - 2 \text{CS}_\text{g} \\
m_\phi^2<0,\, m_\psi>0:\qquad& - \frac1{4\pi} XdX - \frac1{2\pi} XdY \;.
\eea
On the second line we have $U(1)_2$. This specifies the local counterterms for background fields in the gapped phases.

\subsection[Duality $U(1)_{-1}$ with 2 $\psi$ $\;\longleftrightarrow\;$ $U(1)_2$ with 2 $\phi$ and $V_\text{EP}$]{Duality \matht{U(1)_{-1}} with 2 \matht{\psi} \matht{\;\longleftrightarrow\;} \matht{U(1)_2} with 2 \matht{\phi} and \matht{V_\text{EP}}}
\label{sec: SO excluded}

Combining two copies of (\ref{Abdu-c}) we obtain the duality of Lagrangians
\begin{multline}
i\bar\psi_1 \Dslash_{a+X} \psi_1 + i \bar\psi_2 \Dslash_a \psi_2 + \frac1{2\pi} adY - \frac1{4\pi} YdY \qquad\longleftrightarrow \\
|D_{b+Y} \phi_1|^2 + |D_b \phi_2|^2 - V\big( |\phi_1|^2, |\phi_2|^2 \big) + \frac2{4\pi} bdb + \frac1{2\pi} bd(X+Y) \;.
\end{multline}
The two theories respect the spin/charge relation and could be defined on non-spin manifolds promoting $a$ to a spin$_c$ connection. This is the duality
\be
\label{duality U(1)2 2 phi}
U(1)_{-1} \text{ with 2 $\psi$} \qquad\longleftrightarrow\qquad U(1)_2 \text{ with 2 $\phi$ and $V_\text{EP}$} \;.
\ee
Notice that the theory on the LHS is also dual to $SU(2)_1$ with 2 $\phi$ \cite{Hsin:2016blu}.

The theory on the LHS has manifest $SO(3)_X \times O(2)_Y$ symmetry: $SO(3)_X$ is the electric symmetry with maximal torus $U(1)_X$, and there is a magnetic symmetry $U(1)_Y$ that, combined with a suitable $\bZ_2^Y$ charge-conjugation symmetry%
\footnote{If we define $\bZ_2^\cC$ charge conjugation as in Appendix \ref{app: charge conjugation}, we find that it leaves invariant $v^z \equiv \bar\psi_1 \psi^1 - \bar\psi_2 \psi^2$ and $v^x \equiv \bar\psi_1 \psi^2 + \bar\psi^2 \psi^1$, but it inverts $v^y \equiv i \bar\psi_1\psi^2 - i \bar\psi_2\psi^1$, and therefore it does not commute with $SO(3)_X$. We can combine $\cC$ with a rotation of $SO(3)_X$ such that all three operators are inverted: now this action commutes with $SO(3)_X$. We denote such an action by $\bZ_2^Y$.}
\be
\bZ_2^Y: \quad a \to -a-X \;,\quad X \to X \;,\quad Y \to -Y \;,\quad \psi_1 \to i \psi_2^c \;,\quad \psi_2 \to -i \psi_1^c \;,
\ee
gives $O(2)_Y$. The symmetry is isomorphic to $\big( U(2)/\bZ_2 \big) \rtimes \bZ_2^\cC$, where $\bZ_2^\cC$ is the standard charge conjugation (see Appendix \ref{app: charge conjugation}), and there is an anomaly \cite{Benini:2017dus}. Upon symmetry-invariant mass deformation, the theory flows to the $S^1$ NLSM for positive fermion mass, and to $U(1)_{-2}$ for negative fermion mass.

The theory on the RHS has only $O(2)_X \times O(2)_Y$ manifest symmetry, because of the potential $V_\text{EP}$ studied in Section~\ref{sec: potential}. The condition $-2<\lambda<0$ guarantees that the same phase diagram as on the LHS is reproduced. We can describe the $O(2)^2$ symmetry as follows. There is a magnetic symmetry, that we call $U(1)_X$, with current $J_\mu^{(X)} = \epsilon_{\mu\nu\rho}F^{\nu\rho}$. There is an electric global symmetry $U(1)^\text{(elect)}$ that gives charge $1$ to $\phi_1$ and $0$ to $\phi_2$. We combine it with the magnetic symmetry and define $U(1)_Y$ with current $J_\mu^{(Y)} = J_\mu^\text{(elect)} + J_\mu^{(X)}$. Next there is a $\bZ_2^Y$ symmetry that exchanges $\phi_1 \leftrightarrow \phi_2$ (including the background, it shifts $b \to b+Y$ and $Y \to -Y$). Clearly it does not affect the magnetic charge and so it commutes with $U(1)_X$. However consider a gauge-invariant operator with magnetic charge $x$: it is obtained from a bare monopole and, because of the CS term, should be dressed with $2x$ fields $\phi^*_I$, together with an arbitrary string of $\phi_I^* \phi^{\phantom{*}}_J$. It is easy to see that $\bZ_2^Y$ inverts the charge under $U(1)_Y$ of such a gauge-invariant operator. Finally, we can define a $\bZ_2^X$ charge-conjugation symmetry acting as
\be
\bZ_2^X:\quad b \to -b-Y \;,\quad X \to -X \;,\quad Y \to Y \;,\quad \phi_1 \to \phi_2^* \;,\qquad \phi_2 \to \phi_1^* \;.
\ee
This is a $\bZ_2$ on gauge invariants. Such a symmetry inverts the magnetic charge and---as one can easily check---leaves invariant the charge under $U(1)_Y$ of gauge-invariant operators. We conclude that the manifest symmetry is $O(2)_X \times O(2)_Y$, with $O(2)_X = U(1)_X \rtimes \bZ_2^X$ and $O(2)_Y = U(1)_Y \rtimes \bZ_2^Y$. Note that both the generators of $\bZ_2^X$ and $\bZ_2^Y$ lead to an anomaly, while their product does not.

We identify the symmetry factor $O(2)_Y$ on the two sides of the duality, and $O(2)_X$ on the RHS with a subgroup of $SU(2)_X$ on the LHS. The coupling to background fields matches, as well as the anomaly when restricted to $O(2)^2$. A consequence of the duality is that, if the theory flows to a fixed point, $O(2)_X$ on the RHS is enhanced to $SO(3)_X$ in the IR. Let us consider some operators and their duals:
\be
\begin{array}{c|cc|c|cc|c}
U(1)_{-1} \text{ with 2 $\psi$} & U(1)_X & \bZ_2^X & SO(3)_X & U(1)_Y & \bZ_2^Y & U(1)_2 \text{ with 2 $\phi$ and $V_\text{EP}$} \\
\hline \rule{0pt}{1.1em}
\bar\psi_1\psi_1 + \bar\psi_2 \psi_2 & 0 & + & \rep{1} & 0 & + & - |\phi_1|^2 - |\phi_2|^2 \\
\bar\psi_1\psi_1 - \bar\psi_2 \psi_2 & 0 & - & \rep{3} & 0 & - & |\phi_1|^2 - |\phi_2|^2 \\
\bar\psi_1 \gamma_\mu \psi_1 - \bar\psi_2 \gamma_\mu \psi_2 & 0 & - & \rep{3}' & 0 & + & \epsilon_{\mu\nu\rho} F^{\nu\rho} \\
\bar\psi_2\psi_1 \;\oplus\; \bar\psi_2 \gamma_\mu \psi_1 & 1 & \raisebox{-1.3em}[0pt][0pt]{\begin{tikzpicture} \draw [<->,thick] (0,0)--(0,.7); \end{tikzpicture}} & \rep{3} \oplus \rep{3}' & 0 & -/+ & \cM \phi_1^* \phi_2^* \\
\bar\psi_1\psi_2 \;\oplus\; \bar\psi_1 \gamma_\mu \psi_2 & -1 & & \rep{3} \oplus \rep{3}' & 0 & -/+ & \wb\cM \phi_1 \phi_2 \\
\cN & 0 & + & \rep{1} & 1 & \raisebox{-1.3em}[0pt][0pt]{\begin{tikzpicture} \draw [<->,thick] (0,0)--(0,.7); \end{tikzpicture}} & \phi_1 \phi_2^* \\
\wb\cN & 0 & + & \rep{1} & -1 & & \phi_1^* \phi_2
\end{array}
\ee
The spin 1 operators with $U(1)_X \times U(1)_Y$ charges $(\pm1, 0)$ can enhance the symmetry to $SO(3)_X \times O(2)_Y$, and the spectrum forms representations of $SO(3)_X$ (we have indicated two triplets by \rep{3} and $\rep{3}'$).%
\footnote{Note that $\cM \phi_1^* \phi_2^*$ corresponds, in radial quantization, to two modes of the scalars on $S^2$ with one unit of magnetic flux. Each mode has spin $\frac12$, thus the symmetric and antisymmetric contractions give a triplet and a singlet under spacetime rotations, respectively.}

Let us consider mass deformations of $U(1)_2$ with 2 $\phi$ and $V_\text{EP}$. If we deform by a positive mass term for both scalars, we get $U(1)_2$. By duality, this is the same as the LHS theory deformed by a negative mass terms for both fermions. If we deform by a negative mass term for both scalars, we get an $S^1$ NLSM from the spontaneous breaking $O(2)_Y \to \bZ_2^Y$ (see Section~\ref{sec: potential}). This agrees with a positive fermion-mass deformation of the LHS theory.

We can also consider a mass term for a single field. In the fermionic theory we get $U(1)_{-\frac32}$ with $\psi$, or $U(1)_{-\frac12}$ with $\psi$. In the scalar theory a positive mass leads to $U(1)_2$ with $\phi$, while a negative mass leads to the $O(2)$ Wilson-Fisher fixed point. In both cases we have a match.

What happens if we take $U(1)_2$ with 2 $\phi$ on the RHS, namely the theory with the maximally symmetric potential $V = \big( |\phi_1|^2 + |\phi_2|^2 \big)^2$ (\ie{} $\lambda=0$)? In this case the theory has a manifest $O(2)_X \times SO(3)_Y$ global symmetry. Although this seems similar to what we discussed before, now the symmetry-invariant negative mass-squared deformation $- |\phi_1|^2 - |\phi_2|^2$ leads to an $S^2$ NLSM (with a Wess-Zumino term), and is not dual to $U(1)_{-1}$ with 2 $\psi$. In fact, the different duality
\be
U(1)_2 \text{ with 2 $\phi$} \quad\longleftrightarrow\quad SU(2)_0 \text{ with 2 $\psi$} \quad\longleftrightarrow\quad U(1)_{-2} \text{ with 2 $\phi$}
\ee
was proposed in \cite{Komargodski:2017keh}.

\

The duality (\ref{duality U(1)2 2 phi}) can be written as
\be
\label{special SO(2) duality}
SO(2)_{-1} \text{ with 2 $\psi$} \qquad\longleftrightarrow\qquad SO(2)_2 \text{ with 2 $\phi$} \;,
\ee
with the understanding that on the RHS the quartic scalar potential (\ref{potential with lambda}) has $\lambda<0$, as opposed to the cases of Section \ref{sec: SO duality} (see the discussion in Section \ref{sec: potential}). Notice that in the $SO$ dualities, electric and magnetic symmetries are exchanged \cite{Aharony:2013kma, Aharony:2015mjs, Aharony:2016jvv} which is precisely what happens here.

This duality was not recognized in \cite{Aharony:2016jvv}. The reason is that the operator map between quadratic mass terms differs from the other $SO$ dualities, and this is crucial to find agreement between the two phase diagrams. In the $SO$ description, we use four real scalar fields $\varphi_{\alpha I}$ transforming as a vector of both the gauge ($\alpha =1,2$) and global ($I=1,2$) $SO(2)$'s. The scalar theory has the following quartic potential:
\be
V_\text{EP} = \Big( 1 - \frac\lambda2\Big) \, \big( \varphi_{\alpha I} \varphi_{\alpha I} \big)^2 + \lambda \, \varphi_{\alpha I} \varphi_{\alpha J} \varphi_{\beta J} \varphi_{\beta I} \;,
\ee
as in (\ref{potential with lambda}) and (\ref{V_EP}). We can relate the $U$ and $SO$ descriptions by mapping
\bea
\phi_1 &= \frac1{\sqrt2} \big[ (\varphi_{11} + i \, \varphi_{21}) + i (\varphi_{12} + i \, \varphi_{22}) \big] \\
\phi_2 &= \frac1{\sqrt2} \big[ (\varphi_{11} + i \, \varphi_{21}) - i (\varphi_{12} + i \, \varphi_{22}) \big] \;.
\eea
Then the operator map easily follows from the $U$ description:
\bea
\bar\psi_1\psi_1 + \bar\psi_2\psi_2 &\quad\longleftrightarrow\quad - |\phi_1|^2 - |\phi_2|^2 = - \varphi_{\alpha I} \varphi_{\alpha I} \\
\bar\psi_1\psi_1 - \bar\psi_2\psi_2 &\quad\longleftrightarrow\quad |\phi_1|^2 - |\phi_2|^2 = - \varphi_{\alpha I} \varphi_{\beta J} \epsilon^{\alpha\beta} \epsilon^{IJ} \;.
\eea
In particular, giving mass to a single fermion $\psi_I$ does \emph{not} correspond to giving mass to a single scalar $\varphi_I$, as instead happens in the other $SO$ dualities.

\subsection[Duality $U(1)_0$ with 2 $\psi$ $\;\longleftrightarrow\;$ $U(1)_0$ with 2 $\phi$ and $V$]{Duality \matht{U(1)_0} with 2 \matht{\psi} \matht{\;\longleftrightarrow\;} \matht{U(1)_0} with 2 \matht{\phi} and \matht{V}}
\label{sec: scalar QED}

To conclude, we consider QED with two scalars, namely $U(1)_0$ with $\phi_i$ and $i=1,2$. Dualities of this theory and related ones have already been proposed in \cite{Motrunich:2003fz, Xu:2015lxa, Karch:2016sxi, Hsin:2016blu, Benini:2017dus, Wang:2017txt}, and here we would like to add some details.

The faithfully-acting global symmetry is%
\footnote{It has been proposed in \cite{Wang:2017txt} that the symmetry might be enhanced to $SO(5) \times \bZ_2^\cT$ at the critical point.}
\be
SO(3)_X \times O(2)_Y \times \bZ_2^\cT \;,
\ee
where $O(2)_Y = U(1)_M \rtimes \bZ_2^\cC$ is the product of the magnetic symmetry and a charge-conjugation symmetry. Time reversal will not play a role in our analysis, and we will ignore it for now. The gauge-invariant scalar operators that are quadratic in the matter fields, $\phi^*_i \phi^{\phantom{*}}_j$, transform in the $\rep{3} \oplus \rep{1}$ representation of $SO(3)_X$. The quartic gauge-invariant scalar operators are in the $\rep{5} \oplus \rep{3} \oplus \rep{1}$ of $SO(3)_X$.

If we insist on $SO(3)_X \times O(2)_Y$ symmetry, there is only one quadratic and one quartic operator we can turn on:
\be
\label{rel ops SO(3) O(2)}
\cO_\rep1 = \phi_i^* \phi_i^{\phantom{*}} = |\phi_1|^2 + |\phi_2|^2
\ee
and $(\cO_\rep1)^2$. Tuning $\cO_\rep1$, we assume to reach a CFT $\cT_0$ with (at least) $SO(3)_X \times O(2)_Y \times \bZ_2^\cT$ global symmetry.

If, instead, we relax the symmetry to $O(2)_X \times O(2)_Y$ (where $O(2)_X \subset SO(3)_X$ contains the Cartan of $SO(3)_X$ and the $\bZ_2$ symmetry $\phi_1 \leftrightarrow\phi_2$), then there are one quadratic and two quartic gauge-invariants that preserve it: $\cO_\rep1$, $(\cO_\rep1)^2$ and
\be
\cO_{(\rep5)} = \big( |\phi_1|^2 - |\phi_2|^2 \big)^2 \;.
\ee
Here the notation ${}_{(\rep5)}$ represents a particular component of the multiplet in the \rep5. The relevant deformation $\cO_{(\rep5)}$ induces an RG flow to two phases with $O(2)_X \times O(2)_Y \times \bZ_2^\cT$ global symmetry, separated by the CFT $\cT_0$ (tuning $\cO_\rep1$ to zero).

Let us mention the remaining quadratic and quartic gauge invariants. The quadratic ones are $\cO_\rep1$ and
\be
(\cO_\rep3)_{ij} = \phi_i^* \phi_j - \frac12 \delta_{ij} \cO_\rep1 \;,
\ee
satisfying $(\cO_\rep3)_{ij} \delta^{ij} = 0$. The quartic gauge invariants are $(\cO_\rep1)^2$, $\cO_\rep1 \cO_\rep3$ and
\be
(\cO_\rep5)_{ijkl} = (\cO_\rep3)_{ij} (\cO_\rep3)_{kl} + \frac14 (\delta_{ik} \delta_{jl} - \delta_{il} \delta_{jk}) \cO_\rep1^2 \;,
\ee
satisfying $(\cO_\rep5)_{ijkl} \delta^{ij} = (\cO_\rep5)_{ijkl} \delta^{jk} = 0$. Notice the relation $(\cO_\rep3)_{ij} (\cO_\rep3)_{jk} = \frac14 \delta_{ik} \cO_\rep1^2$ and its trace $\Tr \cO_\rep3 \cO_\rep3 = \frac12 \cO_\rep1^2$, thus one singlet is not independent. To break $SO(3)_X \to U(1)_X$ we can use the tensor $(\sigma_3)_{ij}$ and construct $\cO_{(\rep3)} \equiv (\cO_\rep3)_{ij} (\sigma_3)_{ij} = |\phi_1|^2 - |\phi_2|^2$. On the other hand $\cO_{(\rep5)} \equiv (\cO_\rep5)_{ijkl} (\sigma_3)_{ij} (\sigma_3)_{kl}$ breaks $SO(3)_X \to O(2)_X$.

We can study the various phases of the theory under deformations. While preserving the full $SO(3)_X \times O(2)_Y \times \bZ_2^\cT$ symmetry, we can deform with $m^2 \cO_\rep1$ and obtain
\be
m^2 \cO_\rep1\,: \qquad \begin{cases} S^1 \text{ NLSM} & \text{for } m^2 > 0 \\ S^2 \text{ NLSM} & \text{for } m^2 < 0 \;. \end{cases}
\ee
The $S^1$ NLSM comes from $U(1)_0$ and it corresponds to the magnetic symmetry breaking $O(2)_Y \to \bZ_2^\cC$, while the $S^2$ NLSM is a $\bC\bP^1$ model with vanishing Hopf term and it corresponds to the symmetry breaking $SO(3)_X \to U(1)_X$.

If we relax the symmetry to $O(2)_X \times O(2)_Y \times \bZ_2^\cT$, there is one more relevant deformation: $\cO_{(\rep5)}$. Tuning to zero the quadratic terms, the potential is $V = \lambda_1 \cO_\rep1^2 + \lambda_5 \cO_{(\rep5)}$, which is the same as (\ref{V_EP}) (up to an overall constant, $\lambda = - \frac{2\lambda_5}{\lambda_1 + \lambda_5}$). With this notation, the potential is positive definite for $\lambda_1 > 0$ and $\lambda_1 > - \lambda_5$. At this point we can turn on $m^2 \cO_\rep1$ as well. If $m^2 > 0$, the IR physics is not affected by $\lambda_5$: we still are left with $U(1)_0$ which gives an $S^1$ NLSM. If $m^2 < 0$, the minima of the potential depend on the sign of $\lambda_5$ (see Section~\ref{sec: potential}). Precisely
\be
m^2 \cO_\rep1 + \lambda_5 \cO_{(\rep5)}\,: \qquad \begin{cases} S^1 \text{ NLSM} & \text{for } m^2 > 0 \\ S^1 \text{ NLSM} & \text{for } m^2<0,\; \lambda_5>0 \\ \bZ_2 & \text{for } m^2<0,\; \lambda_5<0 \;. \end{cases}
\ee
Notice that the two $S^1$ NLSMs are acted upon by the two $O(2)$ factors in $O(2)_X \times O(2)_Y$, respectively.

Employing the Abelian dualities (\ref{Abelian dualities with background})-(\ref{Abelian dualities with background t-rev}) we can find dualities of $U(1)_0$ with 2 $\phi$ and $V_\text{EP}$, corresponding to $\lambda_5>0$. The reason is that the simple dualities produce a UV potential $V = |\phi_1|^4 + |\phi_2|^4$ corresponding to $\lambda_1 = \lambda_5 > 0$ (or $\lambda=-1$). Such a potential will run, however the RG flow will not cross the divider $\lambda_5 = 0$ and thus, assuming that a fixed point $\cT_+$ exists, the latter will lie somewhere at $\lambda_5>0$ ($\lambda<0$).

Combining (\ref{Abdu-a}) and (\ref{Abdurev-a}) we obtain the duality of Lagrangians
\begin{multline}
|D_{b+X}\phi_1|^2 + |D_b \phi_2|^2 - V\big( |\phi_1|^2, |\phi_2|^2 \big) + \frac1{2\pi} bdY \\
\longleftrightarrow\qquad |D_{c+Y} \sigma_1|^2 + |D_c \sigma_2|^2 - V\big( |\phi_1|^2, |\phi_2|^2 \big) + \frac1{2\pi} cdX \;.
\end{multline}
This is a \emph{self-duality} of $U(1)_0$ with 2 $\phi$ and $V_\text{EP}$ acting as $X \leftrightarrow Y$ on the background fields. Each side has $O(2)_X \times O(2)_Y \times \bZ_2^\cT$ symmetry, and the self-duality is an extra element $\bZ_2^\cD$ that exchanges the two $O(2)$ factors. The operator map includes
\bea
|\phi_1|^2 + |\phi_2|^2 \qquad&\longleftrightarrow\qquad - \big( |\sigma_1|^2 + |\sigma_2|^2 \big) \;,\qquad\qquad &
\phi_1\phi_2^* \qquad&\longleftrightarrow\qquad \cN_+ \\
|\phi_1|^2 - |\phi_2|^2 \qquad&\longleftrightarrow\qquad |\sigma_1|^2 - |\sigma_2|^2 \;,\qquad &
\phi_2\phi_1^* \qquad&\longleftrightarrow\qquad \cN_- \;.
\eea

On the other hand, combining (\ref{Abdu-b}) and (\ref{Abdurev-b}) we obtain the duality of Lagrangians
\begin{multline}
|D_{b+X}\phi_1|^2 + |D_b\phi_2|^2 - V\big( |\phi_1|^2, |\phi_2|^2 \big) + \frac1{2\pi} bdY \qquad\longleftrightarrow \\
i \bar\psi_1 \Dslash_a \psi_1 + i \bar\psi_2 \Dslash_{a+X-Y} \psi_2 + \frac1{4\pi} ada - \frac1{2\pi} adY - \frac1{2\pi} XdY + \frac1{4\pi} YdY + 2 \text{CS}_\text{g} \;.
\end{multline}
The theory on the RHS can be defined on non-spin manifolds promoting $a$ to be a spin$_c$ connection.
We have found the duality
\be
U(1)_0 \text{ with 2 $\phi$ and $V_\text{EP}$} \qquad\longleftrightarrow\qquad U(1)_0 \text{ with 2 $\psi$}
\ee
between scalar QED (with a symmetry-breaking potential $V_\text{EP}$) and fermionic QED. In turn, QED with two fermions has its own self-duality \cite{Xu:2015lxa, Hsin:2016blu, Benini:2017dus}. To make contact with \cite{Hsin:2016blu, Benini:2017dus} we perform the following transformation of the background gauge fields:%
\footnote{The transformation of gauge fields (\ref{background map}) may seem not invertible, however the symmetry group in the variables $\wt X, \wt Y$ is $\big( U(1)_{\wt X} \times U(1)_{\wt Y} \big)/\bZ_2$ and so the well-defined gauge fields are $2\wt X$, $2\wt Y$ and $\wt X-\wt Y$.}
\be
\label{background map}
X - Y = -2 \wt X \;,\qquad\qquad X+Y = - 2 \wt Y \;.
\ee
This gives
\be
i \bar\psi_1 \Dslash_a \psi_1 + i \bar\psi_2 \Dslash_{a-2\wt X} \psi_2 + \frac{ada}{4\pi} + \frac{ad(\wt Y-\wt X)}{2\pi} + \frac{(\wt X-\wt Y)d(\wt X-\wt Y)}{4\pi} - \frac{2\wt Yd\wt Y}{4\pi} + \frac{2\wt Xd\wt X}{4\pi} + 2 \text{CS}_\text{g}
\ee
which precisely agrees with (4.3) in \cite{Benini:2017dus} (tilded quantities are background fields there), except for the background counterterm $2\wt Xd\wt X/4\pi$. Such a term could be removed on both sides of the duality, however this would lead to a Lagrangian in which the background CS terms are not properly quantized for $\big(U(1)_{\wt X} \times U(1)_{\wt Y} \big)/\bZ_2$ and this is just a reflection of the underlying 't~Hooft anomaly.

We can compare massive deformations of the scalar QED with symmetry-breaking potential $V_\text{EP}$ and of fermionic QED, in the presence of background fields, and make contact with \cite{Benini:2017dus}. The operator map is
\bea
- |\phi_1|^2 + |\phi_2|^2 \qquad&\longleftrightarrow\qquad \bar\psi_1 \psi_1 + \bar\psi_2 \psi_2 \\
- \big( |\phi_1|^2 + |\phi_2|^2\big) \qquad&\longleftrightarrow\qquad \bar\psi_1 \psi_1 - \bar\psi_2 \psi_2 \;,
\eea
as it follows from the derivation in terms of Abelian dualities. Deforming (the potential of) scalar QED by $|\phi_1|^2 + |\phi_2|^2 = \cO_\rep1$, the scalars are massive and we are left with $\cL = \frac1{2\pi} bdY$, which is an $S^1$ NLSM from the breaking of $U(1)_Y$. This matches%
\footnote{Since the symmetry is broken, the counterterms associated to it are ambiguous. In the description in terms of a free gauge field $b$, this appears as the freedom to shift $b$ by background gauge fields.}
fermionic QED deformed by $- \bar\psi_1 \psi_1 + \bar\psi_2 \psi_2$. Deforming scalar QED by $-\cO_\rep1$ the scalars condense, breaking both the gauge symmetry and the flavor symmetry associated to $X$. This gives an $S^1$ NLSM, which can be described by a free photon $\tilde b$ as $\cL = \frac1{2\pi} \tilde b dX$. This matches fermionic QED deformed by $\bar\psi_1 \psi_1 - \bar\psi_2 \psi_2$. Deforming scalar QED by  $- |\phi_1|^2 + |\phi_2|^2 = - \cO_{(\rep3)}$, $\phi_1$ condenses while $\phi_2$ is massive. The gauge symmetry is broken, while the global symmetry is not (we have ``color-flavor locking''): setting $b+X = 0$ we are left with $\cL = - \frac1{2\pi} X dY$.
This matches fermionic QED deformed by $\bar\psi_1 \psi_1 + \bar\psi_2 \psi_2$. Deforming scalar QED by $\cO_{(\rep3)}$, instead, $\phi_2$ condenses and $\phi_1$ is massive: setting $b=0$ we are left with $\cL = 0$. This matches fermionic QED deformed by $-\bar\psi_1 \psi_1 - \bar\psi_2 \psi_2)$.

Fermionic QED has enhanced $O(4)$ symmetry (besides time reversal) at its fixed point \cite{Xu:2015lxa, Hsin:2016blu}, thus duality implies that also the scalar QED with symmetry-breaking potential $V_\text{EP}$ at the fixed point $\cT_+$ should have such an enhanced $O(4)$ symmetry. The manifest symmtry along the flow is $O(2)_X \times O(2)_Y$, embedded into $O(4)$ is the following way:
\be
O(4) \supset \mat{ O(2)_X & 0 \\ 0 & O(2)_Y} \;.
\ee
The self-duality of $\cT_+$ is the $\bZ_2^\cD$ that exchanges the two $O(2)$ factors, however to claim the full $O(4)$ symmetry we need to invoke the duality to fermionic QED.


\section{Conclusions}
\label{sec: conclusions}

In this paper we have proposed and analyzed new infinite families of IR dualities between Chern-Simons theories with both scalar and fermionic matter fields in the fundamental representation, for classical gauge groups. The theories have two relevant deformations invariant under all global symmetries---a mass for scalars and a mass for fermions---and we have studied the phase diagram as those masses are varied. We have found interesting (conjecturally) gapless lines, meeting at multi-critical fixed points.

Our analysis of the phase diagram was essentially classical (except for the fact that we used non-perturbative dualities to match the various phases), thus valid for large values of the masses compared with the scale set by the Yang-Mills regulator. In the range of parameters that we discussed, such an analysis has given results consistent with the dualities. For larger values of the numbers $N_s, N_f$ of matter fields, the phase diagrams do not seem to match and we could not claim that a duality exists. However, one could try to assume the existence of quantum phases, not visible classically, and give a consistent picture of the physics which is compatible with the dualities in a wider range of parameters, as done in \cite{Komargodski:2017keh, Gaiotto:2017tne, Gomis:2017ixy}.

When $N_s = N_f$ the matter content of the theories discussed in this paper becomes ``supersymmetric''. We do not have a gaugino, however the gaugino is massive in SUSY CS theories and could be integrated out. Yet, our theories are not supersymmetric because the interactions are not. For instance, the global symmetry contains two independent factors acting on the scalar and on the fermions, while there is only one factor acting on both in supersymmetric theories. Hence, it would be interesting to understand better the relations between the dualities discussed here and those of supersymmetric theories. For the cases with a single scalar and fermion, this has been considered in \cite{Jain:2013gza, Gur-Ari:2015pca, Kachru:2016aon}.


\section*{Acknowledgments} 

I am indebted to Ofer Aharony and Nathan Seiberg for uncountable discussions on this subject, as well as comments on the manuscript. I also thank Sergio Benvenuti, Clay Cordova, Po-Shen Hsin  and Shiraz Minwalla for helpful comments and suggestions. This work is supported in part by the MIUR-SIR grant RBSI1471GJ ``Quantum Field Theories at Strong Coupling: Exact Computations and Applications'', and by the IBM Einstein Fellowship at the Institute for Advanced Study.

\appendix


\section{Summary of other dualities}
\label{app: summary}

In this Appendix we summarize the level-rank dualities of spin-TQFTs and (some of) the dualities with a single matter species.

\subsection{Level-rank dualities}
\label{app: level-rank dualities}

The level-rank dualities of Chern-Simons spin-TQFTs that are relevant for our work are \cite{Naculich:1990pa, Mlawer:1990uv, Verstegen:1990at, Witten:1993xi, Naculich:2007nc, Hsin:2016blu, Aharony:2016jvv}:
\bea
\label{level-rank dualities}
SU(N)_k \times U(0)_1 \qquad&\longleftrightarrow\qquad U(k)_{-N} \times U(Nk)_1 \\[.2em]
U(N)_{k, k \pm N} \times U(0)_1 \qquad&\longleftrightarrow\qquad U(k)_{-N, -N \mp k} \times U(Nk \pm 1)_1 \\[.2em]
USp(2N)_k \times U(0)_1 \qquad&\longleftrightarrow\qquad USp(2k)_{-N} \times U(2Nk)_1 \\[.2em]
SO(N)_k \times SO(0)_1 \qquad&\longleftrightarrow\qquad SO(k)_{-N} \times SO(Nk)_1 \;.
\eea
In these dualities the second factor of each theory is a trivial spin-TQFT whose quantization on any Riemann surface gives a one-dimensional Hilbert space, and whose partition function on any spin Euclidean three-manifold is a phase represented by a classical Lagrangian \cite{Seiberg:2016rsg, Seiberg:2016gmd}. In particular
\be
U(N)_1 \;\leftrightarrow\; SO(2N)_1 \;\leftrightarrow\; \cL = -2N\text{CS}_g \qquad\text{and}\qquad SO(N)_1 \;\leftrightarrow\; \cL = -N\text{CS}_g \;.
\ee
We sometimes use the notation $U(-N)_1 \equiv U(N)_{-1}$ and $SO(-N)_1 \equiv SO(N)_{-1}$. The gravitational CS term is defined as $\int_{M = \partial X} \text{CS}_g = \frac1{192\pi} \int_X \Tr R \wedge R$. The trivial spin-TQFTs contain a transparent line with spin $1/2$, and their partition function depends on the spin structure of spacetime.%
\footnote{In the unitary case the dualities can be generalized to non-spin manifolds with the help of a spin$_c$ connection \cite{Seiberg:2016rsg, Seiberg:2016gmd}, but we will not do so in this paper.}
The trivial spin-TQFT factors thus remind us that the dualities are valid (in general) for spin theories, and represent gravitational counterterms that we will match across the dualities.

In the unitary case, we can couple the TQFTs to a background $U(1)$ gauge field $B$ and keep track of its counterterms. The first duality in (\ref{level-rank dualities}) is
written as \cite{Hsin:2016blu}
\begin{multline}
\label{SU/U level-rank Lagrangian}
\frac k{4\pi} \Tr_N \Big( bdb - \frac{2i}3 b^3\Big) + \frac1{2\pi} cd\big( B - \Tr_N b \big) \qquad\longleftrightarrow\qquad \\
-\frac N{4\pi} \Tr_k \Big( fdf - \frac{2i}3 f^3\Big) + \frac1{2\pi} (\Tr_k f) dB - 2Nk \, \text{CS}_g \;.
\end{multline}
Here $b, f, c$ are dynamical $U(N)$, $U(k)$ and $U(1)$ gauge fields, respectively. When $B=0$, the Lagrange multiplier $c$ forces $b$ to be an $SU(N)$ gauge field. We could perform field redefinitions $b \to - b^\sT$ and/or $f \to - f^\sT$, whose only effect is to change sign to $\Tr_N b$ and $\Tr_k f$, respectively. If we add $\frac1{2\pi} BdC$ on both sides and make $B$ dynamical, we obtain the time reversal of the same duality. If, instead, we add $\frac1{2\pi} BdC \pm \frac1{4\pi} BdB$ and make $B$ dynamical, we obtain the other two unitary level-rank dualities:
\begin{multline}
\frac k{4\pi} \Tr_N \Big( bdb - \frac{2i}3 b^3\Big) \pm \frac1{4\pi} (\Tr_N b)d(\Tr_Nb) + \frac1{2\pi} (\Tr_Nb)dC \qquad\longleftrightarrow\qquad \\
-\frac N{4\pi} \Tr_k \Big( fdf - \frac{2i}3 f^3\Big) \mp \frac1{4\pi} (\Tr_k f)d(\Tr_k f) + \frac1{2\pi} (\Tr_k f) dC \mp \frac1{4\pi} CdC - 2(Nk\pm1) \text{CS}_g \;.
\end{multline}
Here $b,f$ are dynamical $U(N)$ and $U(k)$ gauge fields, respectively, while we called $C$ the background $U(1)$ gauge field.

\subsection{Dualities with a single matter species}

The dualities with a single matter species are \cite{Aharony:2015mjs, Hsin:2016blu, Aharony:2016jvv}
\bea
\label{dualities one spieces}
SU(N)_k \text{ with $N_s$ $\phi$ } \times U(0)_1 \quad&\longleftrightarrow\quad U(k)_{-N + \frac{N_s}2} \text{ with $N_s$ $\psi$ } \times U\big( k(N-N_s) \big)_1 \\
U(N)_k \text{ with $N_s$ $\phi$ } \times U(0)_1 \quad&\longleftrightarrow\quad SU(k)_{-N + \frac{N_s}2} \text{ with $N_s$ $\psi$ } \times U\big( k(N-N_s) \big)_1 \\
USp(2N)_k \text{ with $N_s$ $\phi$ } \times U(0)_1 \quad&\longleftrightarrow\quad USp(2k)_{-N + \frac{N_s}2} \text{ with $N_s$ $\psi$ } \times U\big( 2k(N-N_s) \big)_1 \\
SO(N)_k \text{ with $N_s$ $\phi_\bR$ } \times SO(0)_1 \quad&\longleftrightarrow\quad SO(k)_{-N + \frac{N_s}2} \text{ with $N_s$ $\psi_\bR$ } \times SO\big( k(N-N_s) \big)_1
\eea
and
\begin{multline}
\label{dualities one spieces U/U}
U(N)_{k,k\pm N} \text{ with $N_s$ $\phi$ } \times U(0)_1 \quad\longleftrightarrow\quad \\
U(k)_{-N + \frac{N_s}2,\, -N \mp k + \frac{N_s}2} \text{ with $N_s$ $\psi$ } \times U\big( k(N-N_s) \pm 1 \big)_1 \;.
\end{multline}
In these dualities we assume $N_s \geq 1$. In the $SU/U$ duality we take $k\geq1$, $N_s \leq N$ as well as $k=0$, $N_s \leq N-1$. In the $U/SU$ duality we take $k\geq1$, $N_s \leq N$. In the $U/U$ duality we take $k\geq1$, $N_s \leq N$ as well as $k=0$, $N_s \leq N-1$. In the $USp$ duality we take $k\geq 0$, $N_s \leq N$. In the $SO$ duality we take $k=1$ and $N_s \leq N-2$, or $k=2$ and $N_s \leq N-1$, or $k\geq 3$ and $N_s \leq N$, or $k=N = N_s=2$ (this case is discussed in Section~\ref{sec: SO excluded}).%
\footnote{Dualities with a broader range of parameters have been discussed in \cite{Komargodski:2017keh}.}

We can also consider the duality
\be
U(N)_{0,N} \text{ with $N$ $\phi$ } \times U(0)_1 \qquad\longleftrightarrow\qquad \psi \;.
\ee
The theory on the left has $U(N)$ global symmetry, however the duality predicts that only $U(1) \subset U(N)$ acts on the low-energy theory on the right. The map of mass terms is $|\phi|^2 \leftrightarrow - \bar\psi\psi$. The time reversal of that duality is
\be
U(N)_{0,-N} \text{ with $N$ $\phi$} \times U(0)_1 \qquad\longleftrightarrow\qquad \psi \times U(1)_{-1} \;,
\ee
and the map of mass terms is $|\phi|^2 \leftrightarrow \bar\psi\psi$.

In the unitary case, let us write the dualities in Lagrangian form with a $U(1)$ background field $B$ coupled to the baryonic or the topological symmetry. The first duality, namely $SU(N)_k$ with $N_s$ $\phi$ $\times\; U(0)_1 \leftrightarrow U(k)_{-N + \frac{N_s}2}$ with $N_s$ $\psi$ $\times\; U\big( k(N-N_s)\big)_1$ reads \cite{Hsin:2016blu}
\begin{multline}
\label{SU/U duality 1 species background}
|D_b\phi|^2 - |\phi|^4 + \frac{k}{4\pi} \Tr_N \Big( bdb - \frac{2i}3 b^3 \Big) + \frac1{2\pi} cd\big( B -  \Tr_N b \big) \quad\longleftrightarrow \\
i \bar\psi \Dslash_f \psi - \frac{N-N_s}{4\pi} \Tr_k \Big( fdf - \frac{2i}3 f^3 \Big) + \frac1{2\pi} (\Tr_k f) dB - 2(N-N_s) k \, \text{CS}_g \;,
\end{multline}
where $b,f$ are dynamical $U(N)$ and $U(k)$ gauge fields, respectively. Because of the Lagrange multiplier $c$, on the LHS the baryonic operators $\phi^N$ are coupled to $B$ with charge $+1$. It is easy to check that deforming the two sides of (\ref{SU/U duality 1 species background}) with $m_\phi^2>0$ and $m_\psi<0$, or with $m_\phi^2<0$ and $m_\psi>0$, one reproduces the level-rank dualities (\ref{SU/U level-rank Lagrangian}).

We can add $\frac1{2\pi} BdC$ on both sides of (\ref{SU/U duality 1 species background}) and make $B$ dynamical. We obtain%
\footnote{On the RHS we should also redefine $f \to - f^\sT$ as well as $\psi \to \psi^c$, \ie{} use complex conjugate fields.}
\begin{multline}
|D_b\phi|^2 - |\phi|^4 + \frac{k}{4\pi} \Tr_N \Big( bdb - \frac{2i}3 b^3 \Big) + \frac1{2\pi} (\Tr_N b)dC \quad\longleftrightarrow \\
i \bar\psi \Dslash_f \psi - \frac{N-N_s}{4\pi} \Tr_k \Big( fdf - \frac{2i}3 f^3 \Big) + \frac1{2\pi} ad\big(C -  \Tr_k f \big) - 2(N-N_s) k \, \text{CS}_g \;.
\end{multline}
This is the duality $U(N)_k$ with $N_s$ $\phi$ $\times\; U(0)_1 \leftrightarrow SU(k)_{-N+\frac{N_s}2}$ with $N_s$ $\psi$ $\times\; U\big( k(N-N_s)\big)_1$. We can write its time-reversed version, which after some redefinitions reads
\begin{multline}
\label{SU/U duality 1 species background Trev}
i \bar\psi \Dslash_b \psi + \frac{k}{4\pi} \Tr_N \Big( bdb - \frac{2i}3 b^3 \Big) + \frac1{2\pi} c d\big( B - \Tr_N b \big)  \quad\longleftrightarrow \\
|D_f\phi|^2 - |\phi|^4 - \frac{N}{4\pi} \Tr_k \Big( fdf - \frac{2i}3 f^3 \Big)  + \frac1{2\pi} (\Tr_k f)dB - 2N k \, \text{CS}_g \;.
\end{multline}
This is $SU(N)_{k - \frac{N_f}2}$ with $N_f$ $\psi \leftrightarrow U(k)_{-N}$ with $N_f$ $\phi$ $\times\; U(Nk)_1$.

Finally, starting from (\ref{SU/U duality 1 species background}), adding $\frac1{2\pi} BdC \pm \frac1{4\pi} CdC$ to both sides and making $B$ dynamical, we obtain
\begin{multline}
|D_b\phi|^2 - |\phi|^4 + \frac{k}{4\pi} \Tr_N \Big( bdb - \frac{2i}3 b^3 \Big) \pm \frac1{4\pi} (\Tr_N b)d(\Tr_Nb) + \frac1{2\pi} (\Tr_Nb)dC \quad\longleftrightarrow \\
i \bar\psi \Dslash_f \psi - \frac{N-N_s}{4\pi} \Tr_k \Big( fdf - \frac{2i}3 f^3 \Big) \mp \frac1{4\pi} (\Tr_k f)d(\Tr_k f) + \frac1{2\pi} (\Tr_k f) dC \\
\mp \frac1{4\pi} CdC - 2\big(k(N-N_s)\pm1\big) \text{CS}_g \;,
\end{multline}
which is the duality in (\ref{dualities one spieces U/U}).

\section{Charge conjugation}
\label{app: charge conjugation}

Given a representation $T^a$ satisfying the algebra $[T^a, T^b] = i f^{abc} T^c$, the representation $-(T^a)^\sT$ satisfies the same algebra and is the conjugate representation. Therefore for gauge group $U(N)$ or $SU(N)$ we take the action of $\bZ_2^\cC$ charge conjugation on gauge fields to be
\be
\cC: A_\mu \to - A_\mu^\sT \;.
\ee
The CS terms $\Tr(AdA)$ and $\Tr(A^3)$ are invariant.

We take the action on scalar fields to be complex conjugation,
\be
\cC: \phi \to \phi^* \;.
\ee
As the matter representation is unitary, $A_\mu = A_\mu^\dag$, the scalar kinetic term is invariant. The quadratic gauge invariants
\be
\cM\du{I}{J} = \phi^\dag_I \phi^J
\ee
get transposed under $\cC$, therefore the mass term $\cM\du{I}{I} = \phi^\dag_I \phi^I$, as well as the quartic couplings $(\cM\du{I}{I})^2$ and $\cM\du{I}{J} \cM\du{J}{I}$, are invariant.

To define charge conjugation of fermions we need the charge conjugation matrix $C$ such that%
\footnote{In 3D with Lorentzian signature we can choose $\gamma_0 = \smat{ i & 0 \\ 0 & -i}$, $\gamma_1 = \smat{0 & 1 \\ 1 & 0}$, $\gamma_2 = \smat{0 & -i \\ i & 0}$, and then $C = \gamma_2$. This gives $C = C^\dag = C^{-1} = - C^*$ and $C^*C = -1$.}
$C^{-1} \gamma_\mu C = - \gamma_\mu^\sT$. Then
\be
\cC: \psi \to \psi^c \equiv C \wb\psi^\sT = C \gamma_0^\sT \psi^* \;, \qquad\qquad \wb\psi \to \wb\psi^c = - \psi^\sT C^{-1} \;,
\ee
where recall that $\wb\psi = \psi^\dag \gamma_0$. In the second one we used $C^\dag C = \gamma_0^\dag \gamma_0^{\phantom{\dag}} = \unit$, as they are both positive matrices. It follows that the kinetic term $i\wb\psi \Dslash \psi$ is invariant under $\cC$ up to a total derivative (using that fermions anticommute). The quadratic gauge invariants
\be
\cN\du{A}{B} = \wb \psi_A \psi^B
\ee
get transposed under $\cC$, therefore the mass term $\cN\du{A}{A} = \wb\psi_A \psi^A$ is invariant.

Finally, one easily checks that also the mixed term
\be
\phi_{\alpha I}^* \phi^{\beta I} \wb\psi_{\beta A} \psi^{\alpha A}
\ee
is invariant under $\cC$.

\section{Other notations}
\label{app: spinc}

The following terms are well-defined on a non-spin manifold, provided $B,C$ are standard connections while $A$ is a spin$_c$ connection \cite{Seiberg:2016rsg}:
\be
\frac1{2\pi} BdC \;,\qquad \frac1{4\pi} BdB + \frac1{2\pi} BdA \;,\qquad \frac1{4\pi} AdA + 2 \text{CS}_\text{g} \;,\qquad 16 \text{CS}_\text{g} \;.
\ee
In the first term, $B,C$ could be the same connection.

Because of our choice of regularization, integrating out a (complex) fermion with positive or negative mass gives
\be
\psi \;,\; \cL = i \bar\psi \Dslash_A \psi \qquad \stackrel{m_\psi}{\longrightarrow}\qquad \begin{cases} U(0)_1 \;,\; \cL=0 & m_\psi > 0 \\ U(1)_1 \;,\; \cL = - \frac1{4\pi} AdA - 2\text{CS}_g \quad & m_\psi<0 \end{cases}
\ee
where $A$ is a non-dynamical background field (a spin$_c$ connection on non-spin manifolds). Again because of our choice of regularization, the time reversal of a fermion is
\be
i \bar\psi \Dslash_A \psi \qquad\stackrel{T}{\longleftrightarrow}\qquad i \bar\psi \Dslash_A \psi + \frac1{4\pi} AdA + 2 \text{CS}_g \;.
\ee
The appearance of the term $\frac1{4\pi} AdA$ is because on the RHS there is $U(1)_{-\frac12}$ coupled to $A$ and the level should be inverted. The appearance of the gravitational coupling is necessary on non-spin manifolds. We can check that this is consistent with mass deformations, recalling that the fermion mass term is odd under parity:
\be
\begin{array}{rc|rc}
m_\psi>0: & \varnothing & m_\psi<0: & \varnothing \\
m_\psi<0: & - \frac1{4\pi} AdA - 2 \text{CS}_g & m_\psi>0: & \frac1{4\pi} AdA + 2\text{CS}_g
\end{array}
\ee
In our concise notation we write
\be
\psi \qquad\stackrel{T}{\longleftrightarrow}\qquad \psi \times U(1)_{-1} \;.
\ee


\bibliographystyle{ytphys}
\bibliography{Non_SUSY_dualities}

\providecommand{\href}[2]{#2}\begingroup\raggedright\begin{thebibliography}{10}

\bibitem{Peskin:1977kp}
M.~E. Peskin, ``{Mandelstam 't Hooft Duality in Abelian Lattice Models},''
\href{http://dx.doi.org/10.1016/0003-4916(78)90252-X}{{\em Annals Phys.}
  {\bfseries 113} (1978) 122}.

\bibitem{Dasgupta:1981zz}
C.~Dasgupta and B.~I. Halperin, ``{Phase Transition in a Lattice Model of
  Superconductivity},''
\href{http://dx.doi.org/10.1103/PhysRevLett.47.1556}{{\em Phys. Rev. Lett.}
  {\bfseries 47} (1981) 1556--1560}.

\bibitem{Barkeshli:2014ida}
M.~Barkeshli and J.~McGreevy, ``{Continuous transition between fractional
  quantum Hall and superfluid states},''
  \href{http://dx.doi.org/10.1103/PhysRevB.89.235116}{{\em Phys. Rev.}
  {\bfseries B89} (2014) 235116},
\href{http://arxiv.org/abs/1201.4393}{{\ttfamily arXiv:1201.4393
  [cond-mat.str-el]}}.

\bibitem{Son:2015xqa}
D.~T. Son, ``{Is the Composite Fermion a Dirac Particle?},''
  \href{http://dx.doi.org/10.1103/PhysRevX.5.031027}{{\em Phys. Rev.}
  {\bfseries X5} (2015) 031027},
\href{http://arxiv.org/abs/1502.03446}{{\ttfamily arXiv:1502.03446
  [cond-mat.mes-hall]}}.

\bibitem{Wang:2015qmt}
C.~Wang and T.~Senthil, ``{Dual Dirac Liquid on the Surface of the Electron
  Topological Insulator},''
  \href{http://dx.doi.org/10.1103/PhysRevX.5.041031}{{\em Phys. Rev.}
  {\bfseries X5} (2015) 041031},
\href{http://arxiv.org/abs/1505.05141}{{\ttfamily arXiv:1505.05141
  [cond-mat.str-el]}}.

\bibitem{Potter:2015cdn}
A.~C. Potter, M.~Serbyn, and A.~Vishwanath, ``{Thermoelectric transport
  signatures of Dirac composite fermions in the half-filled Landau level},''
  \href{http://dx.doi.org/10.1103/PhysRevX.6.031026}{{\em Phys. Rev.}
  {\bfseries X6} (2016) 031026},
\href{http://arxiv.org/abs/1512.06852}{{\ttfamily arXiv:1512.06852
  [cond-mat.str-el]}}.

\bibitem{Wang:2016gqj}
C.~Wang and T.~Senthil, ``{Composite fermi liquids in the lowest Landau
  level},'' \href{http://dx.doi.org/10.1103/PhysRevB.94.245107}{{\em Phys.
  Rev.} {\bfseries B94} (2016) 245107},
\href{http://arxiv.org/abs/1604.06807}{{\ttfamily arXiv:1604.06807
  [cond-mat.str-el]}}.

\bibitem{Aharony:2011jz}
O.~Aharony, G.~Gur-Ari, and R.~Yacoby, ``{$d{=}3$ Bosonic Vector Models Coupled
  to Chern-Simons Gauge Theories},''
  \href{http://dx.doi.org/10.1007/JHEP03(2012)037}{{\em JHEP} {\bfseries 03}
  (2012) 037},
\href{http://arxiv.org/abs/1110.4382}{{\ttfamily arXiv:1110.4382 [hep-th]}}.

\bibitem{Giombi:2011kc}
S.~Giombi, S.~Minwalla, S.~Prakash, S.~P. Trivedi, S.~R. Wadia, and X.~Yin,
  ``{Chern-Simons Theory with Vector Fermion Matter},''
  \href{http://dx.doi.org/10.1140/epjc/s10052-012-2112-0}{{\em Eur. Phys. J.}
  {\bfseries C72} (2012) 2112},
\href{http://arxiv.org/abs/1110.4386}{{\ttfamily arXiv:1110.4386 [hep-th]}}.

\bibitem{Aharony:2012nh}
O.~Aharony, G.~Gur-Ari, and R.~Yacoby, ``{Correlation Functions of Large $N$
  Chern-Simons-Matter Theories and Bosonization in Three Dimensions},''
  \href{http://dx.doi.org/10.1007/JHEP12(2012)028}{{\em JHEP} {\bfseries 12}
  (2012) 028},
\href{http://arxiv.org/abs/1207.4593}{{\ttfamily arXiv:1207.4593 [hep-th]}}.

\bibitem{Jain:2013gza}
S.~Jain, S.~Minwalla, and S.~Yokoyama, ``{Chern Simons duality with a
  fundamental boson and fermion},''
  \href{http://dx.doi.org/10.1007/JHEP11(2013)037}{{\em JHEP} {\bfseries 11}
  (2013) 037},
\href{http://arxiv.org/abs/1305.7235}{{\ttfamily arXiv:1305.7235 [hep-th]}}.

\bibitem{Intriligator:1996ex}
K.~A. Intriligator and N.~Seiberg, ``{Mirror symmetry in three-dimensional
  gauge theories},'' \href{http://dx.doi.org/10.1016/0370-2693(96)01088-X}{{\em
  Phys. Lett.} {\bfseries B387} (1996) 513--519},
\href{http://arxiv.org/abs/hep-th/9607207}{{\ttfamily arXiv:hep-th/9607207
  [hep-th]}}.

\bibitem{deBoer:1996mp}
J.~de~Boer, K.~Hori, H.~Ooguri, and Y.~Oz, ``{Mirror symmetry in
  three-dimensional gauge theories, quivers and D-branes},''
  \href{http://dx.doi.org/10.1016/S0550-3213(97)00125-9}{{\em Nucl. Phys.}
  {\bfseries B493} (1997) 101--147},
\href{http://arxiv.org/abs/hep-th/9611063}{{\ttfamily arXiv:hep-th/9611063
  [hep-th]}}.

\bibitem{Aharony:1997bx}
O.~Aharony, A.~Hanany, K.~A. Intriligator, N.~Seiberg, and M.~J. Strassler,
  ``{Aspects of $\mathcal{N}{=}2$ supersymmetric gauge theories in
  three-dimensions},''
  \href{http://dx.doi.org/10.1016/S0550-3213(97)00323-4}{{\em Nucl. Phys.}
  {\bfseries B499} (1997) 67--99},
\href{http://arxiv.org/abs/hep-th/9703110}{{\ttfamily arXiv:hep-th/9703110
  [hep-th]}}.

\bibitem{Aharony:1997gp}
O.~Aharony, ``{IR duality in $d{=}3$ $\mathcal{N}{=}2$ supersymmetric
  $USp(2N_c)$ and $U(N_c)$ gauge theories},''
  \href{http://dx.doi.org/10.1016/S0370-2693(97)00530-3}{{\em Phys. Lett.}
  {\bfseries B404} (1997) 71--76},
\href{http://arxiv.org/abs/hep-th/9703215}{{\ttfamily arXiv:hep-th/9703215
  [hep-th]}}.

\bibitem{Giveon:2008zn}
A.~Giveon and D.~Kutasov, ``{Seiberg Duality in Chern-Simons Theory},''
  \href{http://dx.doi.org/10.1016/j.nuclphysb.2008.09.045}{{\em Nucl. Phys.}
  {\bfseries B812} (2009) 1--11},
\href{http://arxiv.org/abs/0808.0360}{{\ttfamily arXiv:0808.0360 [hep-th]}}.

\bibitem{Benini:2011mf}
F.~Benini, C.~Closset, and S.~Cremonesi, ``{Comments on 3d Seiberg-like
  dualities},'' \href{http://dx.doi.org/10.1007/JHEP10(2011)075}{{\em JHEP}
  {\bfseries 10} (2011) 075},
\href{http://arxiv.org/abs/1108.5373}{{\ttfamily arXiv:1108.5373 [hep-th]}}.

\bibitem{Karch:2016sxi}
A.~Karch and D.~Tong, ``{Particle-Vortex Duality from 3d Bosonization},''
  \href{http://dx.doi.org/10.1103/PhysRevX.6.031043}{{\em Phys. Rev.}
  {\bfseries X6} (2016) 031043},
\href{http://arxiv.org/abs/1606.01893}{{\ttfamily arXiv:1606.01893 [hep-th]}}.

\bibitem{Murugan:2016zal}
J.~Murugan and H.~Nastase, ``{Particle-vortex duality in topological insulators
  and superconductors},'' \href{http://dx.doi.org/10.1007/JHEP05(2017)159}{{\em
  JHEP} {\bfseries 05} (2017) 159},
\href{http://arxiv.org/abs/1606.01912}{{\ttfamily arXiv:1606.01912 [hep-th]}}.

\bibitem{Seiberg:2016gmd}
N.~Seiberg, T.~Senthil, C.~Wang, and E.~Witten, ``{A Duality Web in 2+1
  Dimensions and Condensed Matter Physics},''
  \href{http://dx.doi.org/10.1016/j.aop.2016.08.007}{{\em Annals Phys.}
  {\bfseries 374} (2016) 395--433},
\href{http://arxiv.org/abs/1606.01989}{{\ttfamily arXiv:1606.01989 [hep-th]}}.

\bibitem{Aharony:2015mjs}
O.~Aharony, ``{Baryons, monopoles and dualities in Chern-Simons-matter
  theories},'' \href{http://dx.doi.org/10.1007/JHEP02(2016)093}{{\em JHEP}
  {\bfseries 02} (2016) 093},
\href{http://arxiv.org/abs/1512.00161}{{\ttfamily arXiv:1512.00161 [hep-th]}}.

\bibitem{Hsin:2016blu}
P.-S. Hsin and N.~Seiberg, ``{Level/rank Duality and Chern-Simons-Matter
  Theories},'' \href{http://dx.doi.org/10.1007/JHEP09(2016)095}{{\em JHEP}
  {\bfseries 09} (2016) 095},
\href{http://arxiv.org/abs/1607.07457}{{\ttfamily arXiv:1607.07457 [hep-th]}}.

\bibitem{Metlitski:2016dht}
M.~A. Metlitski, A.~Vishwanath, and C.~Xu, ``{Duality and bosonization of
  (2+1)-dimensional Majorana fermions},''
  \href{http://dx.doi.org/10.1103/PhysRevB.95.205137}{{\em Phys. Rev.}
  {\bfseries B95} (2017) 205137},
\href{http://arxiv.org/abs/1611.05049}{{\ttfamily arXiv:1611.05049
  [cond-mat.str-el]}}.

\bibitem{Aharony:2016jvv}
O.~Aharony, F.~Benini, P.-S. Hsin, and N.~Seiberg, ``{Chern-Simons-matter
  dualities with $SO$ and $USp$ gauge groups},''
  \href{http://dx.doi.org/10.1007/JHEP02(2017)072}{{\em JHEP} {\bfseries 02}
  (2017) 072},
\href{http://arxiv.org/abs/1611.07874}{{\ttfamily arXiv:1611.07874
  [cond-mat.str-el]}}.

\bibitem{Komargodski:2017keh}
Z.~Komargodski and N.~Seiberg, ``{A Symmetry Breaking Scenario for QCD$_3$},''
\href{http://arxiv.org/abs/1706.08755}{{\ttfamily arXiv:1706.08755 [hep-th]}}.

\bibitem{Karch:2016aux}
A.~Karch, B.~Robinson, and D.~Tong, ``{More Abelian Dualities in 2+1
  Dimensions},'' \href{http://dx.doi.org/10.1007/JHEP01(2017)017}{{\em JHEP}
  {\bfseries 01} (2017) 017},
\href{http://arxiv.org/abs/1609.04012}{{\ttfamily arXiv:1609.04012 [hep-th]}}.

\bibitem{Jensen:2017dso}
K.~Jensen and A.~Karch, ``{Bosonizing three-dimensional quiver gauge
  theories},'' {\em JHEP} {\bfseries 11} (2017) 018,
\href{http://arxiv.org/abs/1709.01083}{{\ttfamily arXiv:1709.01083 [hep-th]}}.

\bibitem{Gomis:2017ixy}
J.~Gomis, Z.~Komargodski, and N.~Seiberg, ``{Phases Of Adjoint QCD$_3$ And
  Dualities},''
\href{http://arxiv.org/abs/1710.03258}{{\ttfamily arXiv:1710.03258 [hep-th]}}.

\bibitem{Gur-Ari:2015pca}
G.~Gur-Ari and R.~Yacoby, ``{Three Dimensional Bosonization From
  Supersymmetry},'' \href{http://dx.doi.org/10.1007/JHEP11(2015)013}{{\em JHEP}
  {\bfseries 11} (2015) 013},
\href{http://arxiv.org/abs/1507.04378}{{\ttfamily arXiv:1507.04378 [hep-th]}}.

\bibitem{Kachru:2016aon}
S.~Kachru, M.~Mulligan, G.~Torroba, and H.~Wang, ``{Nonsupersymmetric dualities
  from mirror symmetry},''
  \href{http://dx.doi.org/10.1103/PhysRevLett.118.011602}{{\em Phys. Rev.
  Lett.} {\bfseries 118} (2017) 011602},
\href{http://arxiv.org/abs/1609.02149}{{\ttfamily arXiv:1609.02149 [hep-th]}}.

\bibitem{Benini:2017dus}
F.~Benini, P.-S. Hsin, and N.~Seiberg, ``{Comments on global symmetries,
  anomalies, and duality in (2+1)d},''
  \href{http://dx.doi.org/10.1007/JHEP04(2017)135}{{\em JHEP} {\bfseries 04}
  (2017) 135},
\href{http://arxiv.org/abs/1702.07035}{{\ttfamily arXiv:1702.07035
  [cond-mat.str-el]}}.

\bibitem{Wang:2017txt}
C.~Wang, A.~Nahum, M.~A. Metlitski, C.~Xu, and T.~Senthil, ``{Deconfined
  quantum critical points: symmetries and dualities},''
  \href{http://dx.doi.org/10.1103/PhysRevX.7.031051}{{\em Phys. Rev.}
  {\bfseries X7} (2017) 031051},
\href{http://arxiv.org/abs/1703.02426}{{\ttfamily arXiv:1703.02426
  [cond-mat.str-el]}}.

\bibitem{Chen:2017lkr}
J.-Y. Chen, J.~H. Son, C.~Wang, and S.~Raghu, ``{Exact Boson-Fermion Duality on
  a 3D Euclidean Lattice},''
  \href{http://dx.doi.org/10.1103/PhysRevLett.120.016602}{{\em Phys. Rev.
  Lett.} {\bfseries 120} (2018) 016602},
\href{http://arxiv.org/abs/1705.05841}{{\ttfamily arXiv:1705.05841
  [cond-mat.str-el]}}.

\bibitem{Gaiotto:2017tne}
D.~Gaiotto, Z.~Komargodski, and N.~Seiberg, ``{Time-Reversal Breaking in
  QCD$_4$, Walls, and Dualities in 2+1 Dimensions},''
\href{http://arxiv.org/abs/1708.06806}{{\ttfamily arXiv:1708.06806 [hep-th]}}.

\bibitem{Jensen:2017xbs}
K.~Jensen and A.~Karch, ``{Embedding three-dimensional bosonization dualities
  into string theory},'' \href{http://dx.doi.org/10.1007/JHEP12(2017)031}{{\em
  JHEP} {\bfseries 12} (2017) 031},
\href{http://arxiv.org/abs/1709.07872}{{\ttfamily arXiv:1709.07872 [hep-th]}}.

\bibitem{Armoni:2017jkl}
A.~Armoni and V.~Niarchos, ``{Phases of QCD$_3$ from Non-SUSY Seiberg Duality
  and Brane Dynamics},''
\href{http://arxiv.org/abs/1711.04832}{{\ttfamily arXiv:1711.04832 [hep-th]}}.

\bibitem{Cordova:2017vab}
C.~Cordova, P.-S. Hsin, and N.~Seiberg, ``{Global Symmetries, Counterterms, and
  Duality in Chern-Simons Matter Theories with Orthogonal Gauge Groups},''
\href{http://arxiv.org/abs/1711.10008}{{\ttfamily arXiv:1711.10008 [hep-th]}}.

\bibitem{Aharony:toappear}
O.~Aharony, S.~Jain, and S.~Minwalla, ``{Flows, Fixed Points and Duality in
  Chern-Simons-Matter Theories}.'' To appear.

\bibitem{Seiberg:2016rsg}
N.~Seiberg and E.~Witten, ``{Gapped Boundary Phases of Topological Insulators
  via Weak Coupling},'' \href{http://dx.doi.org/10.1093/ptep/ptw083}{{\em PTEP}
  {\bfseries 2016} no.~12, (2016) 12C101},
\href{http://arxiv.org/abs/1602.04251}{{\ttfamily arXiv:1602.04251
  [cond-mat.str-el]}}.

\bibitem{Witten:2015aba}
E.~Witten, ``{Fermion Path Integrals And Topological Phases},''
  \href{http://dx.doi.org/10.1103/RevModPhys.88.035001}{{\em Rev. Mod. Phys.}
  {\bfseries 88} (2016) 035001},
\href{http://arxiv.org/abs/1508.04715}{{\ttfamily arXiv:1508.04715
  [cond-mat.mes-hall]}}.

\bibitem{Motrunich:2003fz}
O.~I. Motrunich and A.~Vishwanath, ``{Emergent photons and new transitions in
  the $O(3)$ sigma model with hedgehog suppression},''
  \href{http://dx.doi.org/10.1103/PhysRevB.70.075104}{{\em Phys. Rev.}
  {\bfseries B70} (2004) 075104},
\href{http://arxiv.org/abs/cond-mat/0311222}{{\ttfamily arXiv:cond-mat/0311222
  [cond-mat]}}.

\bibitem{Xu:2015lxa}
C.~Xu and Y.-Z. You, ``{Self-dual Quantum Electrodynamics as Boundary State of
  the three dimensional Bosonic Topological Insulator},''
  \href{http://dx.doi.org/10.1103/PhysRevB.92.220416}{{\em Phys. Rev.}
  {\bfseries B92} no.~22, (2015) 220416},
\href{http://arxiv.org/abs/1510.06032}{{\ttfamily arXiv:1510.06032
  [cond-mat.str-el]}}.

\bibitem{Radicevic:2015yla}
{\DJ}.~Radi\v{c}evi\'c, ``{Disorder Operators in Chern-Simons-Fermion
  Theories},'' \href{http://dx.doi.org/10.1007/JHEP03(2016)131}{{\em JHEP}
  {\bfseries 03} (2016) 131},
\href{http://arxiv.org/abs/1511.01902}{{\ttfamily arXiv:1511.01902 [hep-th]}}.

\bibitem{Wu:1976ge}
T.~T. Wu and C.~N. Yang, ``{Dirac Monopole Without Strings: Monopole
  Harmonics},''
\href{http://dx.doi.org/10.1016/0550-3213(76)90143-7}{{\em Nucl. Phys.}
  {\bfseries B107} (1976) 365}.

\bibitem{Shenker:2011zf}
S.~H. Shenker and X.~Yin, ``{Vector Models in the Singlet Sector at Finite
  Temperature},''
\href{http://arxiv.org/abs/1109.3519}{{\ttfamily arXiv:1109.3519 [hep-th]}}.

\bibitem{Barkeshli:2017rzd}
M.~Barkeshli and M.~Cheng, ``{Time-reversal and spatial reflection symmetry
  localization anomalies in (2+1)D topological phases of matter},''
\href{http://arxiv.org/abs/1706.09464}{{\ttfamily arXiv:1706.09464
  [cond-mat.str-el]}}.

\bibitem{Aharony:2013kma}
O.~Aharony, S.~S. Razamat, N.~Seiberg, and B.~Willett, ``{3$d$ dualities from
  4$d$ dualities for orthogonal groups},''
  \href{http://dx.doi.org/10.1007/JHEP08(2013)099}{{\em JHEP} {\bfseries 08}
  (2013) 099},
\href{http://arxiv.org/abs/1307.0511}{{\ttfamily arXiv:1307.0511 [hep-th]}}.

\bibitem{Naculich:1990pa}
S.~G. Naculich, H.~A. Riggs, and H.~J. Schnitzer, ``{Group Level Duality in WZW
  Models and Chern-Simons Theory},''
\href{http://dx.doi.org/10.1016/0370-2693(90)90623-E}{{\em Phys. Lett.}
  {\bfseries B246} (1990) 417--422}.

\bibitem{Mlawer:1990uv}
E.~J. Mlawer, S.~G. Naculich, H.~A. Riggs, and H.~J. Schnitzer, ``{Group level
  duality of WZW fusion coefficients and Chern-Simons link observables},''
\href{http://dx.doi.org/10.1016/0550-3213(91)90110-J}{{\em Nucl. Phys.}
  {\bfseries B352} (1991) 863--896}.

\bibitem{Verstegen:1990at}
D.~Verstegen, ``{Conformal embeddings, rank level duality and exceptional
  modular invariants},''
\href{http://dx.doi.org/10.1007/BF02100278}{{\em Commun. Math. Phys.}
  {\bfseries 137} (1991) 567--586}.

\bibitem{Witten:1993xi}
E.~Witten, ``{The Verlinde algebra and the cohomology of the Grassmannian},''
\href{http://arxiv.org/abs/hep-th/9312104}{{\ttfamily arXiv:hep-th/9312104
  [hep-th]}}.

\bibitem{Naculich:2007nc}
S.~G. Naculich and H.~J. Schnitzer, ``{Level-rank duality of the U(N) WZW
  model, Chern-Simons theory, and 2-D qYM theory},''
  \href{http://dx.doi.org/10.1088/1126-6708/2007/06/023}{{\em JHEP} {\bfseries
  06} (2007) 023},
\href{http://arxiv.org/abs/hep-th/0703089}{{\ttfamily arXiv:hep-th/0703089
  [HEP-TH]}}.

\end{thebibliography}\endgroup
\end{document}